%% file: hcv.tex
\documentclass{aa}  

\usepackage{graphicx}
\usepackage{txfonts}
\usepackage{numprint}
\usepackage{amsmath}
\usepackage{lscape}
\usepackage{xcolor}
\usepackage{enumitem}

\newcommand\HST{\emph{HST}}
\newcommand\Gaia{\emph{Gaia}}
\newcommand\SExtractor{\texttt{SExtractor}}

\begin{document}

   \title{The Hubble Catalog of Variables (HCV)\thanks{Full Tables 9 and 10 are only available at the CDS via anonymous ftp to cdsarc.u-strasbg.fr (130.79.128.5) or via http://cdsarc.u-strasbg.fr/viz-bin/qcat?J/A+A/vol/page}}

   \author{A.~Z.~Bonanos\inst{1}
        \and M.~Yang\inst{1}
        \and K.~V.~Sokolovsky\inst{1,2,3}
        \and P.~Gavras\inst{4,1}
        \and D.~Hatzidimitriou\inst{1,5}
        \and I.~Bellas-Velidis\inst{1}
        \and G.~Kakaletris\inst{6}
        \and D.~J.~Lennon\inst{7,8}
        \and A.~Nota\inst{9}
        \and R.~L.~White\inst{9}
        \and B.~C.~Whitmore\inst{9} \and \\
         K.~A.~Anastasiou\inst{5}
        \and M.~Ar\'evalo\inst{4}
        \and C.~Arviset\inst{8}
        \and D.~Baines\inst{10}
        \and T.~Budavari\inst{11}
        \and V.~Charmandaris\inst{12,13,1}
         \and C.~Chatzichristodoulou\inst{5}
         \and E.~Dimas\inst{5}
         \and J.~Dur\'an\inst{4}
        \and I.~Georgantopoulos\inst{1}
        \and A.~Karampelas\inst{14,1}
       \and N.~Laskaris\inst{15,6}
        \and S.~Lianou\inst{1}
         \and A.~Livanis\inst{5}
        \and S.~Lubow\inst{9}
         \and  G.~Manouras\inst{5}
        \and M.~I.~Moretti\inst{16,1}
        \and E.~Paraskeva\inst{1,5}
        \and E.~Pouliasis\inst{1,5}
        \and A.~Rest\inst{9,11}
        \and J.~Salgado\inst{10}
        \and P.~Sonnentrucker\inst{9}
        \and Z.~T.~Spetsieri\inst{1,5}
        \and P.~Taylor\inst{9}
        \and K.~Tsinganos\inst{5,1}
          }

   \institute{IAASARS, National Observatory of Athens, Penteli 15236, Greece\\
              \email{bonanos@astro.noa.gr}
          \and
             Department of Physics and Astronomy, Michigan State University, East Lansing, MI 48824, USA
          \and
             Sternberg Astronomical Institute, Moscow State University, Universitetskii~pr.~13, 119992~Moscow, Russia
          \and  
            RHEA Group for ESA-ESAC, Villanueva de la Ca\~nada, 28692 Madrid, Spain
          \and
             Department of Physics, National and Kapodistrian University of Athens, Panepistimiopolis, Zografos 15784, Greece
          \and 
            Athena Research and Innovation Center, Marousi 15125, Greece
         \and   
             Instituto de Astrofísica de Canarias, E-38205 La Laguna, Tenerife, Spain
          \and  
            ESA, European Space Astronomy Centre, Villanueva de la Canada, 28692 Madrid, Spain
        \and
             Space Telescope Science Institute, Baltimore, MD 21218, USA
         \and
         Quasar Science Resources for ESA-ESAC, Villanueva de la Ca\~nada, 28692 Madrid, Spain
        \and
            The Johns Hopkins University, Baltimore, MD 21218, USA
         \and Institute of Astrophysics, FORTH, Heraklion 71110, Greece
        \and Department of Physics, Univ. of Crete, Heraklion 70013, Greece
        \and 
            American Community Schools of Athens, Halandri 15234, Greece
         \and
          Greek Research and Technology Network - GRNET, Athens 11523, Greece
         \and 
            INAF-Osservatorio Astronomico di Capodimonte,  Napoli 80131, Italy
             }

   \date{Received June 5, 2019; accepted August 5, 2019}

 
  \abstract
   {}
    {Over its lifetime and despite not being a survey telescope, the \emph{Hubble Space Telescope (HST)} has obtained multi-epoch observations by multiple, diverse observing programs, providing the opportunity for a comprehensive variability search aiming to uncover new variables. We have therefore undertaken the task of creating a catalog of variable sources based on archival \emph{HST} photometry. In particular, we have used version 3 of the {\em Hubble} Source Catalog (HSC), which relies on publicly available images obtained with the WFPC2, ACS, and WFC3 instruments on board the \emph{HST}.}
   {We adopted magnitude-dependent thresholding in median absolute deviation (a robust measure of light curve scatter) combined with sophisticated preprocessing techniques and visual quality control to identify and validate variable sources observed by {\em Hubble} with the same instrument and filter combination five or more times.}
   {The {\em Hubble} Catalog of Variables (HCV) includes 84,428 candidate variable sources (out of 3.7 million HSC sources that were searched for variability) with $V \leq 27$~mag; for 11,115 of them the variability is detected in more than one filter. The data points in the light curves of the variables in the HCV catalog range from five to 120 points (typically having less than ten points); the time baseline ranges from under a day to over 15 years; while $\sim$8\% of all variables have amplitudes in excess of 1~mag. Visual inspection performed on a subset of the candidate variables suggests that at least 80\,\% of the candidate variables that passed our automated quality control are true variable sources rather than spurious detections resulting from blending, residual cosmic rays, and calibration errors.}
   {The HCV is the first, homogeneous catalog of variable sources created from the highly diverse, archival \HST\ data and currently is the deepest catalog of variables available. The catalog includes variable stars in our Galaxy and nearby galaxies, as well as transients and variable active galactic nuclei. We expect that the catalog will be a valuable resource for the community. Possible uses include searches for new variable objects of a particular type for population analysis, detection of unique objects worthy of follow-up studies, identification of sources observed at other wavelengths, and photometric characterization of candidate progenitors of supernovae and other transients in nearby galaxies. The catalog is available to the community from the ESA {\em Hubble} Science Archive (eHST) at the European Space Astronomy Centre (ESAC) and the Mikulski Archive for Space Telescopes (MAST) at Space Telescope Science Institute (STScI).}


\maketitle

%

%

\section{Introduction}
\label{sec:intro}

Diverse astrophysical processes related to stellar evolution, supermassive black holes, and propagation of light through curved space-time manifest themselves in optical variability. 
Standard candles such as Cepheid variables \citep{2001ApJ...553...47F,2017SSRv..212.1817S} and Type~Ia supernovae (SNe; \citealt{2018ApJ...853..126R}) are the crucial elements of the distance ladder and important probes of Cosmology in the local Universe. Eclipsing binaries \citep{2013Natur.495...76P, 2019Natur.567..200P}, RR~Lyrae \citep{2017SSRv..212.1743D}, and Mira variables \citep{2018ApJ...857...67H}  in the local Universe as well as Type~II SNe \citep{2018SSRv..214...32C} at larger distances verify and improve the distances derived from Cepheids and SNe~Ia.
%
For an overview of stellar variability types we refer the reader to the classification scheme\footnote{\url{http://www.sai.msu.su/gcvs/gcvs/iii/vartype.txt}} of the General Catalog of Variable Stars (GCVS; \citealt{2017ARep...61...80S}), as well as the books by \cite{1990vest.book.....H} and \cite{2015pust.book.....C}.

A number of current time-domain surveys explore optical
(DES -- \citealt{2016MNRAS.460.1270D}; SkyMapper -- \citealt{2017PASA...34...30S}; Evryscope -- \citealt{2014SPIE.9145E..0ZL})
and near-IR variability 
(VVV -- \citealt{2010NewA...15..433M}; VMC -- \citealt{2011A&A...527A.116C})
across large areas of the sky in search for microlensing events 
(MOA -- \citealt{2001MNRAS.327..868B}; MACHO -- \citealt{2005IAUS..225..357B}; EROS -- \citealt{2007A&A...469..387T}; OGLE -- \citealt{2015AcA....65....1U}), transiting exoplanets
(HATNet -- \citealt{2004PASP..116..266B}; SuperWASP -- \citealt{2006PASP..118.1407P}; 
MASCARA -- \citealt{2017A&A...601A..11T}; NGTS -- \citealt{2018MNRAS.475.4476W}), minor bodies of the solar system
(CSS -- \citealt{2009ApJ...696..870D}; Pan-STARRS -- \citealt{2014ApJ...795...44R}; ATLAS -- \citealt{2018AJ....156..241H}), Galactic and extragalactic transients
(ASAS-SN -- \citealt{2014ApJ...788...48S,2017PASP..129j4502K}; 
ZTF -- \citealt{2019PASP..131a8002B}),
often combining multiple scientific tasks within one survey. The space-based planet-searching missions, such as CoRoT \citep{2009A&A...506..411A}, Kepler/K2 \citep{2010Sci...327..977B, 2010ApJ...713L..79K}, and TESS \citep{2015ApJ...809...77S} have identified thousands of exoplanets. The \Gaia\ \citep{2016A&A...595A...1G} astrometric survey identifies transients \citep{2012arXiv1210.5007W} and provides time-domain information for the entire sky. 
These surveys also collect a wealth of information on variable stars in our Galaxy 
\citep{2011AJ....141..166H,2018AJ....155...39O,2018MNRAS.477.3145J,2018AJ....156..241H}.

The \emph{Hubble Space Telescope} (\HST) also provides time-domain information, as it has been observing the sky for over 25 years and has visited some regions of the sky multiple times over its lifetime. It thus offers the opportunity to search for variable objects at a range of magnitudes that are difficult to reach with ground-based telescopes. The magnitude depth, along with the superb resolution achieved by \HST\ and the long time-baseline of its operation are the features that make such a variable source catalog unique. The {\em Hubble} Source Catalog \citep[HSC;][]{2016AJ....151..134W} has recently provided photometric measurements of all sources detected from a homogeneous reduction and analysis of archival images from the \HST, thereby enabling such a variability search. Motivated by all of the above, we have undertaken the task of identifying variable sources among the sources in the HSC, aiming to exploit this Level 2 {\em Hubble} data product, and create a higher level product, the ``Hubble Catalog of Variables''. This work presents the results of this effort, named the ``HCV project'', which was undertaken by a team at the National Observatory of Athens and funded by the European Space Agency over four years, starting in 2015.

Table~\ref{tab:surveys} puts the HCV catalog in the context of current and future deep time-domain surveys, listing the filters, magnitude limit, number of sources, epochs, and time baseline. It should be noted that the HCV is not a volume or magnitude limited survey itself, as it relies on individual, largely inhomogeneous, sets of observations\footnote{Statistical analyses based on the HCV catalog should take this into account, as any conclusions will be limited to the sources of the HCV and cannot be generalised for the source population under study.}. The magnitude limit listed is the reported single-exposure detection limit of each survey. Variability analysis is typically possible only for sources well above the detection limit. The listed number of epochs is either a typical one for the survey or the lowest number of observations used for variability search (e.g., a minimum of five epochs is adopted for the HCV). The number of sources, epochs, and the corresponding time baseline vary from source to source within a survey and many of the surveys are still ongoing, so the numbers reported in Table~\ref{tab:surveys} are indicative. For ongoing surveys, we list the numbers corresponding to the current data release (e.g., there are 108 million sources in the latest release of the HSC, which is the input for the HCV catalog), while for the Large Synoptic Survey Telescope (LSST) the numbers correspond to the planned ten-year survey. It is clear that the HCV catalog is considerably deeper than other contemporary surveys, while having a comparable number of sources, despite the fact that it covers a tiny fraction of the sky compared to the other surveys listed in Table~\ref{tab:surveys}. Source confusion in crowded fields of nearby galaxies is another important parameter when comparing HSC to ground-based surveys: many of the HSC sources cannot be accurately measured from the ground even if they are sufficiently bright.

\begin{table}
\caption{Selected deep optical time-domain surveys.}             
\label{tab:surveys}      
{\centering                          
\begin{tabular}{@{~}c@{~~~}c@{~~~}c@{~~~}r@{~~~}r@{~~~}r@{~}}        
\hline\hline                 
Name & Filters & Limit & Sources      & Epochs & Baseline \\    
     &         &   (mag)    & $\times10^6$ &        & (years)  \\
\hline                        
SDSS\,S82 & $ugriz$ &  $r\sim21.5$ &  $4$  &  134 &  8  \\
CRTS   & clear      &  $V\sim21.5$ & $500$ & 300 & 7  \\
OGLE  &  $VI$ & $I\sim21.7$  & $500$  &  300 & 25  \\
ATLAS &  $oc$    &   $r\sim18$   & $142$  & 100 & 2  \\
Gaia  &  G ${\rm G}_{\rm BP}$ ${\rm G}_{\rm RP}$ & G$\sim21$ &  $1700$ & 12 & 2 \\
ZTF   &  $gri$   &  $r\sim20.5$  &   $1000$         &   300  &  1  \\
PS1   &  $grizy$ &  $r\sim21.8$  &   $3000$  &   60      &  3  \\
HCV   & various  &  $V\sim26$  & $108$  &  5    & 23  \\
LSST  &  $ugrizy$ & $r\sim24.5$  &   $18000$ &   1000    & 10  \\
\hline                                   
\end{tabular}
}
\small
References: 
SDSS\,S82 \citep{2008MNRAS.386..887B}; 
CRTS \citep{2009ApJ...696..870D};
OGLE \citep{2015AcA....65....1U};
ATLAS \citep{2018AJ....156..241H};
Gaia \citep{2018A&A...616A...1G};
ZTF \citep{2019PASP..131a8002B};
PS1 \citep{2016arXiv161205560C};
HCV \citep[this work and][]{2016AJ....151..134W};
LSST \citep{2019ApJ...873..111I}.
\end{table}

The time domain and variability properties of astronomical sources provide a wealth of information that can be very useful, for example, for characterizing the fundamental properties of stars, or for identifying particular types of sources from a large dataset. Objects showing variations in flux may be associated with variable stars in our own Galaxy, stars in nearby galaxies, or distant active galactic nuclei (AGN), or possibly transient events such as novae and SNe. The HCV aims to extend our knowledge of variable stars to fainter magnitudes and crowded regions of stellar clusters and distant galaxies, which are inaccessible by ground-based surveys.

\subsection{The {\em Hubble} Source Catalog}
\label{sec:hscintro}

The \HST\ obtains exceptionally deep imaging thanks to the low sky background (free from airglow, scattering, and absorption in the atmosphere of the Earth), a sharp and consistent PSF, and a wide field of view compared to ground-based adaptive optics instruments \citep{2005aoel.book.....L}. 
The \HST\ instruments are sensitive to ultraviolet (UV) light not accessible from the ground and to infrared (IR) radiation that is heavily contaminated by airglow and atmospheric absorption. Since its launch in 1990, a variety of instruments have been installed during five 
astronaut servicing missions. 
Imaging instruments in the UV and optical include the initial Wide Field and Planetary Camera, followed by the Wide Field and Planetary Camera~2 (WFPC2; 1993--2009), the 
Advanced Camera for Surveys (ACS, 2002--present), and the Wide Field Camera~3 (WFC3, 2009--present) in the optical. In the near-IR, the Near 
Infrared Camera and Multi-Object Spectrometer (NICMOS, 1997--1999, 2002--2008) pioneered IR 
studies using {\em Hubble}. NICMOS was succeeded by the much more powerful IR channel of WFC3 in 2009.

The {\em Hubble} Legacy Archive (HLA; \citealt{2006ASPC..351..406J}) aims to increase the scientific output from the \HST\ by providing online access to advanced data products from its imaging instruments. The most advanced form of these data products are lists of objects detected in visit-combined images\footnote{An \HST\ visit is a series of one or more consecutive exposures of a target source interrupted by the instrumental overheads and Earth occultations, but not repointing to another target.  While exposures may be taken at several different positions, all exposures in a visit rely on the same guide star as a pointing reference.}.

Cosmic ray hits limit the practical duration of an individual exposure with a CCD. 
Primary cosmic rays of Galactic origin together with protons trapped in the inner Van~Allen belt 
create a hostile radiation environment in low Earth orbit \citep{1997RadR..148S...3B}, 
compared to the one faced in ground-based CCD observations where the primary sources of particles 
are the secondary cosmic-ray muons and natural radioactivity \citep{2002ExA....14...45G}. 
In a 1800\,s \HST\ exposure, between 3 to 9\,\% of pixels will be affected by cosmic ray hits 
depending on the level of particle background and the instrument used \citep{2008wfpi.book.....M,2012wfci.book.....D}.
To combat the effects of cosmic ray contamination, most \HST\ observing programs split observations into multiple exposures. The HLA relies on the \texttt{AstroDrizzle} code
\citep{2012AAS...22013515H} to stack individual exposures obtained within one visit and produce images free of cosmic rays. The \texttt{AstroDrizzle} code corrects for geometrical distortion in the instruments and also handles the case where the image pointing center is dithered to different positions during the visit (which is a commonly used strategy to eliminate the effects of bad pixels and improve the sampling in the combined image).

The \SExtractor\ code \citep{1996A&AS..117..393B} is used to detect sources on these images, perform aperture photometry and measure parameters characterizing the source size and shape. Most \HST\ observations are performed using multiple filters. To facilitate cross-matching between objects detected with different filters, images in all filters obtained during a given visit are stacked together in a ``white-light'' image. Stacked images in each individual filter are also produced and sampled to the same pixel grid as the white-light image. The \SExtractor\ program is executed in its ``dual-image'' mode to use the white-light image for source detection and the stacked filter images for photometry. Each visit results in a list of sources, with every source having a magnitude (or an upper limit) measurement in each filter used in this visit.

The {\em Hubble} Source Catalog\footnote{The HSC version 1 was released on 2015 February 26, HSC version 2 on 2016 September 30, version 2.1 on 2017 January 25 (the only change was the addition of links to spectra), and HSC version 3 on 2018 July 9.} (HSC;  \citealt{2016AJ....151..134W}, \citealt{2013ASPC..475..203L}) 
combines source lists \citep{2008ASPC..394..481W} generated from individual \HST\ visits into a single master catalog.
The HSC creates a combined source catalog from a diverse set of observations taken with many 
different instruments and filters (by various investigators) after the data proprietary period 
expires. This approach was pioneered by X-ray catalogs such as the WGACAT (derived from pointed 
observations of ROSAT; \citealt{1994IAUC.6100....1W}), the Chandra Source Catalog 
\citep{2010ApJS..189...37E}, the XMM-Newton serendipitous survey \citep{2016A&A...590A...1R}, and 
the catalogs derived from Swift X-ray telescope observations 
\citep{2014ApJS..210....8E,2013A&A...551A.142D}. The same approach was used to create catalogs of 
UV and optical sources detected by the OM and UVOT instruments of XMM-Newton and Swift, 
respectively \citep{2012MNRAS.426..903P,2014Ap&SS.354...97Y}. The more recent All-sky NOAO Source 
Catalog \citep{2018AJ....156..131N} combines public observations taken with the CTIO-4m and 
KPNO-4m telescopes equipped with wide-field mosaic cameras.

In many ways the challenge faced by the HSC project to integrate \HST\ observations is the most daunting of all these missions and observatories.  The field of view of the {\em Hubble} cameras is tiny, with even the ``wide-field'' cameras covering only 0.003 square degrees (less than $10^{-7}$ of the sky). That leads to highly variable sky coverage even in the most commonly used filters. It also makes reference objects from external catalogs such as \Gaia\ relatively rare in the images.  A major complication is that the uncertainty in the pointing position on the sky is much larger than the angular resolution of the \HST\ images, making it necessary to correct for comparatively large pointing uncertainties when matching observations taken at different epochs.

The HSC provides a homogeneous solution to the problem of correcting absolute astrometry for the \HST\ images and catalogs. Typical initial astrometric errors range from 0.5 to 2" (depending on the epoch of the observations), due to uncertainties in the guide star position and in the calibration of the camera's focal plane position and internal geometric distortion (both of which change over time). In some cases much larger errors (up to 100") are found; those are probably attributed to selection of the wrong guide star for pointing by the onboard 
acquisition system. The HSC uses a two-step algorithm to correct the astrometry, first matching to an external reference catalog to correct large shifts, and then using a cross-match between catalogs from repeated \HST\ observations of the region to achieve a fine alignment of the images and catalogs \citep{2012ApJ...761..188B,2016AJ....151..134W}.  The fine alignment algorithm includes features designed to produce good results even in extremely crowded regions such as globular clusters and the plane of the Milky Way.

The current release of the HSC is version 3\footnote{\url{https://archive.stsci.edu/hst/hsc}} (HSC~v3), which includes 542 million measurements of 108 million unique sources detected on images obtained with the WFPC2, ACS/WFC, WFC3/UVIS, and WFC3/IR instruments that were public as of 2017 October~1 (based on source lists from HLA Data Release 10 or DR10\footnote{\url{https://hla.stsci.edu}}).  The observations include measurements using 108 different filters over 23 years (1994--2017) and cover 40.6 square degrees ($\sim0.1$\% of the sky).

The HSC~v3 release contains significant improvements in both the astrometry and photometry compared with earlier releases\footnote{See online documentation for HSC v3.}. The external astrometric calibration is based primarily on \Gaia\ DR1, falling back on the Pan-STARRS, SDSS, and 2MASS catalogs when too few \Gaia\ sources are available. About 2/3 of the images are astrometrically calibrated using \Gaia, and 94\% of the images have external astrometric calibrations. The photometric improvements are mainly the result of an improved alignment algorithm used to match exposures and filters in the HLA image processing. There were also improvements and bug fixes for the sky-matching algorithm and the \SExtractor\ background computation that significantly improved both the photometry and the incidence of spurious detections near the edges of images. Many of the improvements in HSC~v3 were the direct result of testing and analysis by the HCV team at the National Observatory of Athens.


The median relative astrometric accuracy (repeatability of measurements) is  7.6\,mas for the whole catalog, but it varies depending on the instrument, from 5\,mas for WFC3/UVIS to 25\,mas for WFPC2. The absolute astrometric accuracy is determined by the accuracy of the external catalog used as the reference for a given \HST\ field. As \Gaia~DR2 was not available at the time HSC~v3 was created, proper motions of reference stars used to tie the \HST\ astrometry to the external catalog could not be accounted for.

\begin{figure}
\includegraphics[width=0.5\textwidth]{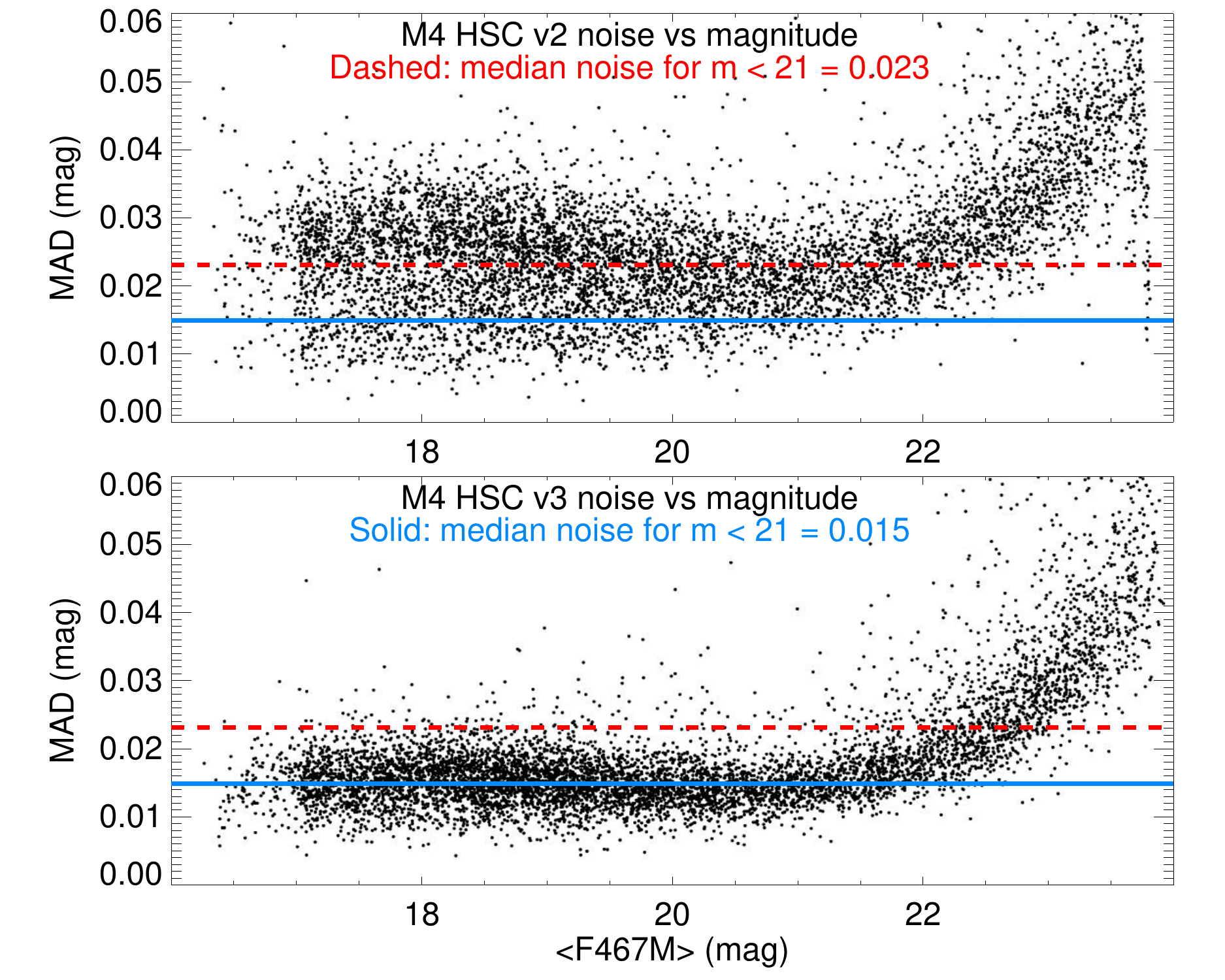}
\caption{Median absolute deviation (MAD; see Section~\ref{sec:hcvalgorithm}) as a function of the median magnitude in filter F467M in globular cluster M4 for HSC~v2 (upper panel) and HSC~v3 (lower panel). The plot includes 5300 objects (black dots) that have more than 50 measurements in the WFC3/UVIS F467M filter. The  red and blue lines represent the median noise for sources with magnitudes brighter than 21 in HSC~v2 and HSC~v3, respectively. This field was affected by image alignment problems that have been corrected in HSC~v3, which resulted in improved accuracy of the photometry.
\label{fig:hsc_v2_v3_photometry}}
\end{figure}

The photometric accuracy of HSC~v3 is limited by the signal-to-noise of the observations, the accuracy of \HST\ magnitude zero-points, the residual sensitivity variations across the field of view of the instrument, due to imperfect flat-fielding and charge transfer efficiency corrections, and, for the fainter sources, the use of aperture rather than PSF-fitting photometry, which mainly affects crowded fields. For objects with adequate signal-to-noise, the photometric accuracy is generally about 1.5--2\%. Figure~\ref{fig:hsc_v2_v3_photometry} demonstrates the accuracy in the field of globular cluster M4 and the improvement in HSC~v3 compared with the previous release, HSC~v2.

\subsection{An overview of variability detection techniques}
\label{sec:vrsearchtech}

The simplest way of finding variable sources is pair-wise image comparison, used since the early days of photographic astronomy \citep{1990vest.book.....H}. The contemporary approach to image comparison, known as the difference image analysis \citep[DIA;][]{1998ApJ...503..325A,2016MNRAS.457..542B,2016ApJ...830...27Z, 2017PASP..129d4501S} is effective in identifying variable sources in crowded fields \citep[e.g.][]{2001AcA....51..317Z,2003ApJ...591L.111B,2003AstL...29..599Z}.
The intrinsic limitation of the two-image technique is that variations in the source brightness between the images need to be large compared to image noise in order to be detected. 

One may use aperture or point-spread function fitting photometry to measure the source brightness on multiple original (or difference) images, constructing the light curve. Using multiple measurements one may identify brightness variations with an amplitude below the noise level of individual measurements. One may test the hypothesis that a given object's brightness is constant given the available photometric measurements and their uncertainties \citep{2005ESASP.576..513E,2006AJ....132..633H,2001A&A...373..576P}. This is the standard variability detection approach in X-ray astronomy, where the uncertainties are well known as they are typically dominated by photon noise \citep{1998ApJ...504..405S}. The hypothesis testing is less effective for optical and near-IR photometry, as the measurement uncertainties are dominated by the poorly-constrained systematic errors for all sources except the ones close to the detection limit. 
The scatter of brightness measurements of a non-variable star may be used to estimate photometric uncertainty \citep{1988AJ.....95..247H,2010AJ....139.1269D} under the assumption that the measurement uncertainties are the same for sources of the same brightness. Relying on this assumption, one may construct various statistical measures of scatter \citep{2008AcA....58..279K,2018MNRAS.476.2813D} or smoothness \citep{1993AJ....105.1813W,1996PASP..108..851S,2014A&A...568A..78M,2018AcA....68...63R} of a light curve to identify variable sources (for a review see \citealt{2017MNRAS.464..274S,2016A&A...586A..36F,2017A&A...604A.121F}). 
Hereafter, we refer to these measures of scatter and smoothness (degree of correlation between consecutive magnitude measurements) as ``variability indices'' \citep[e.g.]{1996PASP..108..851S,2009MNRAS.400.1897S,2016A&A...586A..36F}.
They are also known as ``variability features'' in the machine learning context \citep{2014A&A...566A..43K,2015arXiv150600010N,2018MNRAS.475.2326P}.

Period search is a primary variable star investigation tool and also a very efficient method of variable star identification \citep{2014A&A...566A..43K,2014ApJS..213....9D,2017MNRAS.469.3688D,2018ApJS..237...28C}. While many types of variable stars show periodic or semi-periodic light variations, photometric errors are expected to be aperiodic, or associated with a known periodic process inherent to the observations (diurnal cycle, periodic guiding errors, orbital period of a space borne telescope, etc.). These spurious periodicities can be identified using the window function \citep{1975Ap&SS..36..137D}. The down side of the period search is that it is computationally expensive, requires hundreds of light curve points for the period search to be reliable \citep{2013MNRAS.434.3423G}, and excludes the class of non-periodic variables.

To identify specific types of variable objects such as Cepheids, RR~Lyrae stars, and transiting exoplanets, one may utilize template fitting.
This dramatically increases the search sensitivity to a specific type of variability at the cost of the loss of generality. The sensitivity gain is especially evident for exoplanet transits that typically cannot be identified in ground-based photometry using general-purpose variability detection methods. If templates for multiple variability types are fitted, classification of variable sources is performed simultaneously with their detection \citep{1999AJ....117.1313L,2014A&A...567A.100A}.

%
The output of multiple variability detection tools may be combined using principal component analysis \citep{2018MNRAS.477.2664M}, supervised \citep{2018MNRAS.475.2326P} or unsupervised machine learning \citep{2009MNRAS.400.1897S,2012AJ....143...65S}. Machine learning may be applied to design new variability detection statistics  \citep{2016ApJ...820..138M,2018MNRAS.475.2326P}.

Visual inspection of light curves and images of candidate variables selected using the above methods remains an important quality control tool. It is always applied when one aims to produce a clean list of variable stars \citep[e.g.][]{2016AcA....66..421P,2016AJ....151..110K,2018AJ....155..183S,2018MNRAS.477.3145J} rather than a (more extensive, but contaminated) list of candidate variables \citep[e.g.][]{2018AJ....155...39O,2018AJ....156..241H}. Both types of lists may be useful. Consider two example problems: {\it a)}\,the study of period distributions of W~UMa type binaries (which requires confidence in classification of the studied objects as binaries of this particular type) and {\it b)}\,selection of non-variable stars in a given field (to be used as photometric standards or for microlensing studies). In the latter case, it is more important to have a complete list of variable stars, rather than a clean one.

Visual inspection is needed to control various instrumental effects, which produce light curves that are smooth and/or have an elevated scatter. One of the most important effects is the variable amount of blending between nearby sources \citep[e.g.][]{2011AJ....141..166H}. The degree of blending may vary with seeing (for ground-based observations), or with the position angle of the telescope if its point spread function (PSF) is not rotationally symmetric (e.g., due to the diffraction spikes produced by spiders holding the telescope's secondary mirror). If aperture photometry is performed, light from nearby sources may cause additional errors in the position where the aperture is placed over the source in a given image, which can lead to large errors in the measured source flux. Depending on the optical design of the telescope, slight focus changes may have noticeably different effects on the PSF size and shape depending on the source color \citep[e.g.][]{2014ASPC..490..395S}. The amount of blending may also change if one of the blended sources is variable. Other effects that may corrupt photometry of an individual source include the various detector artifacts (hot pixels, bad columns, cosmic ray hits) or the proximity to the frame edge/chip gap. Uncorrected sensitivity variations across the CCD (due to imperfect flat-fielding and charge transfer inefficiencies) coupled with the source image falling on different CCD pixels at different observing epochs may produce artificial variations in a light curve. If the sensitivity varies smoothly across the CCD chip affecting nearby sources in a similar way, one may try to correct the light curves for these variations using algorithms like SysRem \citep{2005MNRAS.356.1466T}, a trend filtering algorithm \citep{2005MNRAS.356..557K,2009MNRAS.397..558K,2016PASP..128h4504G}, or local zero-point correction (Section~\ref{sec:preproc}).

In this paper, we describe the HCV\footnote{Preliminary reports on the progress of the HCV project were presented by \cite{2017IAUS..325..369G}, \cite{2017EPJWC.15202005S}, \cite{2018ASPC..514..159Y}, and \cite{2018arXiv180304974S}.} system and catalog resulting from a systematic search for variable objects in the HSC~v3. It should be noted that ``HCV'' can either refer to the processing system (i.e., the development of the hardware, software system and pipeline to create the catalog) or the catalog itself. The paper is structured as follows: Section 2 presents an overview of the HCV system developed to identify variable sources in the HSC. Section 3 describes the preprocessing applied to the HSC photometry, while Section 4 describes the algorithm for selecting candidate variables. Section 5 presents the algorithm adopted for validating the candidate variables. Section 6 presents the performance and limitations of the HCV catalog, while Section 7 outlines the statistics of the HCV catalog and highlights some scientific results. A summary is given in Section 8. 

\section{HCV system overview}
\label{sec:hcvsystem}

The HCV processing system aspires to identify all the variable and transient sources in the HSC through simple mathematical techniques, thus producing the HCV catalog.

The sole data input to the HCV pipeline is the HSC, which provides a set of tables containing specific information about the individual sources observed by the \HST\ instruments at different epochs. The HSC is naturally divided into {\bf groups} of sources detected on overlapping \HST\ images \citep{2016AJ....151..134W}. Each group was assigned a unique \texttt{GroupID} identifier. Within the group, observations of the same source are identified and combined into a ``matched source'' to which another unique identifier is attached (\texttt{MatchID}). The observations of a matched source, hereafter simply mentioned as ``a source'', over all available epochs form the input to the HCV pipeline.

The HCV catalog is generated by a pipeline that consists of the following stages of operation:

\begin{itemize}
\item importing and organizing the HSC data in a form that facilitates processing for variability detection, 
\item detection of candidates for variability, after applying specific limits on the data quality and quantity, rejecting inappropriate sources within a group and even groups (see Section~\ref{subsec:limitations}), 
\item validation of the detected candidates using an automated algorithm, 
\item extraction of source and variability index (Sec.~\ref{sec:vrsearchtech}) data for all the processed sources (candidate variables and non-variables), \item curation of candidate variable sources and expert validation,
\item publication of the resulting catalog datasets into publicly available science archives, specifically, the ESA {\em Hubble} Science Archive, eHST (ESAC), and the Mikulski Archive for Space Telescopes, MAST (STScI).
\end{itemize}

In the following subsections we describe the (largely configurable) components of the HCV system, which supports this computationally intensive process and forms a pipeline of distinct data fetching, processing, and depositing.

\subsection{System concept}

\begin{figure*}[!htp]
\includegraphics[width=1.0\textwidth]{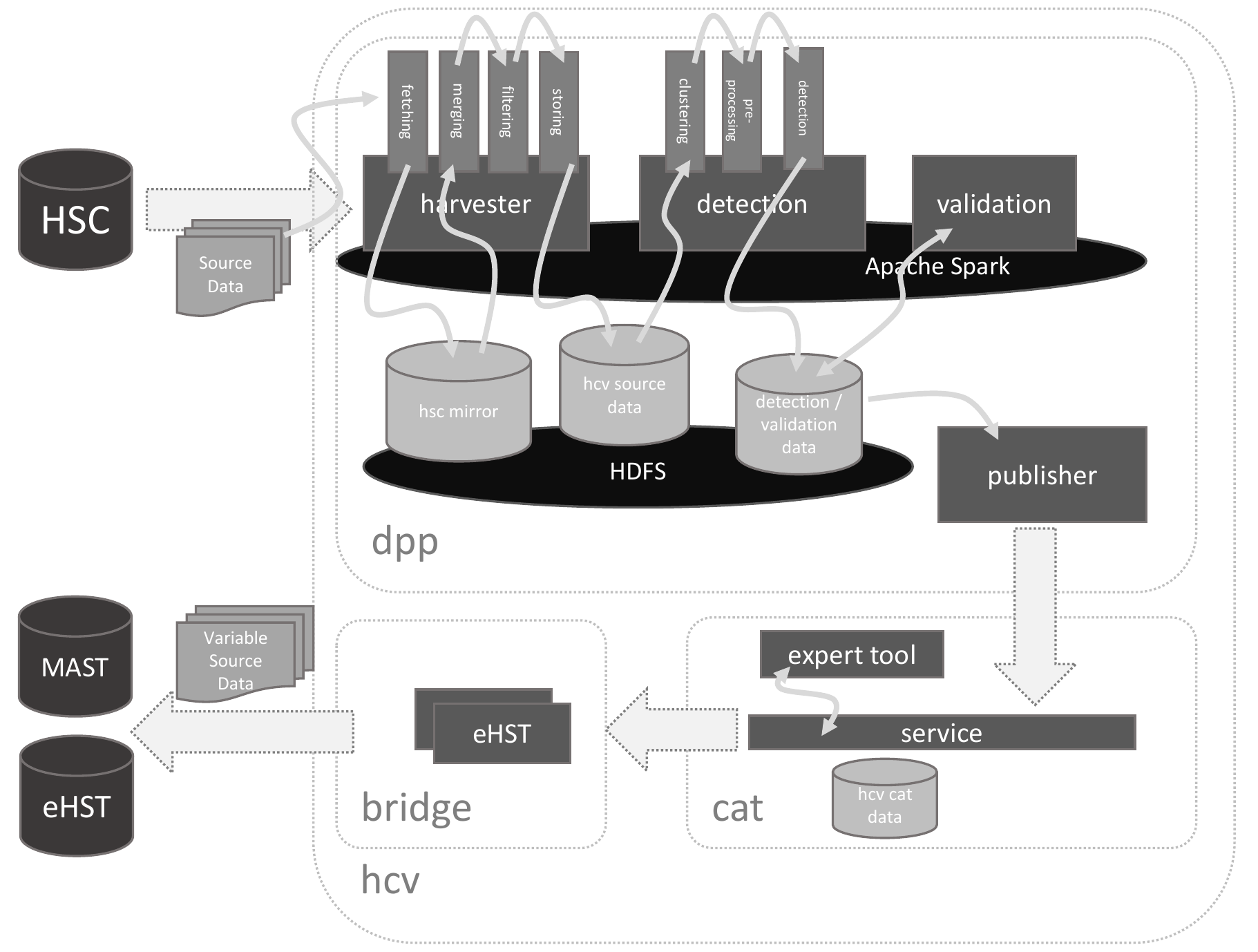}
\caption{HCV system high-level architecture, showing three major components (dpp, cat, bridge) and their elements. These components correspond to the top-level functional subsystems of the HCV system.
\label{hlsystemArchi}}
\end{figure*}

The HCV system, at the highest level, consists of three functional sub-systems: 
(a)\,\texttt{DPP}, the {\bf data processing pipeline (hcv.dpp)}, 
which deals with the computational requirements of the system, 
employing distributed infrastructure for processing and data storage; 
(b)\,\texttt{CAT}, the (mostly) relational data driven {\bf HCV catalog sub-system (hcv.cat)} 
where typical expert driven data management and curation operations are performed; 
(c)\,an {\bf interface to specific science archives (hcv.bridge)}. Furthermore, there is a fourth enabling element, the {\bf infrastructure (hcv.infra)} that manages the security and access, monitoring, logging, and other non-functional aspects of the system. The top-level architecture of the system is illustrated in Figure~\ref{hlsystemArchi} and the major elements of the HCV system and their functions are listed in Table~\ref{tab:hcvsys}.

\begin{table}
    \caption{HCV system functional components.}
    \label{tab:hcvsys}
    \centering
    \begin{tabular}{l|l}
    \hline\hline
        HCV component               & Function \\
     \hline
        {hcv.dpp}               & data processing pipeline \\
        {\it hcv.dpp.harvester}     & interface to the HSC archive tables \\
        {\it hcv.dpp.detection}     & detection of variables  \\
        {\it hcv.dpp.validation}    & validation of candidate variables  \\
        {\it hcb.dpp.publisher}	    & publication of source data to catalog \\ 
        {hcv.cat}	            & HCV catalog subsystem \\
        {\it hcv.cat.service}       & HCV database service layer \\        
        {\it hcv.cat.ui}            & HCV expert tools for curation \\        
        {hcv.bridge}	        & system interfaces to external archives \\
        {\it hcv.bridge.mast}       & STScI archive component \\
        {\it hcv.bridge.ehst}       & ESAC archive component \\
        {hcv.infra}             & infrastructure enabling layer \\
        {\it hcv.infra.management} 	& infrastructure management tools \\
        {\it hcv.infra.monitoring}	& operation monitoring subsystems \\
        {\it hcv.infra.logging}     & logging subsystem \\
        {\it hcv.infra.security}	& authentication / authorization \\
    \hline      
    \end{tabular}
\end{table}

The data processing pipeline {\bf (hcv.dpp)} hosts the computationally and data intensive processes of the HCV system. It utilizes high performance distributed processing and storage technologies and employs highly configurable algorithms for its operations, which may be fine-tuned or even replaced to fit future needs of the HCV system. Its elements are:

\begin{itemize}
    \item {\it hcv.dpp.harvester} - enables access to the external HSC archive. 
    \item {\it hcv.dpp.detection} - applies the variability detection algorithm, estimates parameters that characterize variability (i.e., variability indices), and identifies candidate variables.
    \item {\it hcv.dpp.validation} - provides tools to analyze and verify and/or validate the variable candidates. It applies the variability validation algorithm and validates candidates as variable sources.
    \item {\it hcv.dpp.publisher} - ingests the outcome of processed data into the HCV relational database.
\end{itemize}

The objective of the harvester is to retrieve sources and their metadata from the HSC database and transform those data into a form efficient for further processing, according to the requirements of the HCV data model. The harvester utilizes a compressed, columnar data format and parallel processing of the HSC data. In order to save space, the harvester may opt to completely omit specific portions of the HSC (e.g., single epoch observations). A significant feature of the harvester is its configurability to adapt to changes in HSC data structures.

The detection and validation elements implement the data processing pipeline. Initially, the extremely large groups are split into a number of 
{\bf clusters} (hereafter referred to also as "subgroups"), based on source coordinates.
The sources that are nearby on the sky tend to be assigned to the same cluster. 
This clustering procedure is based on the k-nearest neighbors algorithm (k-NN) 
and is necessary to satisfy the CPU requirements of the core algorithms, 
and enable sufficient parallelization of the process, 
as the element of work is the cluster, which may be an entire HSC group (\texttt{GroupID}). 
Next, the HSC photometry is preprocessed (see Section~\ref{sec:preproc}) in order to remove unreliable measurements and apply local zero-point corrections. The light curves are constructed by retrieving the (corrected) photometric measurements obtained with the same instrument and filter combination at different epochs for all sources in each subgroup. The pipeline then computes the magnitude scatter in each light curve, the variability indices, and applies a magnitude-dependent threshold to select candidate variable sources (Section~\ref{sec:hcvalgorithm}). Finally, it applies an automated validation algorithm to the candidate variables in order to remove obvious false-detections (see Section~\ref{sec:validationalgorithm}). 

The last step of the pipeline, the publisher, implements a Representational State Transfer (REST) web service interface. It ingests all data delivered by the detection and validation components into the HCV database, that is each and every candidate and non-candidate variable source processed. This dataset is the HCV catalog. 

The catalog sub-system {\bf (hcv.cat)} is a typical web application component that enables inspection and validation of the outcome of the pipeline. It is based on fundamentally different technologies and employs a relational database management system to contain its data structures. It consists of two elements:
\begin{itemize}
    \item {\it hcv.cat.service} - provides a REST web service abstraction layer over the HCV database, covering all functionality for data management, such as create-read-update-delete operations, data publication, and authentication/authorization.
    \item {\it hcv.cat.ui} - offers a tool for highly streamlined expert-driven data validation of the catalog data.
\end{itemize}

The last subsystem of the HCV system, the {\bf hcv.bridge}, provides the external science archives with access to the publishable release of the HCV catalog. The latter is exported via the bridge adapters in a fixed open format, {\bf JSON} (JavaScript Object Notation), utilized by the targeted archives.

Furthermore, additional tools are provided to facilitate data inspection and handling, which are particularly useful during algorithm fine-tuning and exploration of the HSC. One of the tools utilizes the temporary outputs of the pipeline supporting the algorithm validation and configuration phase. The tool allows experts to inspect the light curves and other information available for the candidate variables in order to identify issues prior to producing the catalog, as well as to evaluate the success of the pipeline. The second tool operates on the data imported from the pipeline to the database allowing for validation by experts, which is optional. It enables fine grained manipulation, updating of a particular dataset, freezing and/or publishing the catalog and exporting it in native format. 

\subsection{Implementation technologies}
\label{sec:implementation}

The main driver behind our choice of technologies for the HCV system implementation was the large amount of input, 
intermediate and output data (see Table~\ref{tab:hcvsysperf}), and, correspondingly, the large-scale data processing. 
An additional driver was the nature of the source processing. Although there is a discrete sequence of steps, that is HSC ingestion, variable candidate detection and validation, publication to the catalog, and catalog operations, there are processes within most of these steps that can run in parallel. This is because either these sub-stages are independent or because one set of data can be processed independently of another (groups or subgroups). The portability of the system was important in order to allow different infrastructures to be developed and deployed, as well as to avoid ``vendor lock-in''. The ability of the system to utilize resources that are offered to it (e.g., CPU and RAM) is the cornerstone of the scalable design and technologies implemented for the HCV system.

Essentially the whole HCV system relies on Free Open Source Software (FOSS). The following list presents the most essential elements:

\begin{itemize}
    \item {\it Linux} is the operating system of the infrastructure, providing many of the baseline services required for operating the infrastructure.
    \item {\it Hadoop Distributed File System (HDFS)} is the distributed, high-throughput file system employed for storing the non-relational data of the system.
    \item {\it Apache Spark} is the distributed parallel processing platform, which allows the system to carry out its computationally intensive tasks exploiting all resources provided to it. Apart from implementation of Spark-enabled algorithms, DPP uses specific Machine Learning elements
    (clustering) provided in the Spark ecosystem.
    \item {\it Mesos} is the hardware abstraction layer over which Apache Spark operates.
    \item {\it Apache ZooKeeper} is utilized for centralized configuration management and synchronization of services of the system.
    \item {\it Apache Parquet} is the columnar format employed over HDFS to provide storage and access features for DPP processes.
    \item {\it PostgreSQL} is the relational database management system that hosts the HCV catalog data component.
    \item {\it Java} is the platform for the implementation of the components of the system supported by several Java ecosystem technologies such 
as {\it Hibernate}, {\it Spring Framework}, {\it Tomcat} etc.
\end{itemize}

\subsection{Deployment and performance}

The system is deployed on a virtualized Intel x64 architecture, yet there are no particular dependencies on this architecture. It has been successfully operated over {\it XEN}, {\it VMware}, and {\it Hyper-V} hypervisors.

In operational deployment at STScI, the HCV pipeline is provided with four worker nodes, each consisting of 16 virtual cores and 64 GB RAM, and shared HDFS storage of over 10 TB. Those can be easily up-scaled to larger numbers if required. Two additional nodes, one consisting of four virtual cores and 16 GB RAM, the other of eight nodes and 32 GB memory, are dedicated, respectively, 
to (a)\,operation of the infrastructure and several enabling components, such as the code repository and 
(b)\,the hosting of the {\it hcv.cat} and {\it hcv.bridge} subsystems for the handling and publication of the catalog of variables.

Over this infrastructure, the processing of the HSC~v3 was carried out. Performance data are presented in Table~\ref{tab:hcvsysperf}; the total duration of the run is about ten days. We note that the processing times in Table~\ref{tab:hcvsysperf} are indicative, as they heavily depend on network and VM load and have been observed to deviate by more than 100\% during peak hours.

\begin{table}
    \caption{Performance of the HCV system processes in operational deployment at STScI.}
    \label{tab:hcvsysperf}
    \centering
    \begin{tabular}{l|l|l|r|c}
    \hline\hline
        Process & Product & Type & Size & Duration \\
     \hline
        Download    & HSC tables    & CSV-files & 1.9 TB    & 12h~45m \\
        Harvesting  & HCV tables    & parquet   & 700 GB    & 02h~48m \\
        Input       & HCV input     & parquet   & 38 GB     & 00h~20m \\
        Clustering  & HCV input     & parquet   & 38 GB     & 02h~47m \\
        Det.\&Valid.& DPP output    & JSON      & 80 GB     & 17h~40m \\
        Import DB   & CAT dataset   & SQL       & 80 GB     & 7 days \\
        Export CAT  & HCV export    & JSON      & 0.5 GB    & 00h~15m\\
    \hline      
    \end{tabular}
\end{table}

\section{HSC photometry preprocessing}
\label{sec:preproc}

The task of identifying variability in a photometric light curve requires a reliable and clean dataset. As \HST~is not a survey telescope, it performs observations that are specifically designed for each individual project, using diverse filters, exposure times, pointings, and dithering patterns. Consequently, the uniform reduction and photometry provided by the HSC cannot address issues specific to certain datasets as well as a tailored reduction of each dataset. It can thus, inadvertently, introduce systematic effects in the photometry. Furthermore, cosmic ray hits, measurements near the edge of the CCD or in a region with nebulosity etc.\ can also introduce systematic effects. Therefore, before proceeding with the variability search described in Section~\ref{sec:hcvalgorithm}, we apply quality cuts and additional photometric corrections to the input HSC data. The procedure described below was developed from a comprehensive investigation of the ``Control Sample'' fields (see Section~\ref{sec:limitations}) and numerous randomly selected fields from different instruments, initially with data from the HSC~v1, and subsequently with data from the HSC~v2 and HSC~v3, as the new releases became available. The procedure was eventually applied to the whole HSC~v3.

\subsection{Light curve data collection}

We adopted the following procedure to construct light curves of HSC~v3 sources. We used the HSC parameter \texttt{GroupID}, which indicates a group of overlapping white-light images (corresponding to Level 2 "detection" images of the HLA), to select observations of sources in a specific field. We consider only the groups that have at least 300 detected sources (N$_{\rm sources}\geq300$) and only the sources that have at least five detections ($n_{\rm LC}\geq5$) with the same instrument and filter combination. These constraints should ensure the reliable operation of the variability detection algorithm described in Section~\ref{sec:hcvalgorithm}. We also applied cuts on the following HSC parameters \citep[for a detailed description see][]{2016AJ....151..134W}: 

\begin{itemize}
    \item The Concentration Index (\texttt{CI}), defined as the difference between the source magnitude measured in two concentric apertures \citep[see aperture sizes in Table 1 of][]{2016AJ....151..134W}, 
was limited to $\texttt{CI}=\texttt{MagAper1-MagAper2} <5.0$. 
The \texttt{CI} is a measure of the spatial extension of a source and can be used to identify sources potentially affected 
by light from their neighbors (blending, diffraction spikes from bright stars, a diffuse background) or cosmic rays. 
Typically, real extended sources have $\texttt{CI}\approx2-4$~mag, while larger values of \texttt{CI} usually indicate 
problematic photometry or image artifacts.
    \item Magnitude cuts of $\texttt{MagAper2}<31.0$ and $\texttt{MagAuto}<35.0$~mag were used to remove unphysical measurements.
    \item The photometric error \texttt{MagerrAper2} estimated by \texttt{SExtractor}, 
which provides a lower limit on the total photometric uncertainty, 
was used to eliminate uncertain measurements by adopting $\texttt{MagerrAper2}\leq0.2$ mag. This value is a conservative, typical error adopted from the faint end of the magnitude.
    \item \SExtractor\ and HSC flags were constrained to $\texttt{SE\_flags}\leq7$ and $\texttt{Flags}<4$, respectively, 
to exclude objects flagged as truncated, incomplete/corrupted, or saturated. We cannot rely on the \SExtractor\ saturation 
flag, as the CCD saturation limit is not always propagated properly to the white-light images.
    \item Sources with undefined (null) values of the above parameters were also rejected. 
\end{itemize}
After applying all the quality cuts described above, light curves from the same instrument and filter combination 
were constructed for each source, which is identified by its unique \texttt{MatchID}
in the HSC catalog.

\subsection{Light curve filtering and outlier identification}

During the construction of the HCV pipeline, we identified several issues with the HSC photometry
including misalignment between images in a visit stack, background estimation problems
resulting in corrupted photometry of sources close to the image edges (``edge effect''), 
issues regarding local correction, saturation, double detection, and so on. 
Several of these were corrected or improved in the HSC~v3, although some continue to affect the HSC photometry and therefore the HCV catalog (see Section~\ref{subsec:limitations}). 
Additional light curve filtering was therefore required to reduce the false detection rate of candidate variables.

The filtering performed during the preprocessing included the following main steps: 
identification and flagging of photometric outliers, identification and rejection of ``bad'' images that produce many photometric outliers, local magnitude zero-point correction, and identification and rejection of additional unreliable data points that have large synthetic errors (defined below). 
Four parameters were used to evaluate the quality of the HSC photometry:
\begin{enumerate} 
\item \texttt{MagerrAper2}. 
\item \texttt{CI}.
\item The offset distance \texttt{D} of a source from its average position listed in the catalog (``match position''), as an uncleaned cosmic ray or misalignment between images in the white-light stack will change the center of light (pixel-flux-weighted position of the source) and corrupt its photometry. 
\item The difference between the source magnitude measured using the circular (\texttt{MagAper2}) and 
the elliptical aperture (\texttt{MagAuto}\footnote{See the description of automatic aperture magnitudes in \texttt{SExtractor} User's manual at \url{https://www.astromatic.net/software/sextractor}}), \texttt{MagAper2-MagAuto}. 
This difference is expected to be constant for isolated sources. For a close
pair of sources that were not resolved by \texttt{SExtractor}, 
the elliptical aperture may include both sources, while the circular aperture
may include only one.
\end{enumerate}

We assigned a weight to each light curve point that is inversely proportional to the square of the quantity we defined as the ``synthetic error'':
\begin{equation}
Error_{syn}=\sqrt{
 \begin{aligned}
   &\left(\frac{\texttt{MagerrAper2}}{{<\texttt{MagerrAper2}}>}\right)^2 \\ 
   &\qquad + \left(\frac{\texttt{CI}}{<{\texttt{CI}}>}\right)^2 + \left(\frac{D}{<{D}>}\right)^2 \\
   &\qquad + \left(\frac{\texttt{MagAper2}-\texttt{MagAuto}}{<{\texttt{MagAper2}-\texttt{MagAuto}}>}\right)^2
 \end{aligned}
}\qquad ,
\end{equation}
where ${<}X{>}$ indicates the median value of each parameter of the light curve. When parameters were not available (which most commonly occurs when a flux in one of the apertures is negative),
they were set to zero. The synthetic error was used to identify data points that differ from the rest of the measurements for each object. In principle, the synthetic error could be used on its own to identify bad measurements. However, due to the image misalignment problem and uncertainties of the parameters (which may result in some outliers with normal synthetic error), the synthetic error had to be combined with additional outlier-rejection steps. 

A weighted robust linear fit \citep{2002nrca.book.....P} was performed for each light curve, using the synthetic error to set relative weights of the light curve points instead of the photometric error. 
The fit was used to obtain the scatter of measurements, 
the robust sigma ($\sigma'$; a resistant estimate of the dispersion of a distribution, 
which is identical to the standard deviation for an uncontaminated distribution), 
around the best-fit line and mark outlier points deviating by more than $3\sigma'$ from that line. 
The ``potential outliers'', with magnitude measurements within $3\sigma'$ but having their synthetic error $>4\sigma_{syn}$ 
were also marked at this step. 
The $\sigma_{syn}$ was calculated in a similar way to $\sigma'$, but instead of using a linear fit, 
the median value was adopted, since the components are expected to be constant and to be measured in the same way, 
unless external contaminating factors exist. 

By comparing the light curves of all sources in a given subgroup,
we identified visit-combined images having $>20\%$ of their sources marked as outliers. 
These visits were marked as ``bad'' and all measurements obtained during these visits 
(not only the ones marked as outliers) were considered unreliable and discarded from the analysis. 
This was found to be a very efficient way of identifying corrupted images. 
Once the bad visits were removed from the dataset, the robust linear fit was repeated
as the removal of a bad visit may have changed the $\sigma'$ values.

The magnitudes predicted by the robust linear fit for each visit were used to compute the local zero-point corrections \citep[e.g.][]{2014MNRAS.442.2381N}. For each source we used other HSC sources within a radius of 20$^{\prime\prime}$ around it to compute the correction. For each visit we calculated the magnitude zero-point correction for a given source as the median difference between magnitudes predicted by the linear fit and the ones measured for the nearby sources that have measurements in the same visit that are not marked as outliers or potential outliers (i.e., having <$3\sigma'$ and <$4\sigma_{syn}$). This correction should be able to eliminate the photometric zero-point variations from image to image and from one area of the chip to the other 
(as long as the spatial variation of the zero-point is sufficiently smooth), and also the effects of thermal breathing of the telescope, changing of PSF, 
for which the total flux difference from visit to visit can reach up to $\sim$6\% \citep{2017MNRAS.470..948A}. 
It may also partly compensate for any residual charge transfer inefficiencies \citep{2015MNRAS.453..561I} that remain 
after the corrections that were applied at the image processing stage. 

\begin{figure}
    \centering
    \includegraphics[scale=0.5]{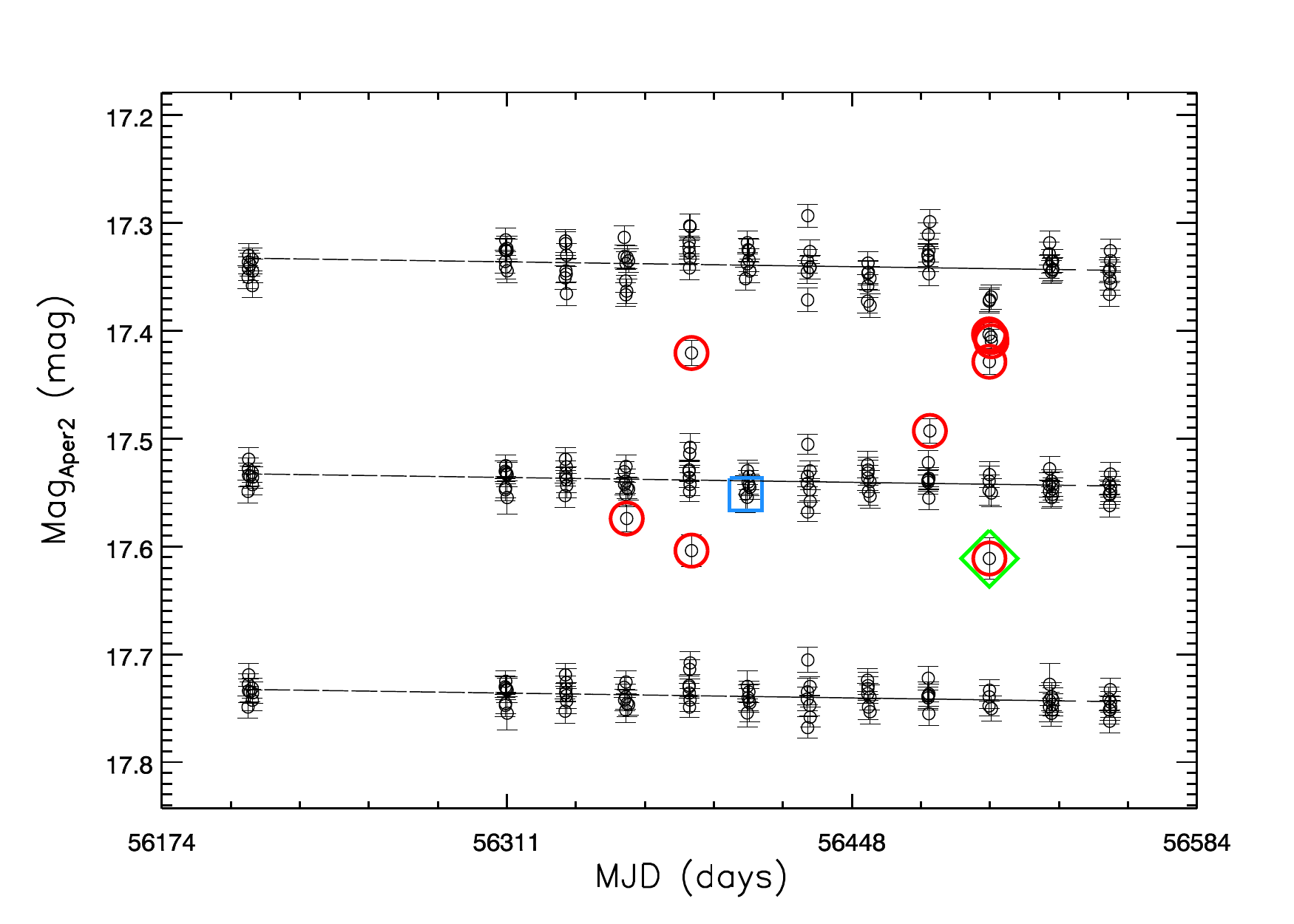}
    \caption{Light curve of a source in the field of M4 (\texttt{MatchID}=27382770; WFC3\_F775W filter), demonstrating the application of the preprocessing procedure. The figure shows the original light curve retrieved from the HSC (top), the local-corrected light curve (middle), and the final cleaned light curve (bottom). For clarity, the light curves are offset by 0.2 mag.  In each step of the procedure, the outliers (>$3\sigma'$; red circles) and potential outliers (<$3\sigma'$ and >$4\sigma_{syn}$; blue squares) are marked. Outliers that also have a large synthetic error (>$4\sigma_{syn}$) are marked by green diamonds.}
    \label{fig:preprocessing_example}
\end{figure}

After the local correction was applied, the robust linear fit was performed again as the local correction (just as the bad visit removal above) may change the $\sigma'$ values. We flagged all the outliers and potential outliers in all light curves and discarded them from further analysis 
in order to reduce the rate of false detections among the candidate variables. While application of the local zero-point correction considerably improved light curve quality, the procedure cannot correct the extreme outliers. These outlier measurements are associated with poor quality images and with cases where photometry of an individual star, rather than a group of nearby stars, is corrupted by an image artifact. Figure~\ref{fig:preprocessing_example} presents an example of the preprocessing procedure applied to a light curve in M4. We note that the outliers were removed by the preprocessing in an iterative process. Also, the systematic offset between different visits was removed, which reduced the light curve scatter.

The data preprocessing techniques used for the production of the HCV catalog can be applied to any other time-domain survey, following a careful evaluation of the dataset. Remaining issues, which correspond to limitations and caveats of the HCV catalog, are described in Section~\ref{subsec:limitations}.

\section{Algorithm for detecting candidate variables}
\label{sec:hcvalgorithm}

Our goal is to recover all variable objects that can in principle be recovered from each dataset in the HSC~v3. 
The efficiency of the HCV pipeline in finding variable objects should be limited by the input data, not by the processing algorithm.

We require a general-purpose variability detection algorithm that is robust to individual outlier measurements, 
applicable to a wide variety of observing (sampling) cadences and efficient in detecting a broad range of variability patterns, including periodic and non-periodic ones, rapidly and slowly varying objects, and transients visible only on a small subset of images of a given field. Taking into account the heterogeneous nature of the input HSC~v3 data, we 
tested various statistical indicators of variability (``variability indices'', Section~\ref{sec:vrsearchtech}), 
which characterize the overall scatter of measurements in a light curve and/or degree of correlation between 
consecutive flux measurements.

\cite{2017MNRAS.464..274S} presented a detailed description and comparison of 18 variability indices proposed in the
literature. These indices were tested on seven diverse sets of ground-based photometric data containing a large number of known variables. Simulated data were also used to investigate the performance of the indices based on the number of points in a light curve. 
The authors concluded that for light curves with a small number of points, the best result is achieved with variability indices 
quantifying scatter (such as the interquartile range and median absolute deviation). This study resulted from the development phase of the HCV variability detection algorithm and the search for the optimal variability indices for the HSC data. We complement this study with simulations based specifically on the HSC data, which are described in Appendix~\ref{sec:simulations}.

We adopted the median absolute deviation (MAD) as the robust variability index for the HCV detection algorithm. 
The MAD is defined as
\begin{equation}
{\rm MAD} = {\rm median}(|m_i-{\rm median}(m_i)|),
\end{equation}  
where $m_i$ is the magnitude of $i$'th point in the light curve.
This index is robust to individual outlier measurements and sensitive to a broad range of variability types. In a five-point light curve, up to two points may be completely corrupted without compromising the MAD value. The interquartile range (IQR), another robust variability index discussed by \cite{2017MNRAS.464..274S}, is less robust to outliers in the extreme case of a five-point light curve: it will be able to tolerate one or zero outlier points depending on the exact implementation. According to our simulations described in Appendix~\ref{sec:simulations}, MAD is more efficient than the IQR in a data set heavily contaminated with outlier measurements when the number of light curve points in small 
($n_{\rm LC} < 10$).

\begin{figure}
    \centering
    \includegraphics[scale=0.55]{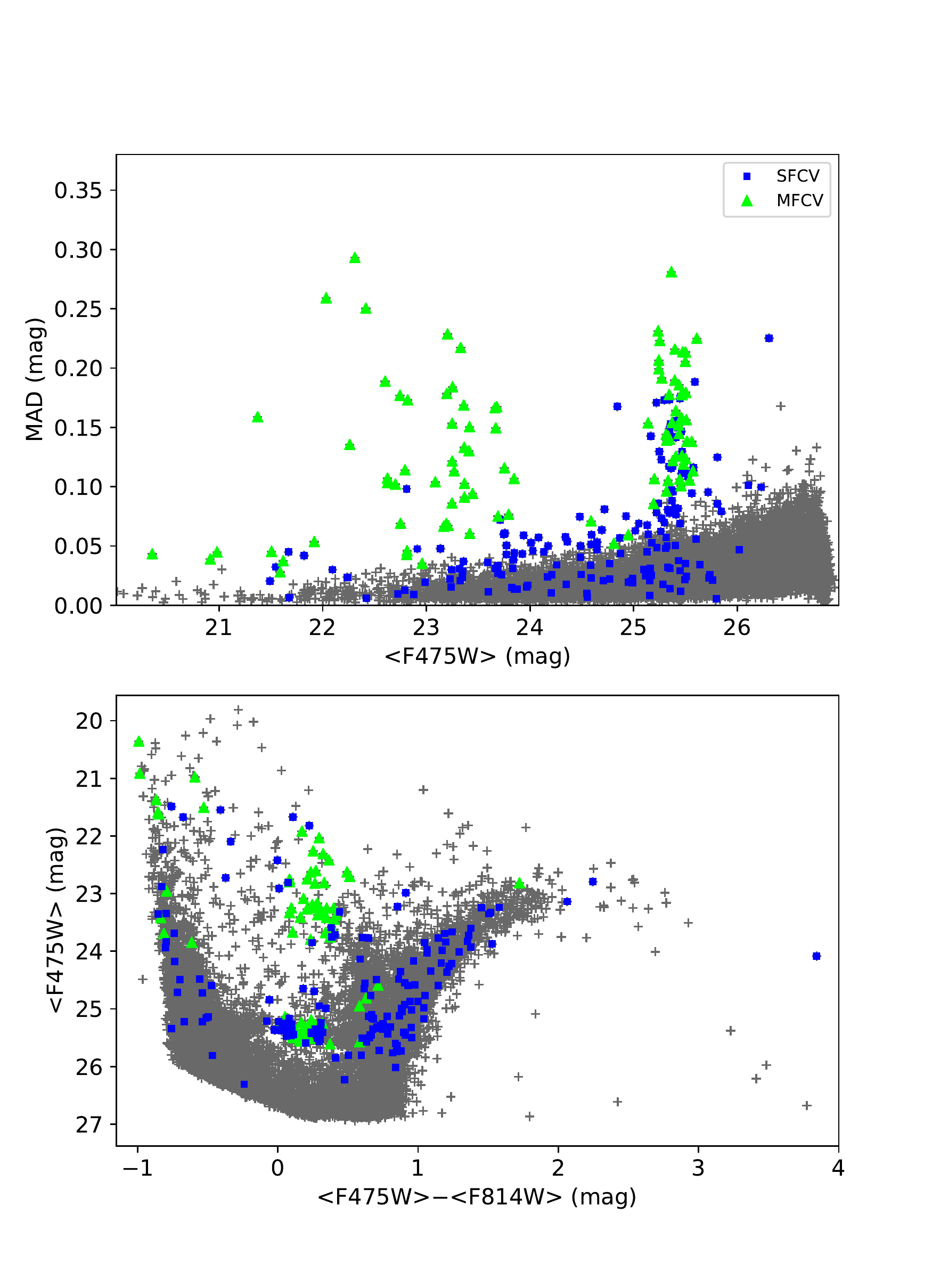}
    \caption{{\it Upper panel}: MAD vs. median ACS\_F475W magnitudes for all sources in the Control Sample field IC~1613, shown with gray crosses. The multi-filter variable candidates (MFVCs) are marked with green triangles, while the single-filter candidates (SFVCs) are marked with blue squares. SFVCs that have low MAD values in the F475W filter were selected as having MAD values above the threshold in the F814W filter. {\it Lower panel}: Color-magnitude diagram for IC~1613 showing the location of the MFVCs and SFVCs in the field. Many of the candidate variables lie on the instability strip.}
    \label{fig:IC1613_MAD}
\end{figure}

For each HSC group and for each filter, the HCV pipeline constructs a diagram of the median magnitude of each source versus the value of MAD for its light curve. 
The candidate variables are identified as the sources having a MAD value above a magnitude-dependent $5\sigma$
threshold. Figure~\ref{fig:IC1613_MAD}  presents an example from the Control Sample field IC~1613 (see also Table~\ref{tab:CS}). The location of the selected variables on the color-magnitude diagram is also shown.

The magnitude-dependent threshold was calculated for each subgroup in each filter as follows. First, the sources were ordered in magnitude. We rejected sources within 0.2\,mag of the faintest source as they may be affected by background estimation inaccuracies and residual cosmic rays. We similarly rejected sources within 0.5\,mag of the brightest star, to avoid saturation problems. The usable range of magnitudes varied depending on the gain setting of the camera. We divided the range into 20 overlapping bins in magnitude and calculated the median and clipped $\sigma$ values of MAD for each bin. The MAD value of each source in the bin was compared to the threshold ${\rm median}({\rm MAD})+5\sigma$ (where the median is computed over the MAD values of all sources within that bin) 
and sources above the threshold were marked as variable. The pipeline continued to the next bin, 
but also included 30\% of the sources from the previous bin so that the bins were overlapping. 
A source located in an overlapped bin was marked as variable if it was above the threshold in at least one of the bins. 
The candidates having large photometric errors (as estimated by \texttt{SExtractor}) were rejected at 
this stage by requiring the value of reduced ${\chi^2>3}$ for the null-hypothesis of the source magnitude being constant. 
The outcome of the detection algorithm was a list of candidate variables, which was input to the validation algorithm described 
in the following section.

The above algorithm failed in the rare cases where the majority of stars in a magnitude bin were actually variable. In such cases, the calculated threshold was too high and some real variables failed to pass it. The dwarf galaxy Eridanus~II (\texttt{GroupID} 1075853) is an extreme example for this situation. Here, no RR~Lyrae variables were detected as they all occupy a narrow magnitude range (being horizontal branch stars, all at the same distance).

\section{Algorithm for validating candidate variables}
\label{sec:validationalgorithm}

The candidate variables that were identified by the application of the variability detection algorithm (Sec.~\ref{sec:hcvalgorithm}) 
were evaluated and further characterized by the DPP using a validation algorithm. 
This algorithm applied a series of criteria leading to variability quality flags assigned to each candidate variable. Furthermore, expert validation was conducted for a random subset of groups to evaluate the performance of the HCV pipeline. Although we implemented and tested several period finding algorithms in the HCV pipeline, they did not yield useful results, given the inhomogeneous cadence and small number of epochs of the HCV light curves (typically having less than ten points), and were therefore turned off in the final version of the pipeline.

Initially, the validation algorithm determined the number of filters for which a variable candidate displayed variability. 
If variability was detected in more than one filters, then the candidate was classified as a ``multi-filter variable candidate'' (MFVC), 
otherwise it was classified as a ``single-filter variable candidate'' (SFVC). 
The SFVCs include two different classes: (a)\,sources for which there are data available only in one instrument and filter combination, 
(b)\,sources for which there are data in more than one instrument and filter combinations, but variability was detected only in one of them. 
One might assume that these variable candidates are less reliable. 
However, the two major classes of variables in the HCV catalog, namely Cepheids and RR Lyrae variables (see Section~\ref{sec:results}) 
have a larger amplitude in the blue than in red, so some of them may not be detected in the redder filters. Also, there are cases where the quality of the photometry is much worse in one instrument and filter combination, which may also affect the variability detection. The remaining sources that showed no significant variability were classified as ``constant sources''. In addition, a ``variability quality flag'' was assigned to each candidate variable. This variability quality flag aims to quantify the variation of the source image characteristics between visits, given that a corrupted photometric measurement may be associated with a noticeable change in the source image shape (e.g., Fig.~\ref{fig:out45740877_WFC3_F775W_img}).

The parameters used to assign the variability quality flag are the concentration index \texttt{CI}, 
the offset distance \texttt{D}-parameter, \texttt{MagerrAper2}, the difference between \texttt{MagAper2} and \texttt{MagAuto}, 
which have been defined in Section~\ref{sec:preproc}, and the peak-to-peak amplitude (p2p) in the light curve. 
For each source and each light curve point there is one value for each of these parameters. We derived the standard deviation for each of these parameters and for each source light curve, 
$\sigma _{par,i}$, where $i$ signifies the $i^{th}$ source. We also constructed the distribution of $\sigma _{par,i}$ for all sources lying in the same magnitude bin as the source in question, 
whether they are variable or not. This distribution is fit with a gaussian function and the mean value and standard deviation 
($\sigma _{dist}$) were calculated. The amount by which $\sigma _{par,i}$ differs from the average value within the magnitude bin is indicative of the quality of the photometric data for the specific source
compared to other sources of similar brightness in this subgroup. For example, if $\sigma_{\texttt{CI},i}$ for a particular source is much higher than average, the source is probably blended. 

Based on these parameters, we constructed a variability quality flag, 
which consists of five letters and quantifies the deviation of each parameter from 
the average behavior within the subgroup. 
Each flag can obtain the values A (highest quality), B, or C (lowest quality). 
In the HCV output, the flags are ordered as follows: 
\texttt{CI}, \texttt{D}, \texttt{MagerrAper2}, \texttt{MagAper2}-\texttt{MagAuto}, p2p. 
The assignment of values A, B, and C depends on the deviation of a value from the average. 
The criteria for the first four parameters are defined as follows:
\begin{itemize}
    \item value A: $|\sigma _{par,i}$– $\overline{\sigma _{par}}|<3\sigma _{dist}$
    \item value B: $3\sigma _{dist}\leq |\sigma _{par,i} $–$  \overline{\sigma _{par}}|<5\sigma _{dist}$
    \item value C: $|\sigma _{par,i}$– $\overline{\sigma _{par}}|\geq5\sigma _{dist}$.
\end{itemize}
For the p2p parameter, the values A, B, C, are defined as follows:
\begin{itemize}
    \item value A: $|\sigma _{p2p,i}$– $\overline{\sigma _{p2p}}|\geq5\sigma _{dist}$
    \item value B: $3\sigma _{dist}\leq |\sigma _{p2p,i}$–$ \overline{\sigma _{p2p}}|<5\sigma _{dist}$
    \item value C: $|\sigma _{p2p,i}$– $\overline{\sigma _{p2p}}|<3\sigma _{dist}$.
\end{itemize}

A comparison of flag values to expert-validated variables (see next Section) shows that 
a candidate is more likely to be a true variable if there are at least 3A's in the quality flag 
and if the \texttt{D}-parameter and \texttt{MagAper2}-\texttt{MagAuto} have a quality flag A. However, there were cases where the pipeline was not able to evaluate one or more of the parameters (denoted by a "dash" in the variability quality flag), 
for example, when \texttt{SExtractor} returned a negative flux or when $\sigma _{dist}$ 
was very small and its value was rounded to zero. The latter occurred when the number of sources per bin was small.

\subsection{Expert validation}

The variable candidates produced by the pipeline were individually evaluated by ``expert users'', 
using an ``expert tool'' interface developed for this purpose. 
Due to the large number of candidates and time constraints of the project, only the multi-filter variable candidates and a random sub-sample of single-filter candidates were visually inspected.
It is noted that the experts flagged 25 subgroups as unreliable because they presented a large number of artifacts (see Section~\ref{subsec:limitations}).
These unreliable groups contained around 50\% of the multi-filter variables identified by the pipeline. 
The results of the expert validation are discussed in Section~\ref{sec:results}. 
The majority ($\sim75\%$) of the remaining multi-filter variables were expert validated by three experts, 
while the single-filter variables were validated by one expert. The expert-validated variables, 
and in particular the multi-filter variable candidates are considered highly reliable. 

The expert users inspected the ``discovery diagram'' (MAD versus magnitude, with the calculated thresholds used for variable selection), the light curves of a specific candidate variable, its location on the color-magnitude diagram (when magnitudes in at least two different filters were available), the variations of \texttt{CI} and \texttt{D} as a function of time, the variability quality flag and the appearance of the candidate on three image ``stamps'' (corresponding to the faintest, brightest and median points in the source light curve) downloaded from the HLA\footnote{The image stamps were accessed
through the public {\it fitscut.cgi} interface at \url{http://hla.stsci.edu/fitscutcgi_interface.html}}. 
Taking into consideration all the different diagnostics, the expert classified a candidate as a ``high-confidence variable'', a ``probable variable'', 
or a ``possible artifact''. It must be noted that the expert validation relies heavily on the inspection of the three stamp images. 
Low amplitude variability (less than 0.5~mag) is difficult to assess by eye, especially when no neighboring comparison sources
are visible on the same stamp. Therefore, it is possible that a low amplitude or an isolated source is not confirmed as a high confidence 
variable by an expert, while it may actually be variable. Therefore, a significant percentage of the ``probable variables'' are likely true variables.

The classifications by the expert users were merged using a simple voting algorithm. 
If one of the three outcomes had a majority vote -- it was accepted as the
final result. If there was one vote for ``high-confidence variable'' and one 
for ``possible artifact'' then the result was ``probable variable''. 
If there were equal votes for ``probable variable'' and ``possible artifact'', or ``probable variable'' and ``high-confidence variable'', 
then we accepted the ``possible artifact'' and ``high-confidence variable'' result, respectively. 
The expert classification is included in the HCV catalog as an additional flag.

The expert validation procedure was very useful in discovering and discarding spurious candidate variables. In most cases, such spurious variables originated from the proximity of the source to very bright stars and their diffraction spikes, or from blended sources, 
objects projected on a highly spatially variable background, extended and diffuse objects (e.g., galaxies) where small noise-induced variations in aperture centering induce false variability, and other issues, such as image misalignment (see Section~\ref{subsec:limitations} for a full list of caveats). 

\section{Performance and limitations}
\label{sec:limitations}

The performance of both the validation and the variability detection algorithms was monitored using ten representative \HST\ fields, constituting the ``Control Sample''. These fields have been previously searched for variables in dedicated, published studies. The Control Sample includes a globular cluster, galaxies of the Local Group, more distant resolved galaxies, and a deep field. The comparison between the detected (by the detection and validation algorithm) and documented (in the literature) variables in the Control Sample led to the identification of possible problems and to necessary revisions and refinement of both algorithms. 

\begin{table*}
\caption{Properties of the HCV Control Sample fields.}
\label{tab:CS}
\setlength{\tabcolsep}{0.05in}
\centering
\begin{tabular}{ccccccrrccc}
\hline\hline
Field & Distance & Instrument &Filters & ${\Delta_{\rm MJD}}^{a}$ &\# visits & $N_{\rm sources}$ & \# known & Recovery & Type of  & Reference$^{b}$\\
Name     &         &        &        &   (days)                        &                      &            &      variables           &  rate (C)&   variables \\
\hline
 
M~4        & 1.86 kpc & WFC3 & F467M,     & 300 & 100 &	8,460      & 38  &  0.32    & RR Lyr,    &  1  \\
           &          &      & F775W      &     &     &           &     &      & EB         &       \\
IC~1613    &  760 kpc & ACS  & F814W,     &  3  &  12 &     23,106 & 182 &  0.71    & RR Lyr,    &  2  \\
           &          &      & F475W      &     &     &           &     &      & Cepheids,  &    \\
           &          &      &            &     &     &           &     &      & EB         &       \\
M31-Halo11 &  770 kpc & ACS  & F814W,     & 40  & 32  &     10,059 & 115 &   0.80$^{c}$  & RR Lyr,    &  3  \\
           &          &      & F606W      &     &     &           &     &      &  Dwarf Cepheids,    &       \\
&          &      &      &     &     &           &     &      &  LPVs,    &       \\
&          &      &      &     &     &           &     &      &  semiregulars    &       \\
M31-Stream &  770 kpc & ACS  & F814W,     & 30  & 16  &      6,792 & 24  &  0.88    & RR Lyr     &  4  \\         
           &          &      & F606W      &     &     &           &     &      &            &       \\
M31-Disk   &  770 kpc & ACS  & F814W,     & 39  & 15  &     10,644 & 23  &  0.83    & RR Lyr     &  4  \\         
           &          &      & F606W      &     &     &           &     &      &            &       \\  
M101-F1    &  6.4 Mpc & ACS  & F814W,     &$>$30& 14  &     58,263 & 411 &  0.80    & Cepheids   &  5  \\
           &          &      & F555W      &     &     &           &     &      &            &       \\      
NGC~4535   &  16 Mpc  & WFPC2& F814W,     & 75  & 14  &     1,032  &  50 &   0.09   & Cepheids,          &  6,7  \\
           &          &      & F435W      &     &     &           &     &      &  supergiants  &       \\       
M~87       &  16.5 Mpc& ACS  & F814W      & 30  & 48  &     15,731 &  32 &   0.63   & Novae      &  8  \\
NGC~1448   &  17.3 Mpc& WFC3 & F350LP,    & 50  & 11  &     9,228  &  54 &  0.44   & Cepheids   &  5 \\
           &          &      & F160W      &     &     &           &     &      &            &       \\
GOODS-S    &         & ACS  &several     &50 to&5 to &     14,278 & 116 &  0.26    & AGN,       &  9\\
           &          & WFC3 &            & 3000& 120 &           &     &      & SNe        &   \\       
\hline
\end{tabular}
\tablefoot{\tablefoottext{a}{The values for the Mean Julian Date (MJD) are approximate as they can differ among sources in the group.}\\
\tablefoottext{b}{(1) \citet{2014MNRAS.442.2381N}, (2) \citet{2010ApJ...712.1259B}, (3) \citet{2004AJ....127.2738B}, (4) \citet{2011AJ....141..171J}, (5) \citet{2016ApJ...830...10H}, (6) \citet{1999ApJ...521..155M}, (7) \citet{2018A&A...618A.185S}, (8) \citet{2016ApJS..227....1S}, (9) \citet{Pouliasis19}.}\\
  \tablefoottext{c}{The recovery rate is 0.90 for RR Lyrae variables alone.}}
\end{table*}

The Control Sample fields were selected from HSC~v1, which was available during the early development phase of the HCV project, on the basis of the following requirements: 
\begin{itemize}
    \item {Availability of accurate astrometry for the known variables}
    \item {Coverage of as wide a range of input data characteristics as possible, namely,
          a wide range in the number of visits, source number densities, numbers and types of known variables, distances, as well as instrument and filter combinations.}
\end{itemize}

Table~\ref{tab:CS} presents the
characteristics of each Control Sample field: the name of the field, the average distance of the sources in the field (when applicable), the instrument(s) and filter(s) used, the time baseline of the data, calculated as the difference in the Modified Julian Dates (MJD) of the start and end of the observations, the median number of visits (since not all sources belonging to the same \texttt{GroupID} have the same number of visits), the total number of sources, the number of published variables, the completeness, C, of the recovered published variables, that is the ratio of the number of detected candidate variables over the number of published variables that are included as sources in the HCV catalog sample, the type of published variables, and the corresponding reference(s).

After selecting the fields appropriate for the assembly of the Control Sample, astrometric corrections were estimated and applied to the published coordinates of the variables in each field, where necessary, in order to make cross-matching with the HSC possible. Indeed, many published \HST\ variables lack proper astrometry, while only pixel coordinates (with or without finding charts) are provided by some authors. For example, we found offsets as large as 5$\arcsec$ between published \citep{2010ApJ...712.1259B} and HSC coordinates for sources in IC~1613. 

Although care has been taken for the Control Sample to be as representative as possible, it is clear that there are several cadence profiles that one may encounter in the HSC, but not in published data. 
In order to better characterize the variability detection efficiency we use simulations injecting artificial variability into real HSC light
curves and then reduce the number of points by randomly removing observations. The simulations are described in Appendix~\ref{sec:simulations}. They show that the efficiency of variability detection increases dramatically with the number of light curve points increasing from five to ten. For the
larger number of points, the efficiency continues to rise, but more slowly
(Fig.~\ref{fig:sim_npoints_WFC3_F775W}, Fig.~\ref{fig:sim_npoints_WFC3_F467M}). 
This result is valid for the situations where the variability timescale is shorter than the time difference between consecutive light curve points.

\subsection{Recovery of known variables}

The Control Sample fields were used to evaluate the performance of the variability detection and validation algorithms and the limitations present. The recovery rates C (defined as the ratio of variables identified by the pipeline over the total number of known variables) presented in Table~\ref{tab:CS} vary from 9\% for NGC~4535 (WFPC2) to 88\% for the M31-Stream (ACS) field. The recovery rate generally depends on the type of variable, the distance of the field studied, and the number of epochs available. WFPC2 systematically yields a lower recovery rate due to the lower data quality. Other conditions affecting the recovery rate include fields that are close enough for proper motions to cause a deterioration of the localization of the sources (e.g., Galactic bulge fields), fields that have several bright stars in their field of view (e.g., globular clusters), and extended sources (distant galaxies), where robust source centering is not possible. 

Generally, high-amplitude periodic variables are more easily detected, depending on the cadence of the observations. The use of visit-combined photometry adopted in the HSC, reduces the effective number of available epochs and often limits the detectability of fast variability, for example, eclipsing binaries (EB; with short duration of eclipses). This is the case for M4, where the majority of the variables are eclipsing binaries (the low recovery rate for the M4 field is also affected by blending issues and the presence of several bright stars with diffraction spikes). The reduction of available epochs may also affect the detection of transients. Additionally, the use of aperture rather than PSF photometry does not yield high quality photometry in more distant and/or crowded fields. A detailed description of the caveats in the HCV catalog is provided in Section~\ref{subsec:limitations}.

The only Control Sample field using WFPC2 is the \HST\ Key Project \citep{2001ApJ...553...47F} galaxy NGC 4535, which is known to host 50 Cepheids \citep{1999ApJ...521..155M}. \citet{2018A&A...618A.185S} performed PSF photometry using DOLPHOT on the archival images of the galaxy, and applied variability indices to recover the 50 known Cepheids and 120 additional candidate variable stars. The HCV catalog includes eight of the known Cepheids and 11 of the additional candidate variables. The differences in recovery rate and identification of variables are due to the fact that the HSC~v3 source lists for WFPC2 are not very deep and that the particular field is crowded. This field demonstrates the limitations of the HCV catalog results in crowded fields observed with WFPC2 (see also Section 6.2). Future releases of the HSC are expected to improve on the quality and depth of the WFPC2 source lists.

\citet{2019A&A...629A...3S} similarly analyzed WFPC2 data of the Key Project galaxies NGC 1326A \citep{Prosser99}, NGC 1425 \citep{Mould2000}, and NGC 4548 \citep[]{Graham1999}, which contain 15, 20, and 24 reported Cepheids, respectively. The study yielded 48 new candidate variables in NGC 1326A, 102 in NGC 1425, and 93 in NGC 4548. The number of variable sources recovered by the HCV catalog in the three galaxies are: six in NGC 1326A, eight in NGC 1425, and 15 in NGC 4548. We note that all variable sources detected by the HCV pipeline were identified as variable in this analysis, although few of the published Cepheids were included in the HCV catalog. We expect that a variability analysis of WFPC2 photometry based on future releases of the HSC will yield much improved results.

\begin{figure}
    \includegraphics[width=0.5\textwidth]{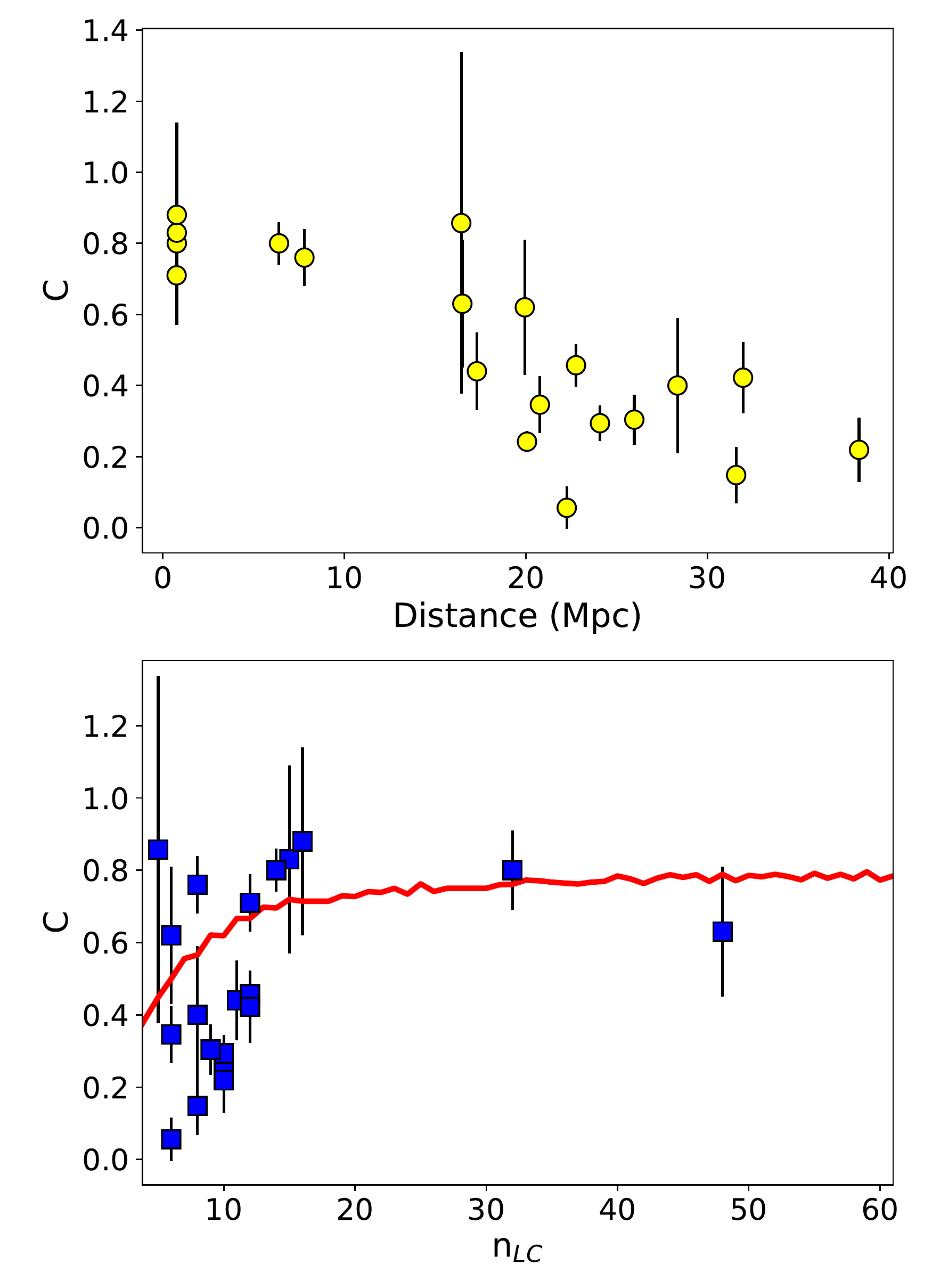}
    \caption{The dependence of recovery rate, C, of known variables on the distance of the host galaxy (upper panel) and on the average number of points in the light curve
    ($n_{\rm LC}$; lower panel) for galaxies in the Control Sample and in the \citet{2016ApJ...830...10H} galaxy sample, observed with the ACS or the WFC3 instruments. The red line indicates the results of the simulations (see text).}
    \label{fig:C-vs-distance}
\end{figure}

It is interesting to compare the HCV catalog success in recovering known variables as a function of distance of the host galaxy. This comparison highlights the limitations of using aperture rather than PSF photometry in the HSC, which mainly affects more distant galaxies, where crowding and blending becomes significant. In Figure~\ref{fig:C-vs-distance} we show the recovery rate C for variables in galaxies in the Control Sample, as well as Cepheids found in the 19 SN Type Ia host galaxies with \HST\ photometry and NGC~4258 analyzed by \citet{2016ApJ...830...10H} as a function of distance (upper panel) and $n_{\rm LC}$ (lower panel). We only considered galaxies observed with the ACS or WFC3 instruments. This comparison is of particular interest as the same original \HST\ data were used in the published catalogs. Errors are computed via error propagation, using the square root of the number of variables. Despite the significant scatter seen in the upper panel of Figure~\ref{fig:C-vs-distance}, there is a clear decrease of the recovery rate as a function of distance. The large scatter is caused by other factors that affect the variable detection process, such as the number of epochs available. The lower panel of  Figure~\ref{fig:C-vs-distance} shows the dependence of the recovery rate C on $n_{\rm LC}$. The recovery rate increases sharply between five and $\simeq$15 points in the light curve and then stabilizes. A similar behavior is displayed by simulated data (red line), described in the Appendix. The simulated recovery rate is somewhat higher than what is observed. This is probably caused by the fact that in the simulations variability is modeled as a simple sine variation with an amplitude randomly selected for each model variable source to be between 0 and 1\,mag. The real light curves are not sinusoidal in shape and the amplitude distribution is not uniform, but is weighted toward lower amplitudes (Figure~\ref{fig:NM_example}).

\subsection{Limitations}
\label{subsec:limitations}

The users should be aware of the following limitations of the HCV catalog:

\begin{enumerate}[itemsep=2pt,parsep=2pt]

\item The HSC pipeline is designed to process the majority of ACS/WFC, WFC3/UVIS, WFC3/IR, and WFPC2 images.
Its main design goal was applicability to a wide variety of input data,
rather than extraction of all possible information from a given data set
(which would require fine-tuning of the analysis procedure for these
specific data). The HSC (and the HCV) pipeline design is a compromise
between the general applicability and quality of the output.

\item The HSC is built from visit-combined images. This means that on one hand, it does not go as deep as a mosaic combining all visits of this specific field could go. On the other hand, the time resolution of the HSC is not a good as it could be, had the individual exposure images been used for photometry.

\item The synthetic error analysis is not capable of removing all the corrupted measurements. Thus, visual inspection of HLA images, light curves, and CMDs is highly recommended when using the HCV catalog, 
at least for the part of the catalog that has not been validated by the experts. 

\item For extended objects, the aperture centering algorithm, which is used to determine \texttt{MagAper2}, does not always yield the same exact pixel for different images of filters. This may result in apparent variability due to the offset of the \texttt{D}-parameter. Also, the aperture sizes used for \texttt{MagAper2} may be too small for the deep fields where false variability may be induced by the changes of the PSF \citep{vill2010}. This should be considered when studying deep fields, such as CANDELS, RELIC, CLASH, etc.


\item During the expert validation of the multi-filter variable candidates, some \texttt{GroupIDs} (or subgroups) were found to exhibit a relatively large fraction (in some cases over 10\%) of variable candidates, whereas, normally, the fraction of variable candidates is around 2-3\%. This is due to corrupted photometry caused by the reasons outlined in this section, or proper motion. For instance, in the ``Sagittarius Window Eclipsing Extrasolar Planet Search'' (SWEEPS) fields,  a fraction of sources have been split into two \texttt{MatchIDs} because of the detection of their large proper motions, as the two major observation periods are separated by $\sim$3000 days. The large amount of false candidates in such fields significantly delays the process of validation, forcing us to not fully expert validate all the unreliable groups. We suggest that users are cautious when exploiting these groups. Here is a list of the 25 unreliable
\texttt{GroupIDs}(\_subgroups): 24555, 33004, 33109, 53275, 56019, 73455, 289829, 439774\_5, 439774\_8, 1024360, 1033498, 1033692, 1039945, 1040910, 1042327, 1043384\_0, 1043756, 1045492, 1045904\_8, 1045904\_57, 1045904\_102, 1045904\_108, 1047823, 1063416, 1073046. It should be noted that version 3.1 of the HSC (released on 2019 June 26, after submission of this paper) provides proper motions for 400,000 sources in the SWEEPS field.

\item Some fields were found to contain a large number of outliers, which are mainly due to saturation and a diffuse background. In the case of saturation, it generally occurs in globular clusters in the Milky Way (e.g., NGC~6397) or nearby galaxies with long exposures (e.g., HIGH-GALACTIC-LATITUDE fields in the SMC), in which a small to large number of saturated stars create spikes and/or halos crossing the whole image and degrade the photometry. For cases with a diffuse background, which is typically related to star formation regions (e.g., 30 Doradus, Westerlund 2) or spiral galaxies with a large amount of gas and dust, the nebulous structures/clumps are sometimes detected by \SExtractor\, since it is designed to measure such extended objects.

\item The quality of photometry is degraded in fields containing large
numbers of bright stars (Milky Way globular clusters like NGC~6397 and the
fields in the SMC) which create diffraction spikes, halos, and saturation trails 
(due to charge bleeding) that together affect a large fraction of the image area.

\item In diffuse background regions with nebulosity (30~Doradus, Westerlund~2) or spiral galaxies, the nebulous clumps are sometimes detected as discrete objects and enter the HSC and our variability analysis.

\item Prior to HSC~v3, we found that there was a general problem of image alignment both for 
the white-light and single-filter visit-combined images, which leads to outliers in the light curves 
and consequently false variables. Those misalignments are generally caused by a failure to reacquire 
the guide stars after an Earth occultation (affecting the white-light image), or a slight drift and/or 
(small) rotation between exposures (affecting the single-filter visit-combined image). 
We recommend a careful look at the color images to make sure the stars are circular. 
For the few fields with a large number of observations in different filters, 
the alignment between different filters and exposures is more difficult. 
Furthermore, there are cases of ``intrinsic'' misalignment, such as moving objects 
(e.g., nearby stars with high proper motions, planets). 
Since HLA DR10 is the first data release that fixes misalignment between both exposures and filters, 
HSC~v3 data from both ACS and WFC3 should have very good alignment.

\item Besides the ``physical misalignment'' issue described above, an algorithm 
failure in the crossmatching of the same source between different filters/exposures 
also occurred occasionally during the construction of the HSC and resulted in outliers in 
the light curve and false variability in the HCV catalog. However, this is fairly rare 
and mostly occurs in crowded fields and/or IR wavelengths (e.g., NGC~4258). 

\item The tremendous improvements in the image processing of the latest release of the HLA, 
and therefore in the photometry and astrometry of the HSC, are mainly available for ACS and WFC3 data, not for WFPC2 data. 
At the moment, the HSC data quality is relatively lower for WFPC2\footnote{The improvement of the WFPC2 source detection is planned for a future HSC release.}, making source detections from WFPC2 data in the HSC and variability detection in the HCV catalog less reliable. Furthermore, the data quality
in the IR is relatively lower compared to the optical bands, that is for WFC3/IR versus the WFC3/UVIS, and ACS/WFC data, due to the more complicated instrumental effects (e.g., cosmic rays and snowballs, on-orbit degradation, image persistence, large background fluctuation, blobs, and optical anomalies).

\item The cadence of observations may affect the detection of variability. When several visits are taken in a relatively concentrated 
time period or relatively equal intervals, which can form a baseline for the preprocessing, the output of the HCV pipeline is reliable. 
However, if there is a big gap (e.g., several months or years) between two observational periods, the instrumental conditions and/or 
sky background subtraction may change and result in false variability. These effects are supposedly corrected by the preprocessing, 
but special attention is also needed when dealing with such cases, especially when there is only single observation after/before the gap.

\item Another problem of the HSC is that sometimes more than one measurements were kept in the database for the same source in a given image. This is likely due to that a source with critical size may appear slightly different in each image and just above the deblending threshold of \SExtractor. Occasionally one source may be split into two, resulting in a double-detection for the same source. Although the cross-matching of HSC white-light sources between images is initially done by using a friends-of-friends (FoF) algorithm with a specified search radius and a Bayes factor, some of the double-detections are not identified and merged during the process. The HCV pipeline has excluded all targets with double-detections.

\item There are still some remaining issues in the data set which are poorly understood. One such issue is that some ``variable candidates'' exhibit step-shaped light curves with a rapid jump or drop, which can reach up to about one magnitude (e.g., NGC~1015, NGC~1448, NGC~2442, NGC~7250, M31-POS44, Leo~A). Those targets are inevitably selected by the pipeline due to the large scatter in the light curves. 
Visual inspection indicates that some of them are extended gaseous structures or image artifacts, while others are real objects. The spatial distribution of those targets is also not fully in accordance with the CCD gaps. 
The exact cause of this phenomenon is currently under investigation. It may be due to the rotation of the telescope, which results in different CTE correction and background subtraction.

Another issue is that in a few fields, some light curves show very similar tendencies or shapes (e.g., NGC~2070, NGC~6388), which may be related to uncorrected systematic effects. Again, this is inevitable due to the complicated observational strategies and instrumental conditions of \HST. Moreover, for some fields, the saturation flag is not fully working (e.g., NGC~1851). For instance, one of the filters combined in the white-light image may be heavily saturated, which causes photometric degradation in other filters. For some deep fields (e.g., fields from RELICS, {\em Hubble} Frontier Fields), we also notice that an unusually large fraction of point sources show variability in their light curves and are selected as variable candidates, while the reason for this is still unknown. Crowding is generally an issue for globular clusters of the Milky Way, where the targets in the central region are largely blended, with corrupted photometry. Finally, due to \HST\ observational strategies, the HSC also includes some measurements from single, long exposures, which are heavily contaminated by cosmic rays.
\end{enumerate}

\section{Results}
\label{sec:results}

\begin{figure}
    \centering
    \includegraphics[width=0.50\textwidth]{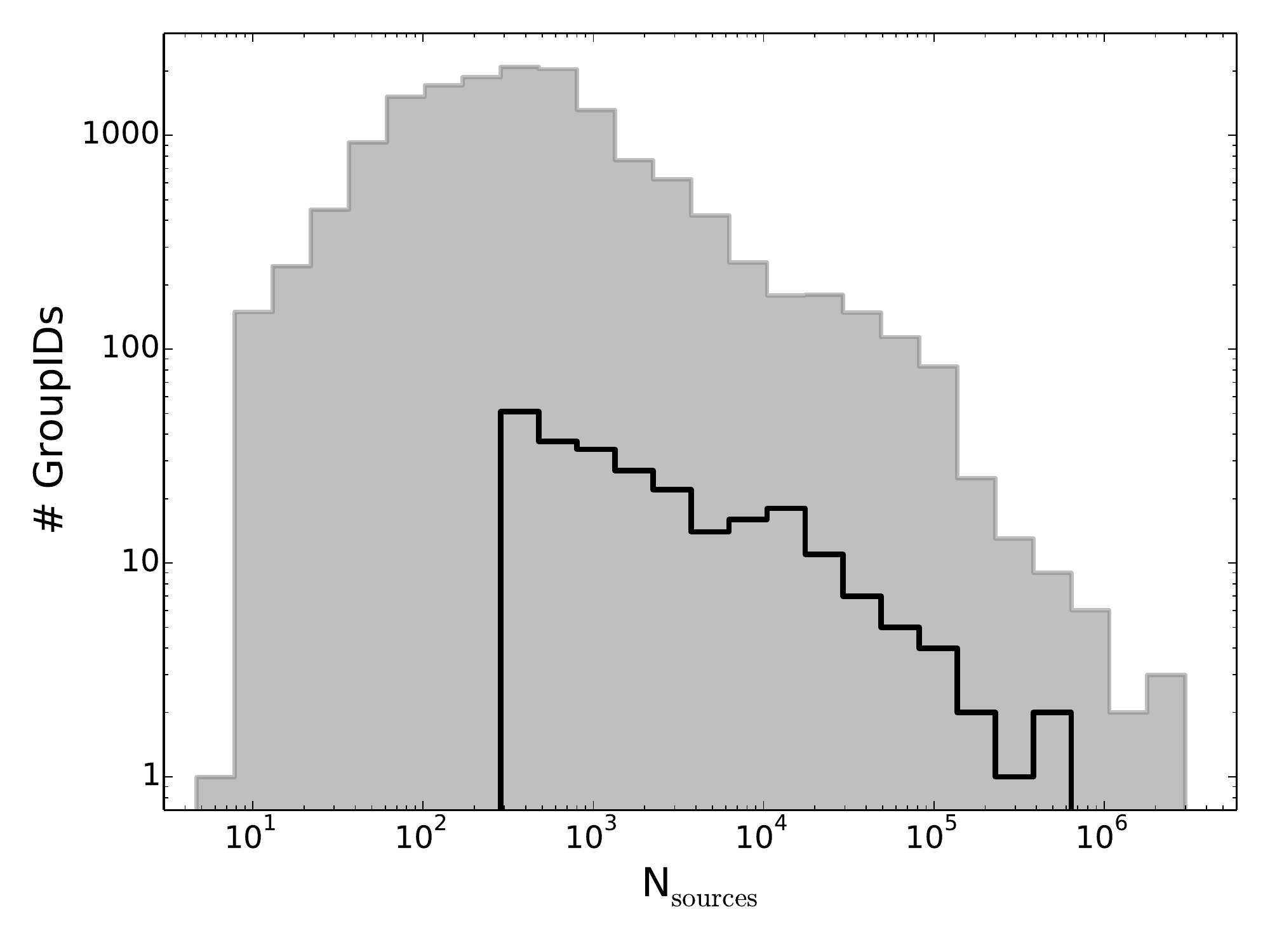}
    \caption{Histogram of the number of sources vs. number of \texttt{GroupIDs} for the entire HSC~v3 (gray) and the HCV catalog (black).}
    \label{fig:HSC_HCV_Number_sources_per_group}
\end{figure}

The HCV processing pipeline identified 84,428 variable candidates by applying the detection and validation algorithm to the sources in the HSC~v3 that passed the preprocessing procedure. Table~\ref{tab:hcvresults} summarizes the results: out of 15,160 \texttt{GroupIDs} in the HSC~v3, 250 \texttt{GroupIDs}\footnote{These were processed in 2,132
subgroups (see Section~\ref{sec:hcvsystem}).} met the selection criteria. Out of the 108 million sources (\texttt{MatchIDs}) in the HSC, 3.7 million sources passed the selection criteria and were processed by the DPP. This corresponds to 3.4\% of the sources and 1.6\% of the \texttt{GroupIDs}. The distribution of the number of sources per \texttt{GroupID} is illustrated in Figure~\ref{fig:HSC_HCV_Number_sources_per_group}. Histograms for the whole HSC~v3 catalog and also for the HCV catalog are plotted. The cut-off at N$_{\rm sources}\geq300$ appears to be near the peak of the distribution of the number of sources in the HSC.

In total, 84,428 sources (i.e., 2.3\% of the sources processed) were automatically flagged by the pipeline as variables, including 73,313 (87\% of the total) single-filter and 11,115 (13\% of the total) multi-filter variable candidates. The expert validation procedure was applied to 49\% of the multi-filter (after excluding unreliable groups rejected by the experts, see Section~\ref{subsec:limitations}) and 11\% of the single-filter candidates, which corresponds to 16\% of the total variable candidates. The results of the expert validation procedure are summarized in Table~\ref{tab:expertvalidation} and are indicative of the reliability of the automatic classification by the DPP. About 40\% of the variables were classified as high-confidence variables, while on average 41\% were classified as probable variables and 19\% as possible artifacts. There is a higher incidence of possible artifacts (22\%) among SFVCs than among MFVCs (16\%), as expected. If we extrapolate to the whole HCV catalog, this implies that $81\%$ of the variable candidates are true variables, while the remaining $19\%$ are artifacts. It should be noted that among multi-filter variables, the success rate is even higher (84\%). These percentages were derived excluding the unreliable groups that were rejected by the experts.

\begin{figure}
    \centering
    \includegraphics[width=0.50\textwidth]{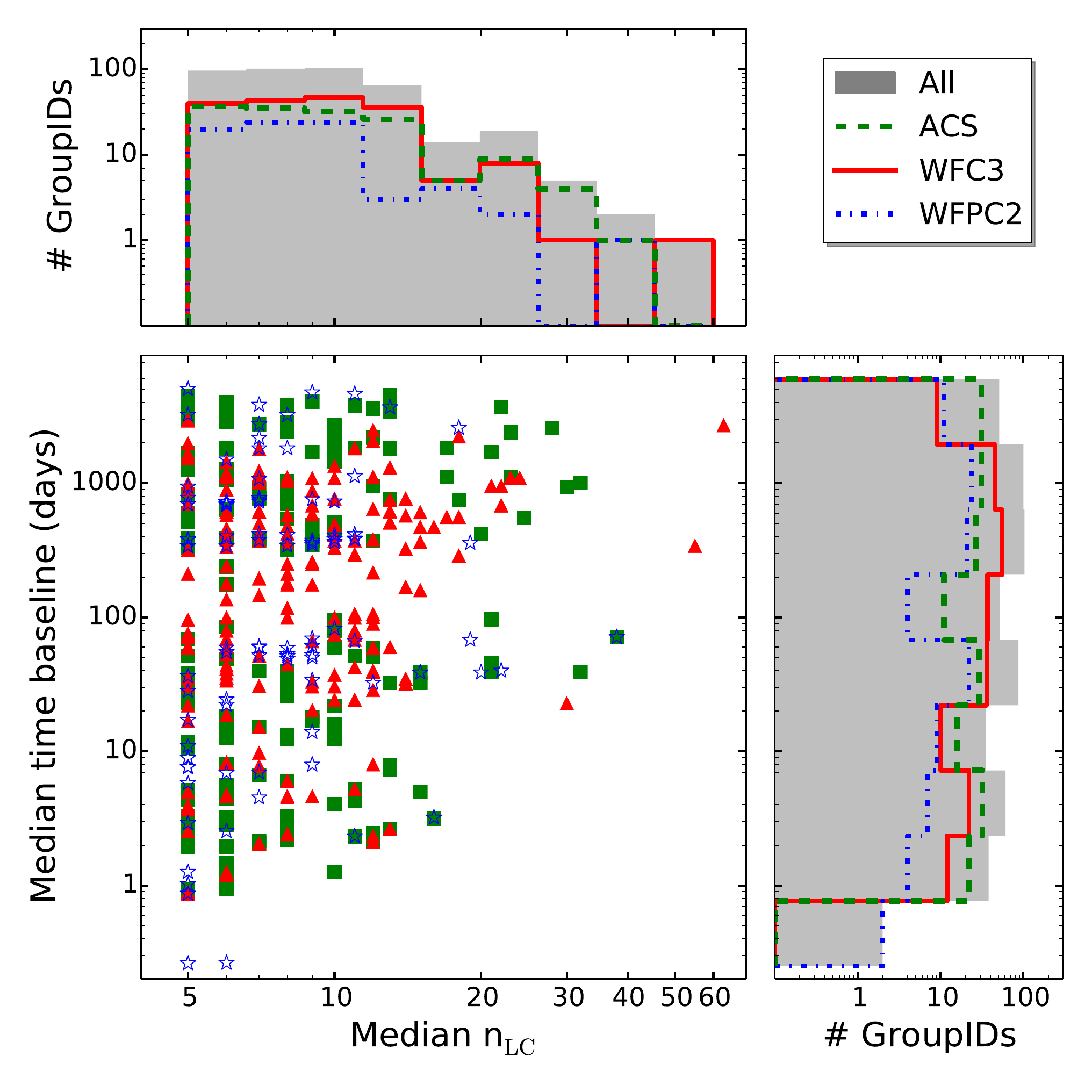}
    \caption{The median time baseline as a function of the median number of data points for all 250 \texttt{GroupIDs} processed by the HCV pipeline, labeled by \HST\ instrument: ACS (green squares), WFC3 (red triangles), and WFPC2 (blue asterisks). The side plots present histograms of the median time baseline and number of data points in light curves, respectively. The histogram lines are color-coded as above, while the gray shaded histogram denotes the total distribution of the three instruments.}
    \label{fig:HCV_Tbas_datapoints}
\end{figure}

In Figure~\ref{fig:HCV_Tbas_datapoints}, we present the median time baseline (i.e., the time difference in days between the first and the last observation of a source) as a function of the median number of data points in the light curve
($n_{\rm LC}$) for all 250 \texttt{GroupIDs} processed by the DPP, color-coded by \HST\ instrument. Clearly, WFC3 and ACS contribute more \texttt{GroupIDs} and therefore more variable candidates to the HCV catalog than WFPC2. The time baseline ranges from under a day to over 15 years with a relatively flat distribution among the \texttt{GroupIDs} and a peak between 200--2000 days. The number of data points in the HCV light curves ranges from five (our cut-off limit) to 120. A typical variable in the HCV catalog will have up to ten data points in its light curve. Given the relatively small number of points in the light curves, the HCV catalog does not provide classifications for the variable sources. 

%
  \begin{figure*}
    \begin{center}
    \begin{tabular}{  c  c }
    \includegraphics[width=0.95\columnwidth]{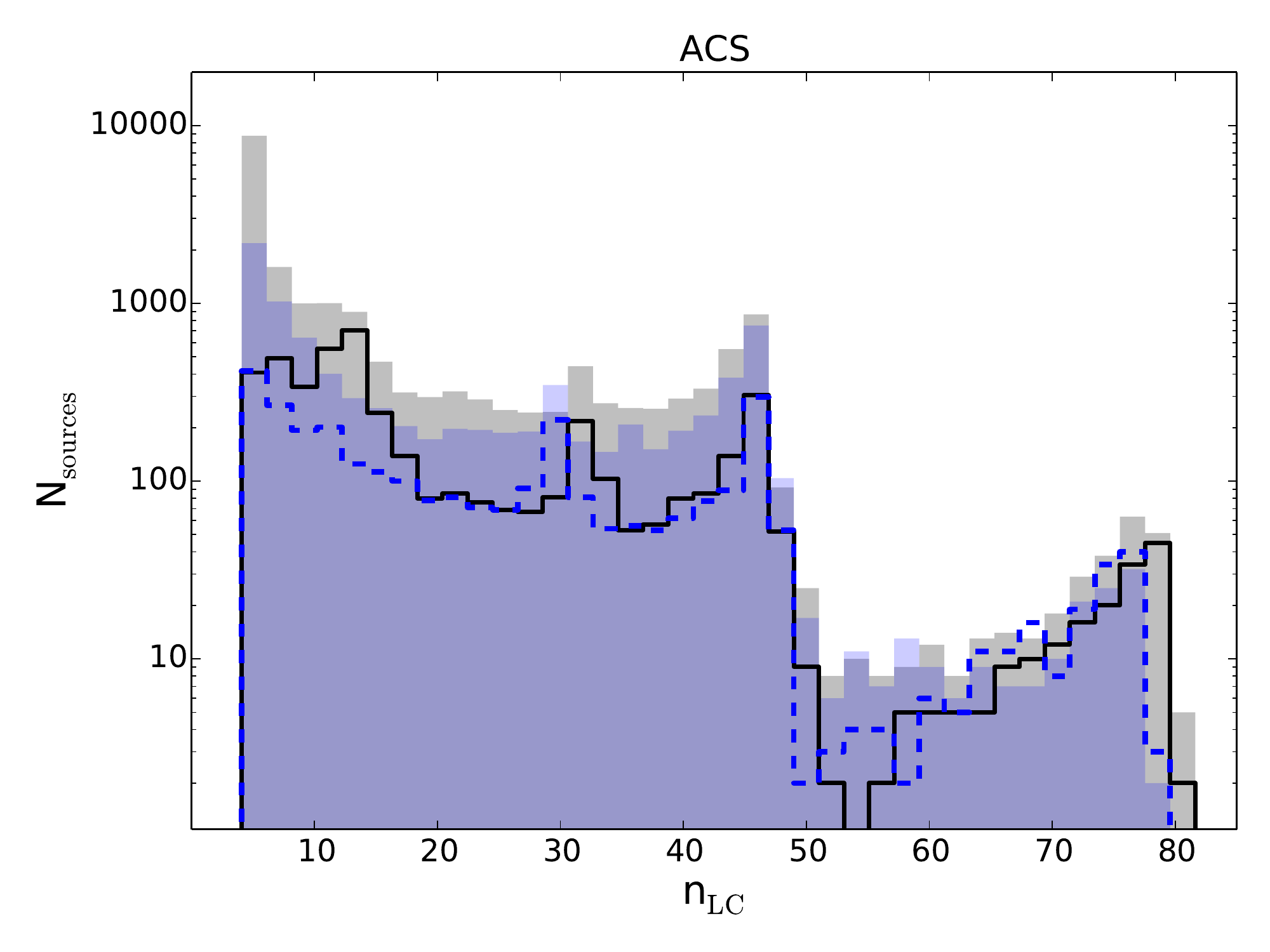} & \includegraphics[width=0.95\columnwidth]{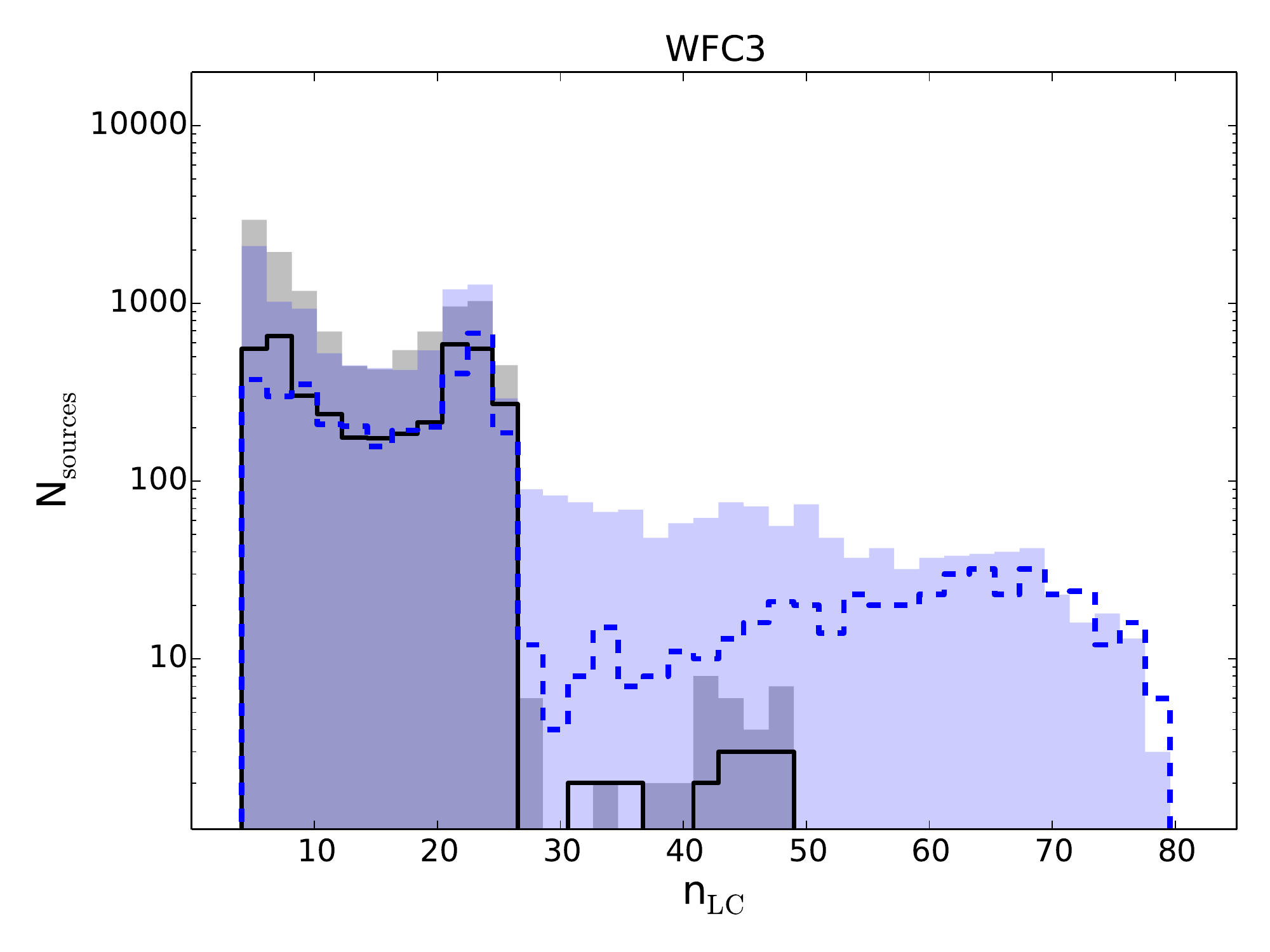}\\
    \includegraphics[width=0.95\columnwidth]{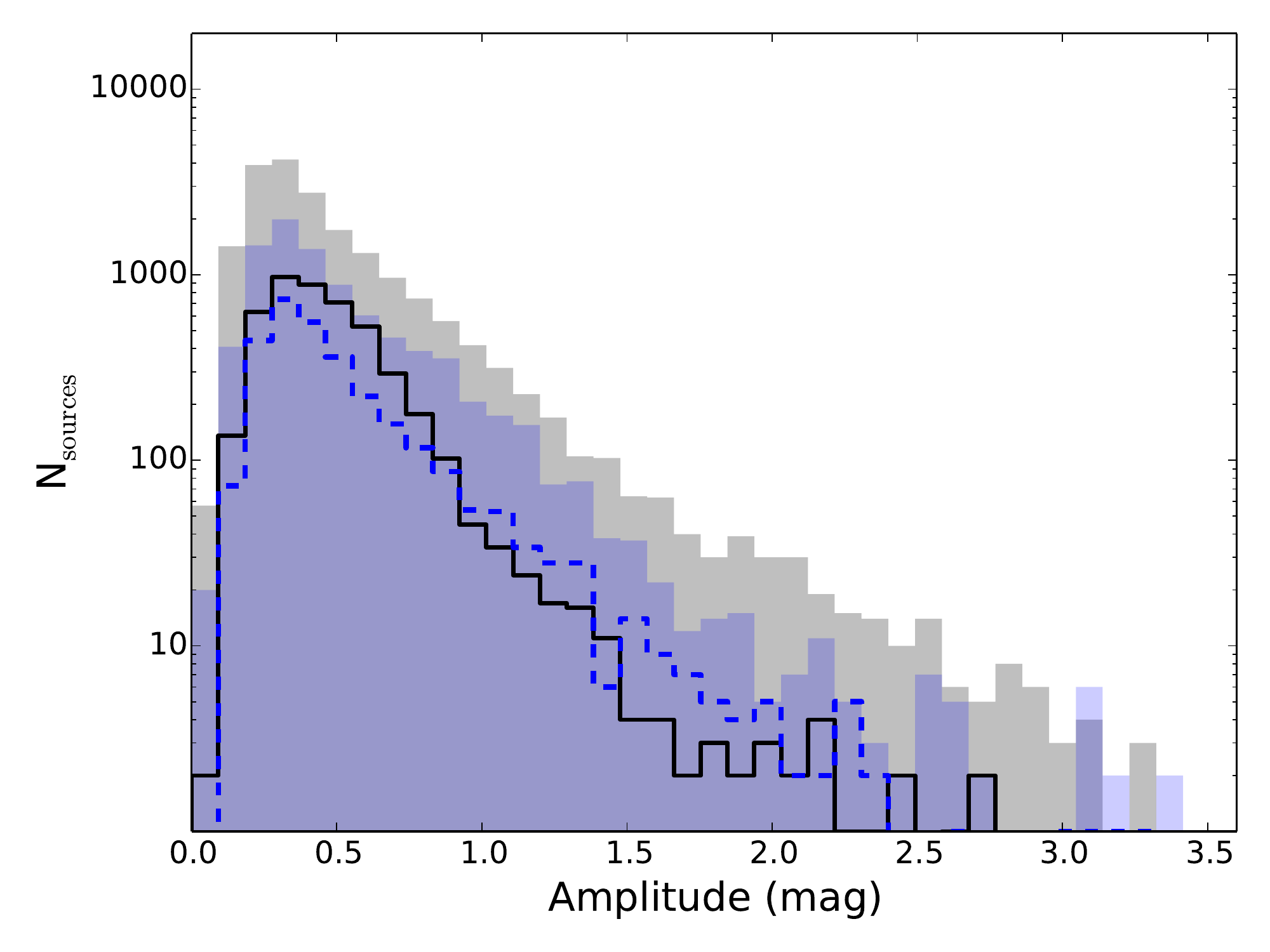}& \includegraphics[width=0.95\columnwidth]{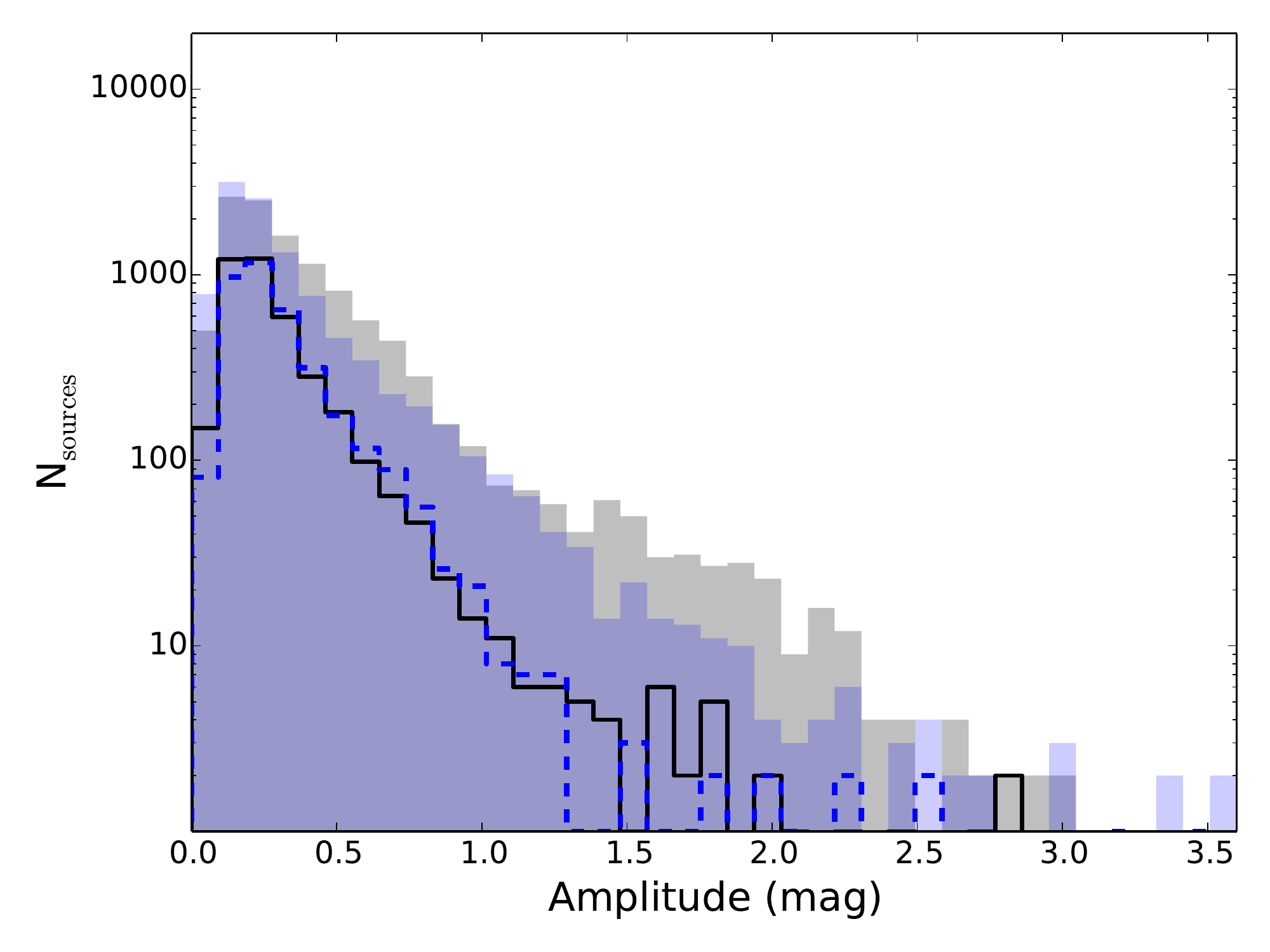}\\
    \includegraphics[width=0.95\columnwidth]{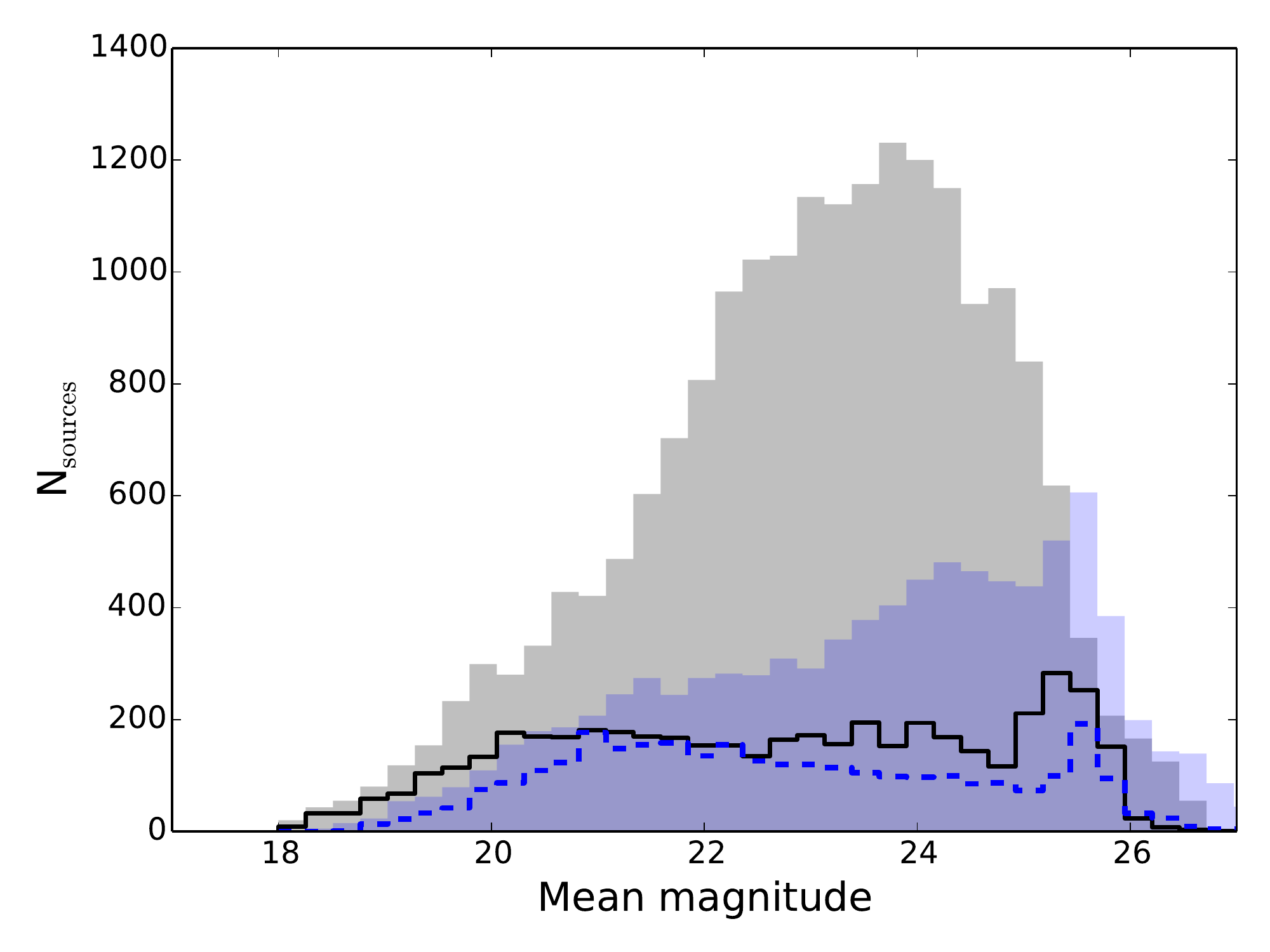}& \includegraphics[width=0.95\columnwidth]{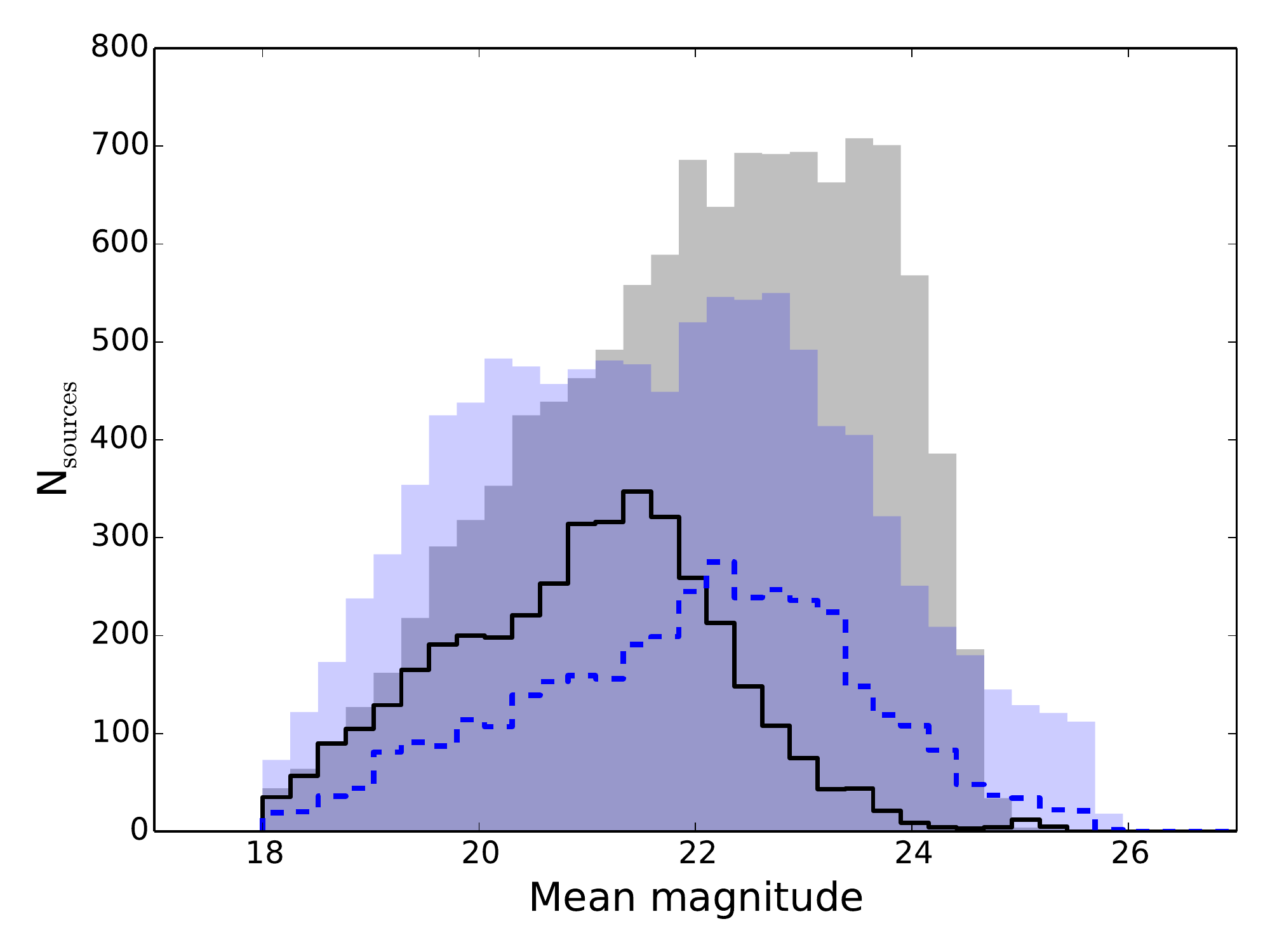}
    \end{tabular}
    \end{center}
    \caption{Histogram of the number of data points in a light curve,
    $n_{\rm LC}$ ({\it upper panel}), the amplitude of variability ({\it middle panel}) and mean magnitudes ({\it lower panel}), for all multi-filter (lines) and single-filter (filled histograms) variable candidates detected by the HCV pipeline in the F814W and F606W filters of ACS ({\it left}) and WFC3 ({\it right}). In F814W (F606W), the SFVCs are shown with a gray (blue) histogram, while the MFVCs with a solid black (dashed blue) line.}
    \label{fig:NM_example}
\end{figure*}

The HCV catalog includes 47 instrument and filter combinations, of which the ACS\_F814W (with 19,550 SFVC and 4,762 MFVC detections in this filter), ACS\_F606W (with 8,807 SFVC and 3,017 MFVC detections in this filter), WFC3\_F814W (with 11,356 SFVC and 3,935 MFVC detections in this filter), and WFC3\_F606W (with 10,457 SFVC and 3,702 MFVC detections in this filter) contain the largest number of candidate variables. In Figure~\ref{fig:NM_example} we present histograms of the number of data points per light curve
($n_{\rm LC}$), the amplitude of variability and the mean magnitudes of the variables for the four most commonly used instrument and filter combinations. The multi-filter and single-filter variable candidates are plotted separately. The distribution of $n_{\rm LC}$ shows several peaks, with the maximum $n_{\rm LC}$ for ACS in the histogram bin with values 12.5--15 for the MFVCs and 5--7.5 for the SFVCs. There is a steep drop for sources with $n_{\rm LC} >45$. For WFC3, the SFVCs peak at 5--7.5, while the MFVCs have a second peak at $\sim$20--25. There is a steep drop at values $n_{\rm LC} >27$ for the MFVCs, while the drop is more gradual for the SFVCs. 
The middle panel shows the distribution of the amplitude, with a peak in the distribution in the bin with values $0.2-0.4$~mag. We note the large number of variables with amplitudes $>1$~mag (see Section~\ref{sec:highampvars}). The lower panel shows the distribution of the mean magnitude of each variable light curve. For ACS, the SFVCs are distributed as expected, increasing in numbers toward fainter magnitudes and dropping after a peak around magnitude 24, where the completeness drops. However, the MFVCs peak around magnitude 25.5, due to the large number of RR Lyrae and Cepheids detected with ACS in nearby galaxies \citep[e.g.][]{2016ApJ...830...10H}, which appear at these magnitudes. Another factor contributing to the almost flat distribution is the fact that both RR Lyrae and Cepheids have a larger amplitude in the blue than in the red, so they often do not appear as MFVCs. In WFC3, the distributions resemble those of a magnitude limited sample, except for the SFVCs in F606W, which display a secondary peak around magnitude 23.

\begin{figure*}
    \centering
    \includegraphics[width=1.0\textwidth]{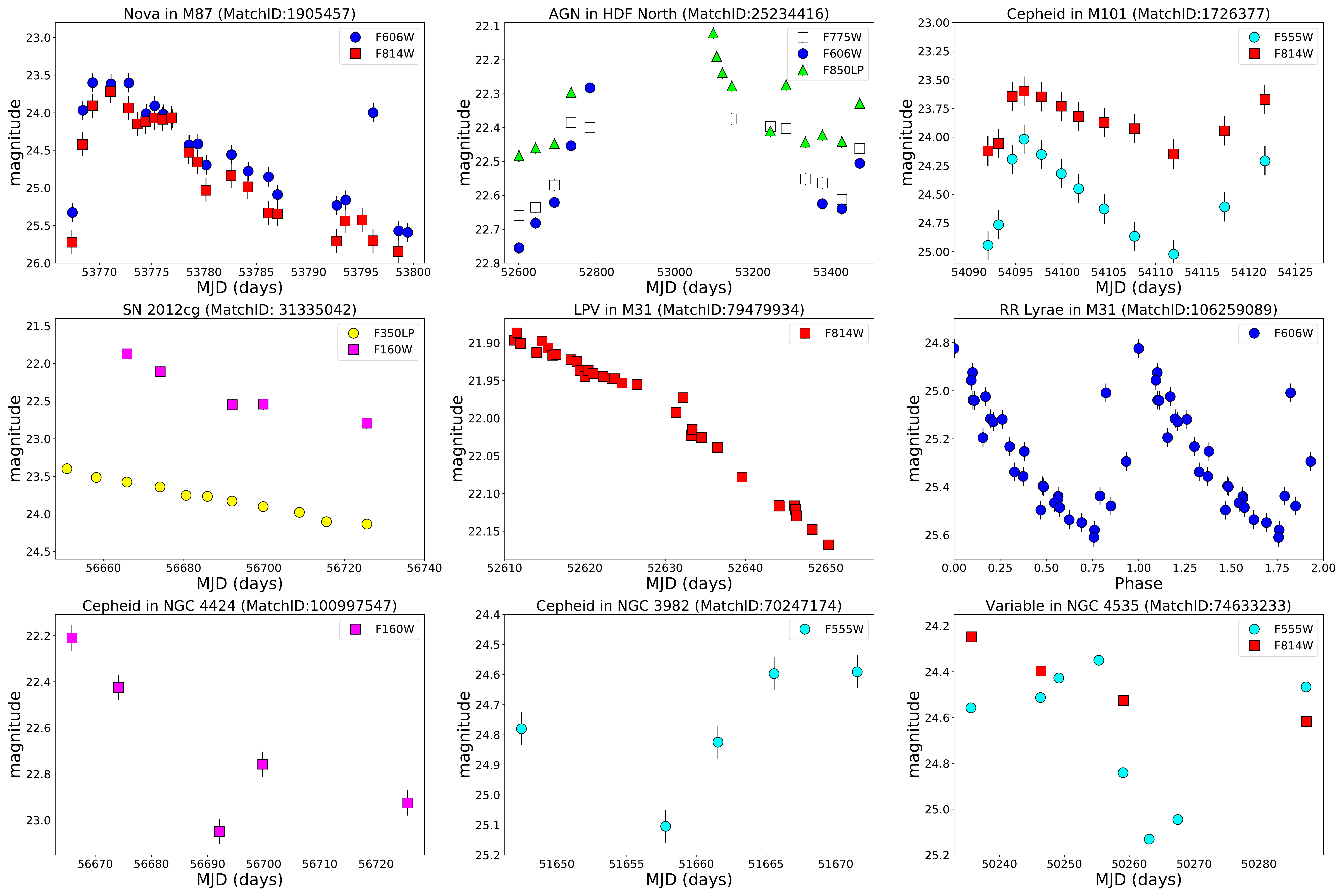}
    \caption{Example light curves for different types of variable sources in the HCV catalog. From left to right and top to bottom, we show the light curves of a classical nova in M87, an AGN in HDF North, a classical Cepheid in M101, SN~2012cg in NGC 4424, an LPV in M31, a RR Lyrae in M31 (phased with P=0.688 days), two Cepheids in NGC~4424 and NGC~3982 with sparse light curves, and a variable in NGC~4535. Error bars are plotted, although they are sometimes smaller than the symbol size. The variables in the first six panels were observed with the ACS instrument, while the variables in the three lower panels were observed with WFPC2.}
    \label{fig:LC_example}
\end{figure*}

We demonstrate the light curve quality and types of variability included in the HCV catalog in Figure~\ref{fig:LC_example}, which presents five multi-filter variables: a classical nova, an AGN, a classical Cepheid, a supernova, and a variable in NGC~4535, as well as four single-filter variables: a long period variable (LPV), a RR Lyrae star, and two classical Cepheids with sparse light curves. The AGN has not been previously reported as a variable. It should be noted that despite the preprocessing applied by the pipeline, outliers do remain, for example, in the nova and RR Lyrae curves (see Section~\ref{sec:limitations}).

\begin{table}
      \begin{minipage}[t]{\columnwidth}
      \caption[]{Results of the HCV data processing pipeline.}
      \label{tab:hcvresults} 
      \begin{tabular}{lrr}
\hline\hline
Item						            & Number		\\	
\hline                                  
\texttt{GroupIDs} in HSC~v3 	        &15,160		\\	
\texttt{GroupIDs} (subgroups) processed by DPP	&250 (2,132) 	\\	
\texttt{GroupIDs} with variable candidates & 244\\
\texttt{GroupIDs} with single-filter variable candidates & 243\\
\texttt{GroupIDs} with multi-filter variable candidates & 127\\
Sources processed by DPP	 			        &3,679,859	\\	
Variable candidates found by DPP	                   	    &84,428		\\	
 \multicolumn{1}{r}{Single-filter variable candidates}   &73,313		\\	
  \multicolumn{1}{r}{Multi-filter variable candidates}   &11,115		\\
\hline
\end{tabular} 
\end{minipage}
\end{table}

\begin{table*}
      \caption{Results of the expert validation for SFVCs and MFVCs. Percentages for the expert validated variables are computed relative to the number of variables in the HCV catalog, while percentages in the following columns are computed relative to the number of expert validated variables.}
      \label{tab:expertvalidation} 
      \begin{tabular}{c|c|c|c|c|c}
\hline\hline
Type of variable& \# HCV catalog & \# expert validated & \# high-confidence  & \# probable & \# possible		\\
candidate& variables& variables & variables & variables &  artifacts		\\	
\hline 
Single-filter & 73,313 & 8,139 (11\%) & 3,323 (41\%) & 3,055 (37\%) & 1,761 (22\%) \\
Multi-filter & 11,115 & 5,394 (49\%) & 2,101 (39\%) & 2,442 (45\%) & 851 (16\%) \\
\hline
Sum & 84,428 & 13,533 (16\%) & 5,424 (40\%) & 5,497 (41\%) & 2,612 (19\%) \\
\hline
\end{tabular} 
\end{table*}

\subsection{The catalog release}

 The catalog is available in its entirety via the ESA {\em Hubble} Science Archive\footnote{\url{http://archives.esac.esa.int/ehst/}} \citep[eHST;][]{2017ASPC..512...97A} and the HCV Explorer\footnote{\url{http://archives.esac.esa.int/hcv-explorer/}} at ESAC. Moreover, the catalog can be queried via the MAST\footnote{\url{https://doi.org/10.17909/t9-m29s-xg91}} interface at STScI. The HCV Explorer is a web visualization tool that will allow the user to access and explore the Hubble Catalog of Variables. The first release of the tool will offer interactive and connected plotting of the variables in the HCV catalog in a region of the sky. By selecting a variable from the finder chart, one can display and download its light curve, view the location of the source on a MAD vs. magnitude diagram, and download the light curve for this source, as well as for non-variable sources in the same \texttt{GroupID} or subgroup. The visualization of the results will help the user to assess whether a candidate variable is reliable or not, particularly in cases where no expert validation has been performed. This paper includes the following release tables: an overview table of the distribution of variables per \texttt{GroupID} (Table~\ref{hcv_tableMFVCs}), the catalog of variables (Table~\ref{hcv_tableHCV}), and the catalog of sources that fall below the 5$\sigma$ MAD detection threshold for variability selection, meaning the ``constant'' stars (Table~\ref{hcv_tableSourcesBelowThreshold}).

  
While a total of 250 \texttt{GroupIDs} were processed by the DPP, there are only 244 \texttt{GroupIDs} with a multi-filter and/or single-filter variable candidate detection. The six \texttt{GroupIDs} without detected variables are: 1423, 4006, 29216, 31829, 1024871, and 1042864. The multi-filter variable candidates are detected in 127 \texttt{GroupIDs}, while the single-filter variable candidates in 243 \texttt{GroupIDs} (see also Table~\ref{tab:hcvresults}). Out of the 244 \texttt{GroupIDs} with variables, 21 \texttt{GroupIDs} (25 subgroups) were flagged as unreliable by the experts (see Section~\ref{sec:limitations}) and therefore the variables included in these \texttt{GroupIDs} were not expert validated. Table~\ref{hcv_tableMFVCs} presents the 244 \texttt{GroupIDs} with detected variables. It lists the coordinates, field name (from the HLA), the initial number of sources that passed to the DPP (to which prior selection criteria were applied), the final number of sources (from which the variable candidate sources were searched for detection), the maximum number of instrument and filter combinations available per \texttt{GroupID}, and the number of multi-filter and single-filter variables in each \texttt{GroupID}. It is ordered by right ascension. The field name ``ANY'' and names containing ``PAR'' refer to observations obtained in parallel to the main science target.

Table~\ref{hcv_tableHCV} is the HCV catalog release. The sources listed correspond to ten entries of the catalog, while the columns show, for each source: 
the equatorial coordinates, 
the \texttt{MatchID}, 
the \texttt{GroupID}, 
the subgroup, 
the pipeline classification flag, 
the expert-validation classification flag, 
the number of existing instrument and filter combinations for the source, 
the name of the instrument and filter combination for which the following data are given: 
the filter detection flag, which indicates whether the variable was detected [``1''] or was not detected [``0''],
the variability quality flag, 
the number of measurements in the light curve $n_{\rm LC}$,
the HSC magnitude m$_\textrm{HSC}$ (i.e., the mean \texttt{MagAper2} of the light curve points),
the corrected magnitude m$_\textrm{HCV}$,
the MAD value,
and the reduced $\chi ^{2}$ value.
For all MFVCs, there are extra columns for each additional instrument and filter combination, in which the source is classified as a variable candidate (provided in the online version of the catalog).

Table~\ref{hcv_tableSourcesBelowThreshold} presents the sources that fall below the 5$\sigma$ detection threshold, 
which include constant sources and possibly, low-amplitude variables below our detection threshold. 
The sources listed correspond to the first ten entries of the catalog, while the columns show, for each source: 
the equatorial coordinates, 
the \texttt{MatchID}, 
the \texttt{GroupID}, 
the subgroup, 
the number of instrument and filter combinations in which individual sources are observed, 
the name of the instrument and filter combination for which the following data are given: 
the number of measurements in the light curve $n_{\rm LC}$,
the HSC magnitude m$_\textrm{HSC}$,
the corrected magnitude m$_\textrm{HCV}$,
the MAD value,
and the reduced $\chi ^{2}$ value. 
A block of information, identical to columns (7) to (12), is added for each additional instrument and filter combination, in which the source has been observed (provided in the online version of the catalog). 

%
%

\begin{figure}
\centering
\includegraphics[width=0.5\textwidth]{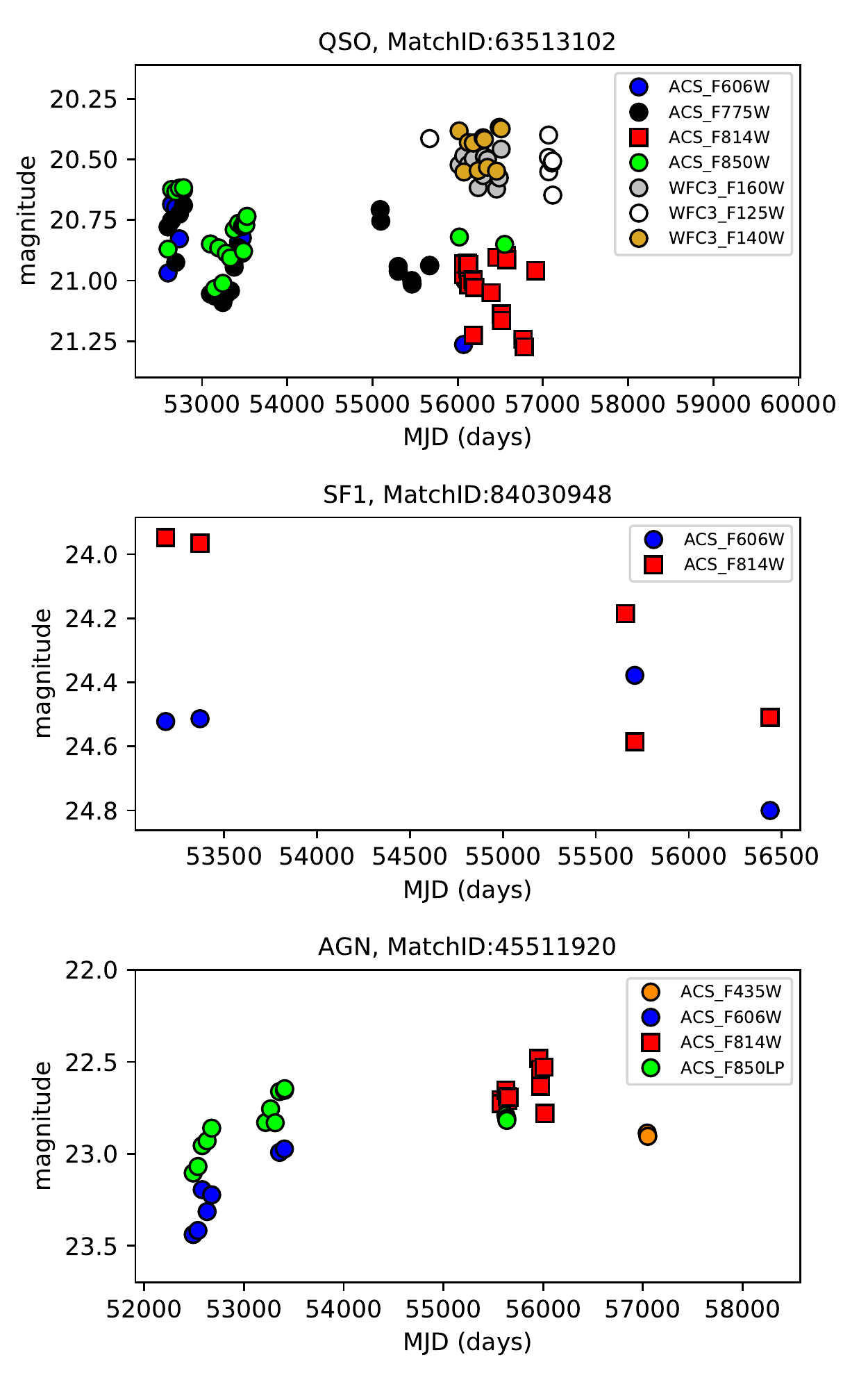}\\ 
\caption{Examples of three variable sources in the CANDELS fields: a quasar in GOODS-North ({\it upper panel}), a Seyfert-1 in GOODS-South ({\it middle panel}), and a new variable AGN in EGS ({\it lower panel}).} \label{candelslight curves}
\end{figure}

\subsection{Comparison of HCV catalog variables with SIMBAD}
In order to estimate the fraction of new variables included in the HCV catalog, we cross-matched the high-confidence (expert-validated) variables with SIMBAD using a conservative search radius of 5", to account for possible errors in the astrometry of published variables. This yielded a total of 2,839 matches out of 5,424 high-confidence variables, implying that 2,585 (48\%) are potentially new variables, as they have no matches in SIMBAD. A careful cross-match with other bibliographical data is necessary to assess whether these sources are indeed newly detected variables. It is also interesting to note that the most common SIMBAD classes for the matched objects were Cepheids and RR Lyrae variables, while several other classes of variables were identified such as eclipsing binaries, long-period variables, novae, supernovae, AGN, etc. Even among the 2,839 matched objects there are several cases where the SIMBAD classification does not indicate variability (e.g., ``star''). Some of these may be mismatches, due to the large search radius. Therefore, there may be several new variables in the sample of ``matched'' objects as well. Projecting these results to the entire HCV catalog, we expect a few thousand new variables.

\subsection{Variable AGN in the HCV catalog}

Variability is a basic characteristic of AGN at all wavelengths, appearing over periods of hours to years \citep{ulrich1997}. Optical variability has therefore been used as a method to identify AGN \citep{Pouliasis19,sara2011,vill2010,vill2012A,decicco2015,falocco2016}; the importance of the method is demonstrated by its ability to identify low-luminosity AGN that even the deepest (currently available) X-ray observations would have missed. The HCV catalog contains many groups with deep observations obtained over multiple visits, making them appropriate for identifying variable AGN (spanning the range from the most luminous point-like quasars to the low-luminosity AGN). Here, we demonstrate the power of the HCV catalog in the ``Cosmic Assembly Near-IR Deep Extragalactic Legacy Survey'' fields \citep[CANDELS;][]{Koekemoer_2011,Grogin_2011}. The HCV catalog contains 621 variable candidates (179 MFVCs and 442 SFVCs) in the five CANDELS fields (GOODS South, GOODS North, COSMOS, EGS, and UDS). Following the expert validation of the sources, we excluded the sources classified as ``possible artifacts'', resulting in 171 variable candidates. 

We cross-matched our variable sources with the SIMBAD database (using a radius of 1") to validate their nature. Table~\ref{table:hcv_AGN} presents the results for the different fields. Excluding the stellar population, 75\% and $\sim$45\% of the ``high-confidence'' and ``probable'' variable sources, respectively, are confirmed AGN or quasars. Regarding the rest of the sources showing variability, but which are classified as 'normal' galaxies in SIMBAD and do not show any X-ray emission, the aforementioned studies have shown that these extended sources are consistent with low-luminosity AGN (LLAGN), which would have been missed by the current depths of X-ray surveys, but are important to complete the AGN demographics. In Figure~\ref{candelslight curves}, we present three example light curves of variable sources in the CANDELS fields that are classified via spectroscopy as AGN or quasars. \texttt{MatchID}=45511920 is reported to be variable for the first time. We thus demonstrate the photometric quality of extended sources and, in particular, variable AGN in the HCV catalog.

\begin{table*}
      \caption[]{HCV variables in the CANDELS fields.}
      \label{table:hcv_AGN} 
      \begin{tabular}{r|c|c|c|c|c||c}
\hline\hline
Item &GOODS-N & GOODS-S & COSMOS & EGS & UDS& Total\\
 &1084534 & 1084533 & 1081922 & 1045196 & 1036556&\\
\hline
Pipeline class. MFVCs/SFVCs & 75/143 (218) & 90/164 (254) & 2/52 (54)  & 10/50 (60)  & 2/33 (35)  & 179/442 (621)\\
Expert validated MFVCs/SFVCs &58/- (58)  & 28/17 (45) & 2/- (2) & 6/31 (37) & -/28 (28) & 95/76 (171)\\
\hline
AGN / Seyfert 1 / Seyfert 2 &11 & 13 & - & 16 & - & 40\\
Quasars &11 & 10 & - & 3 & 2 &26 \\
Stars & 19 & 10 & 1 & - & - & 30\\
Normal galaxies (LLAGN) &18 & 10 & 1 & 6 & 13 & 48 \\
\hline
Total classified sources (SIMBAD) &59 & 43 & 2 & 25 & 15 & 144\\
\hline
\end{tabular} 
\tablefoot{\texttt{GroupIDs} are listed under the field name. Numbers in parentheses indicate the total number of both MFVCs and SFVCs.}
\end{table*}

\subsection{High-amplitude variables or transients in the HCV catalog}
\label{sec:highampvars}

\begin{figure*}
\begin{tabular}{  l c  }
\includegraphics[trim=0cm 0cm 0.cm 0cm, width=1.0\textwidth ]{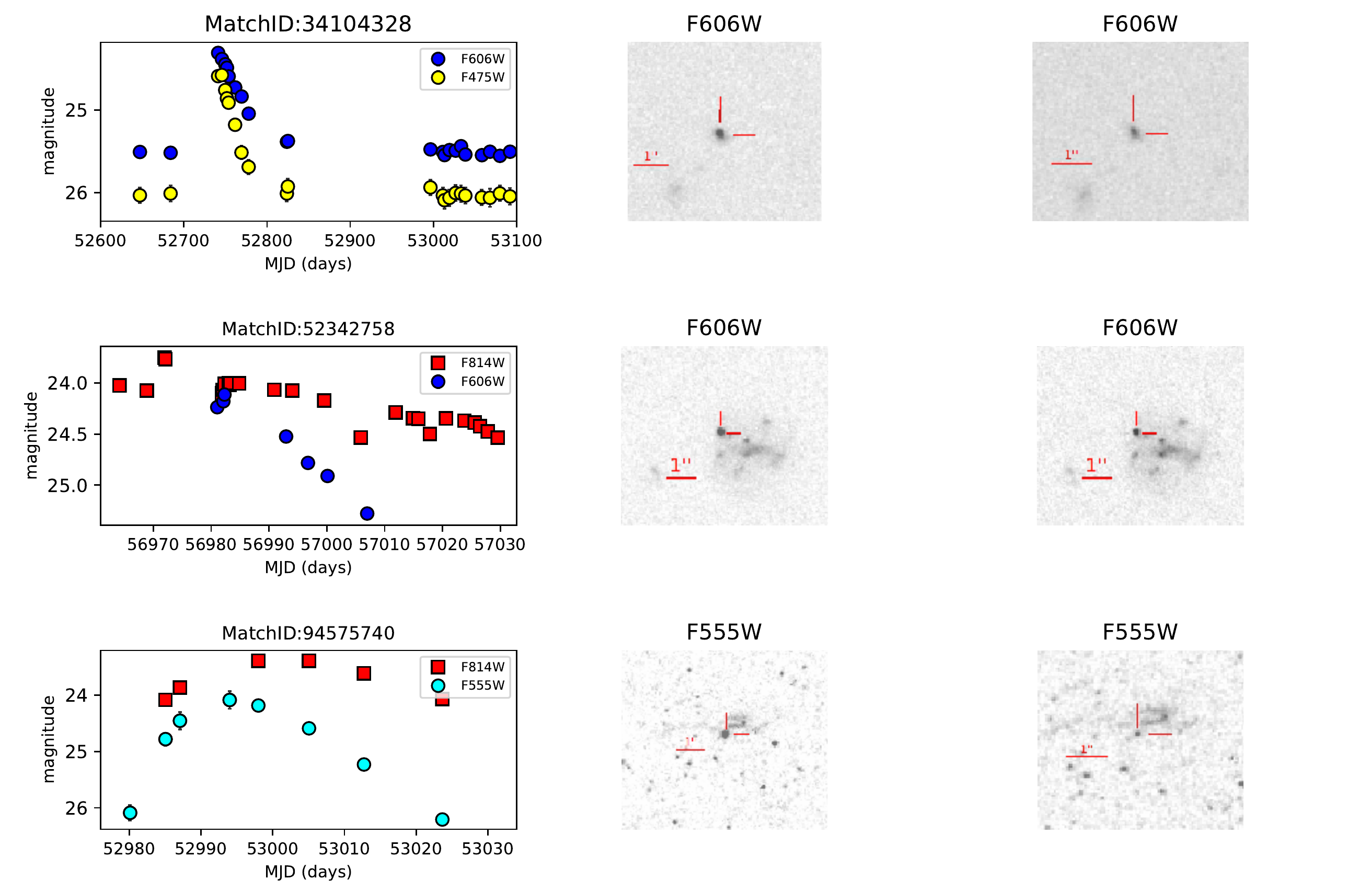}&
\end{tabular}
\caption{HCV catalog light curves and finder charts of the newly-discovered high-amplitude, multi-filter variables in the fields of NGC~3314 (first row), MACSJ1149.5+2223-HFFPAR (second row), and NGC~4258 (third row). The finder charts on the left correspond to the light curve peak, while those on the right to the minimum.}
\label{fig:highampvariables}
\end{figure*}

The HCV catalog contains a large number of high-amplitude variables / transients. The number of variable candidates in the HCV catalog with amplitudes $\geq1$~mag in at least one filter is around 6,500 (Spetsieri et al., in prep.). We have selected three high-amplitude, multi-filter sources that were serendipitously discovered during the expert validation procedure to present below. These sources have not been previously reported in the literature. Light curves and finder charts of these three multi-filter variables are presented in
Figure~\ref{fig:highampvariables} and are described below. 

\begin{enumerate}[itemsep=2pt,parsep=2pt]
    \item {\bf Transient object in the field of NGC~3314\footnote{This object is unrelated to the transient in NGC~3314 reported in the IAU Circular 7388 by \citet{2000IAUC.7388....2K}.}}: 
     This source has a \texttt{MatchID}=34104328 in the HSC~v3 and is located at R.A.: 10:37:15.601 and Dec.: $-$27:40:03.42 (J2000). The HCV light curve of the source contains 24 observations in filters F475W and F606W of ACS (see Figure~\ref{fig:highampvariables}). The object reaches a peak magnitude of 24.32 mag in F606W and 24.5 mag in F475W at MJD=2452741.2064 days and fades by 1 mag in F606W and 1.5 mag in F475W in about 80 days. The F475W$-$F606W color of the host galaxy is 0.5 mag. The nature of this transient is difficult to determine, as there is no redshift information or any estimate of the intrinsic luminosity. However, we excluded the possibility of it being a Tidal Disruption Event (TDE), since TDEs show no color evolution, while this object evolves to the red. Also, the rise and fall timescales exclude the possibility of the object being a Type Ia SN. If we assume that it is associated with NGC~3314 (e.g., a low-mass, low-surface brightness satellite galaxy in the field), the absolute magnitude of this object would be $\sim-12$ mag, which is typical for novae. However, the shape of the light curve resembles a core-collapse SN. We performed fits using SALT2-extended \citep{2007A&A...466...11G,2018PASP..130k4504P} and Nugent templates\footnote{\url{https://c3.lbl.gov/nugent/nugent_templates.html}} \citep{1999ApJ...521...30G}. SN IIn templates \citep[based on blackbody templates from][]{2002ApJ...573..144D} yielded the best fit at z$\sim$1.2. SN IIP templates also provided a good fit, while SN IIL templates a poor fit. Hence, based on the photometric classification, we suggest that this object is a candidate high-redshift, core-collapse SN IIn or SN IIP, which is not associated with NGC~3314.

    \item {\bf Variable in MACSJ1149.5+2223-HFFPAR}:
    
    This source has a \texttt{MatchID}=52342758 in the HSC~v3 and is located at R.A.: 11:49:36.684 $\&$ Dec.: +22:17:14.37 (J2000) in the MACSJ1149.5 parallel field, near a faint galaxy. The HCV light curve of this source is shown in the second panel of Figure~\ref{fig:highampvariables} and contains seven data points in the F606W and 25 in the F814W filter, obtained with ACS. 
    As shown in Figure~\ref{fig:highampvariables} the light curve peaks around magnitude 24 at MJD$\simeq$56972.0934 days in F814W. In F606W the peak is followed by a steep, linear decline of $\sim$1.2 mag within 20 days after the peak, whereas the light curve in F814W drops by $\sim$ 0.8 mag. Unfortunately, there are gaps in the coverage in both filters. We attempted to fit SN templates, which indicate that it might be a high-redshift superluminous SN. The date of maximum according to the fit occurs at MJD=56984.679 days. 

    \item {\bf SN Ia in the field of NGC 4258}: 
    
    This source has a \texttt{MatchID}=94575740 in the HSC~v3 and is located at R.A.: 12:19:25.639 and Dec.: +47:10:30.52 (J2000) in the field of NGC 4258. The HCV light curve includes 12 data points in F555W and six data points in F814W, spanning about 40 days. The light curve shows the early evolution of a possible SN; the peak brightness is followed by a decrease of $\sim$2 mag in F555W and $\sim$1 mag in F814W within 20 days. SALT2-extended provided a good fit for a SN Type Ia, 
    with a maximum at MJD=53000.022 days. 
\end{enumerate}

\section{Summary}

We present the {\em Hubble} Catalog of Variables, a new catalog of variable sources based on photometry from the {\em Hubble} Source Catalog version 3, which includes all available images obtained with WFPC2, ACS, and WFC3 onboard the \emph{Hubble Space Telescope} that were public as of October 2017. The HCV catalog is the first catalog of variables from the \HST. It contains 84,428 variable candidates with $V\leq27$~mag, which were identified among sources having at least five measurements in the same instrument and filter combination, in groups having at least 300 sources. The HCV catalog is the deepest catalog of variables available, reaching on average $\sim4$ mag deeper than current catalogs of variable sources.

The HCV system, developed from scratch for this project, comprises of a data processing pipeline, catalog, and bridge to the science archives at ESAC. The architecture, main elements of the pipeline, implementation technologies, and performance are presented. In brief, we developed a preprocessing algorithm for identifying and removing outlier measurements. We further applied a local zeropoint correction. We used ten ``Control Sample'' fields to evaluate all intermediate steps and refine our selection criteria and thresholds. A total of 250 \texttt{GroupIDs} (or 2,132 subgroups) including 3.7 million sources finally satisfied our selection criteria and were processed through the pipeline.

The data processing pipeline calculated the median absolute deviation and used a 5$\sigma$ threshold to select variable candidates. These were passed to the validation algorithm, which classified the sources as single or multi-filter variable candidates and assigned a variability quality flag to each source. The pipeline identified 84,428 variable sources: 73,313 single-filter and 11,115 multi-filter variable candidates. The data points in the HCV catalog light curves range from five to 120, the time baseline ranges from under a day to over 15 years, while $\sim$8\% of variables have amplitudes in excess of 1~mag. Furthermore, expert validation was performed on 16\% of the total number of variables. The result of this procedure implies that about 80\% of the sources are true variables, while about 20\% may be possible artifacts. It should be noted that classifications of the variable sources are not provided. We finally summarize the challenges present in identifying variables in a large and inhomogeneous set of photometric measurements, which are specific to {\em Hubble}.

We expect the HCV catalog to become a key resource for the community, as it includes variable stars in the Milky Way and nearby galaxies, as well as transients, and variable AGN, which include low-luminosity AGN. Possible uses include searches for new variable objects of a particular type for population analysis, detection of unique objects worthy of follow-up studies, identification of sources observed at other wavelengths, and photometric characterization of candidate progenitors of SNe and other transients in nearby galaxies. The HCV catalog contains many interesting high-amplitude variables, including a core-collapse SN and a Type Ia SN reported here for the first time. A thorough analysis of the high-amplitude variables in the HCV catalog will be presented by Spetsieri et al. (in prep.).

The catalog is available via eHST at ESAC and MAST at STScI. Both platforms allow for the visualization of the results. In principle, the HCV pipeline can be used to generate future versions of the catalog based on future HSC releases.

\begin{acknowledgements}
The HCV project has been funded by the European Space Agency (ESA) under the
``{\em Hubble} Catalog of Variables'' program, contract No. 4000112940. 
This work uses the HSC, based on observations made with
the NASA/ESA \emph{Hubble Space Telescope}, and obtained from the {\em Hubble} Legacy Archive, which is a collaboration between the Space
Telescope Science Institute (STScI/NASA), the Space Telescope European
Coordinating Facility (ST-ECF/ESAC/ESA) and the Canadian Astronomy Data
Centre (CADC/NRC/CSA). 
This work was supported in part by Michigan State University through computational resources provided by the Institute for Cyber-Enabled Research.
This research has made use of the SIMBAD database \citep{wenger2000}, operated at CDS, Strasbourg, France and, also, of NASA's Astrophysics Data System.
We thank Valerio Nascimbeni and Luigi Bedin for providing their \HST\ photometry of the globular cluster M4 for our technical tests. 
We thank Dr.~Georgios Dimitriades and Dr.~Alexandra Kozyreva for advice on the transient in the field of NGC~3314. 
We also thank Lou Strolger and Ronald Downes for their scientific input during this project, 
Raul Guti\'errez-S\'anchez for his help with the HCV Explorer and Bruno Merin for inspiring the work on the visualization tool.
E. Paraskeva acknowledges support by the project ``PROTEAS II'' (MIS 5002515), which is implemented under the Action ``Reinforcement of the Research and Innovation Infrastructure'', funded by the Operational Programme ``Competitiveness, Entrepreneurship and Innovation'' (NSRF 2014-2020) and co-financed by Greece and the European Union (European Regional Development Fund).

\end{acknowledgements}


\bibliographystyle{aa}
\bibliography{hcv}


%

%
%
\twocolumn

\begin{appendix} 
\section{Comparison of variability indices with HSC-based simulations}
\label{sec:simulations}

Before selecting MAD as our primary variability detection statistic
(Sec.~\ref{sec:hcvalgorithm}) we investigated the performance of 18 variability indices using diverse sets of ground-based photometric data \citep{2017MNRAS.464..274S}.
We also explored the possibilities to enhance the variability detection
efficiency by combining multiple indices via the principal component analysis
\citep{2018MNRAS.477.2664M} and machine learning \citep{2018MNRAS.475.2326P}.
While combining multiple indices did show promise, the diverse nature of 
the HSC data (and the resulting difficulty of constructing the training data set) together with the difficulty of implementing the machine learning techniques under our data processing architecture ({\it Java} and {\it Apache Spark}; Sec.~\ref{sec:implementation}; that had to be chosen early in the project) lead us to favor the simple approach of using one general-purpose index for variability detection. 

\begin{figure}
\includegraphics[width=0.5\textwidth]{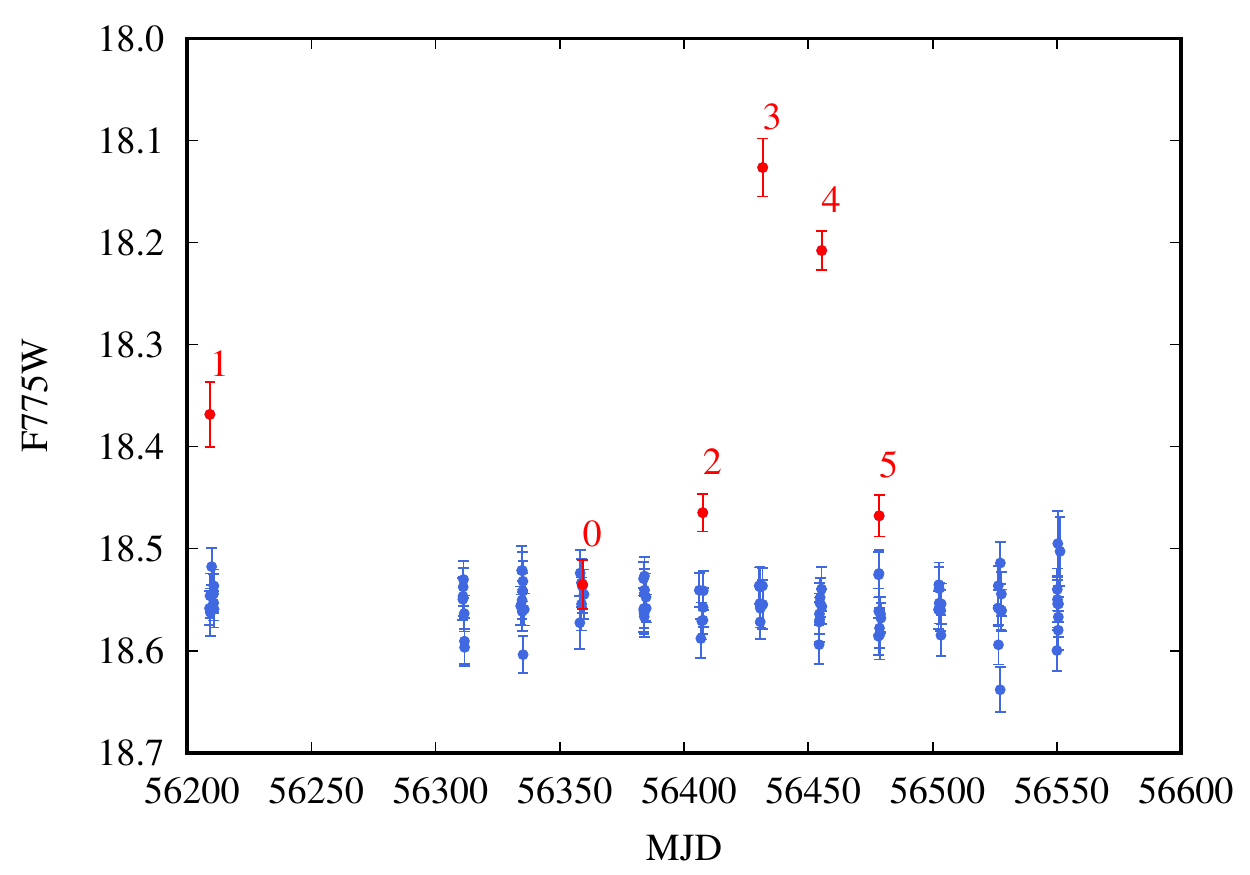}
\caption{Light curve of HSC~v3 source with \texttt{MatchID}=45740877 from the field of the globular cluster M4 (\texttt{GroupID}=33675) observed with WFC3. The numbered red points correspond to measurements obtained from the images presented in Fig.~\ref{fig:out45740877_WFC3_F775W_img}.
\label{fig:out45740877_WFC3_F775W}}
\end{figure}

\begin{figure*}
\includegraphics[width=0.33\textwidth]{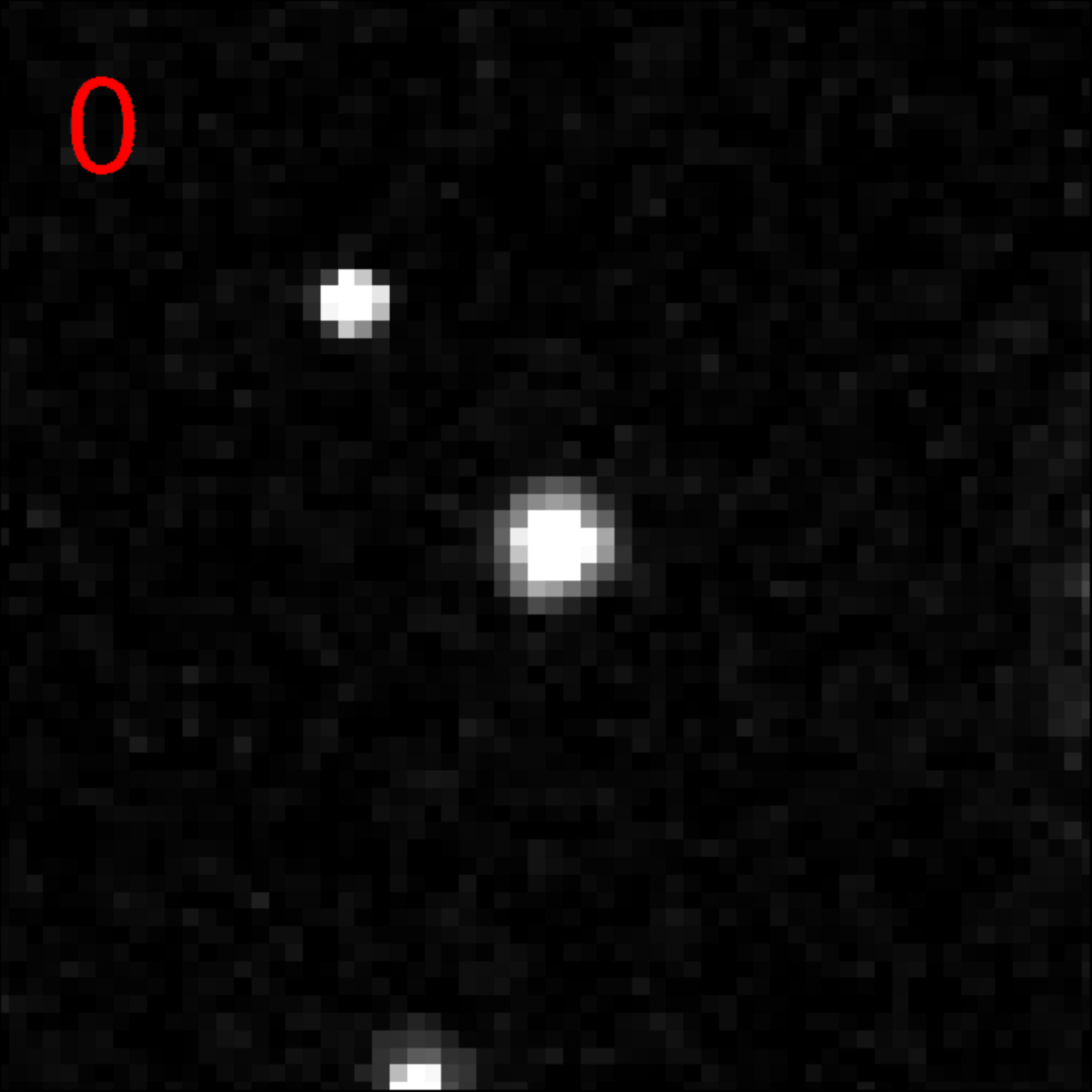}
\includegraphics[width=0.33\textwidth]{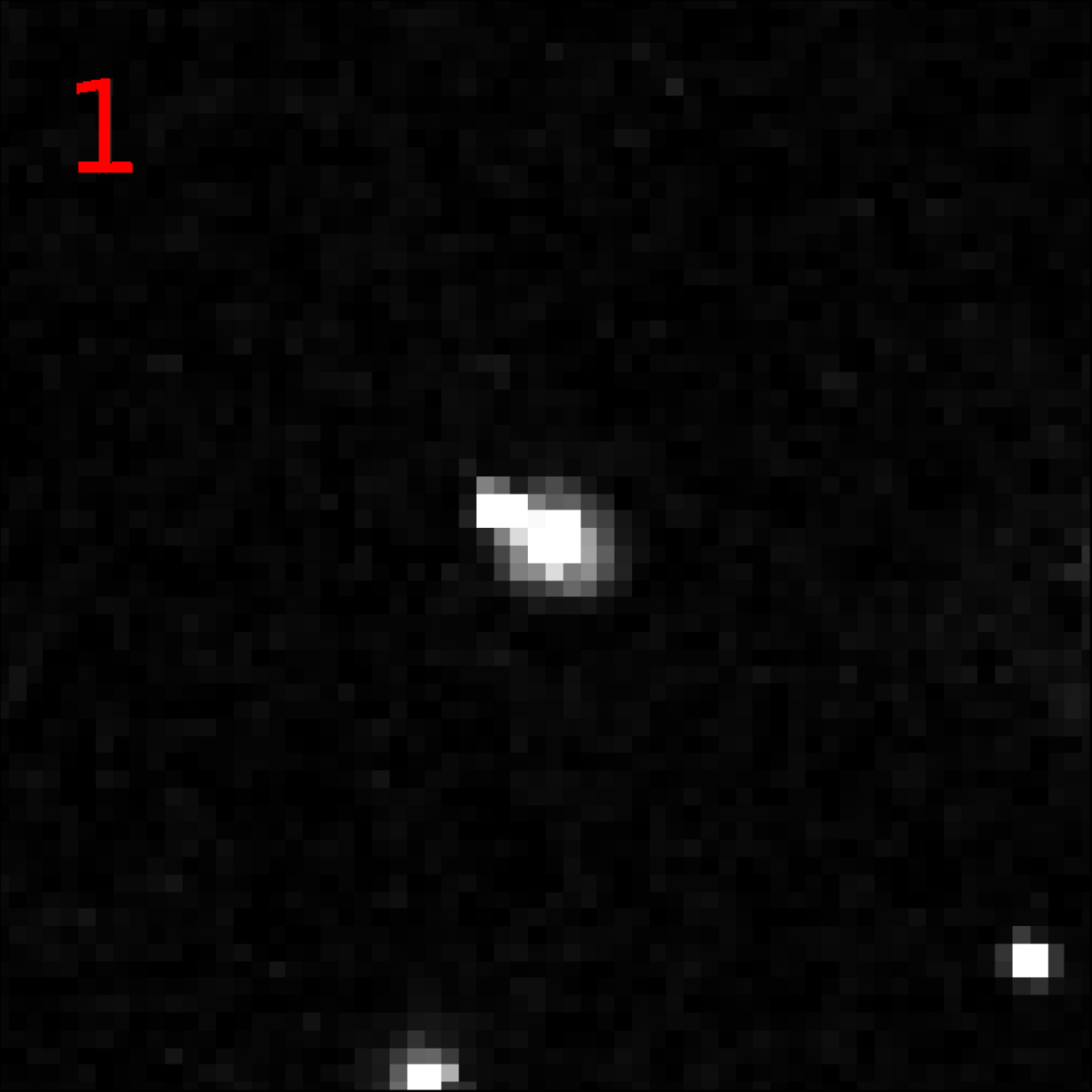}
\includegraphics[width=0.33\textwidth]{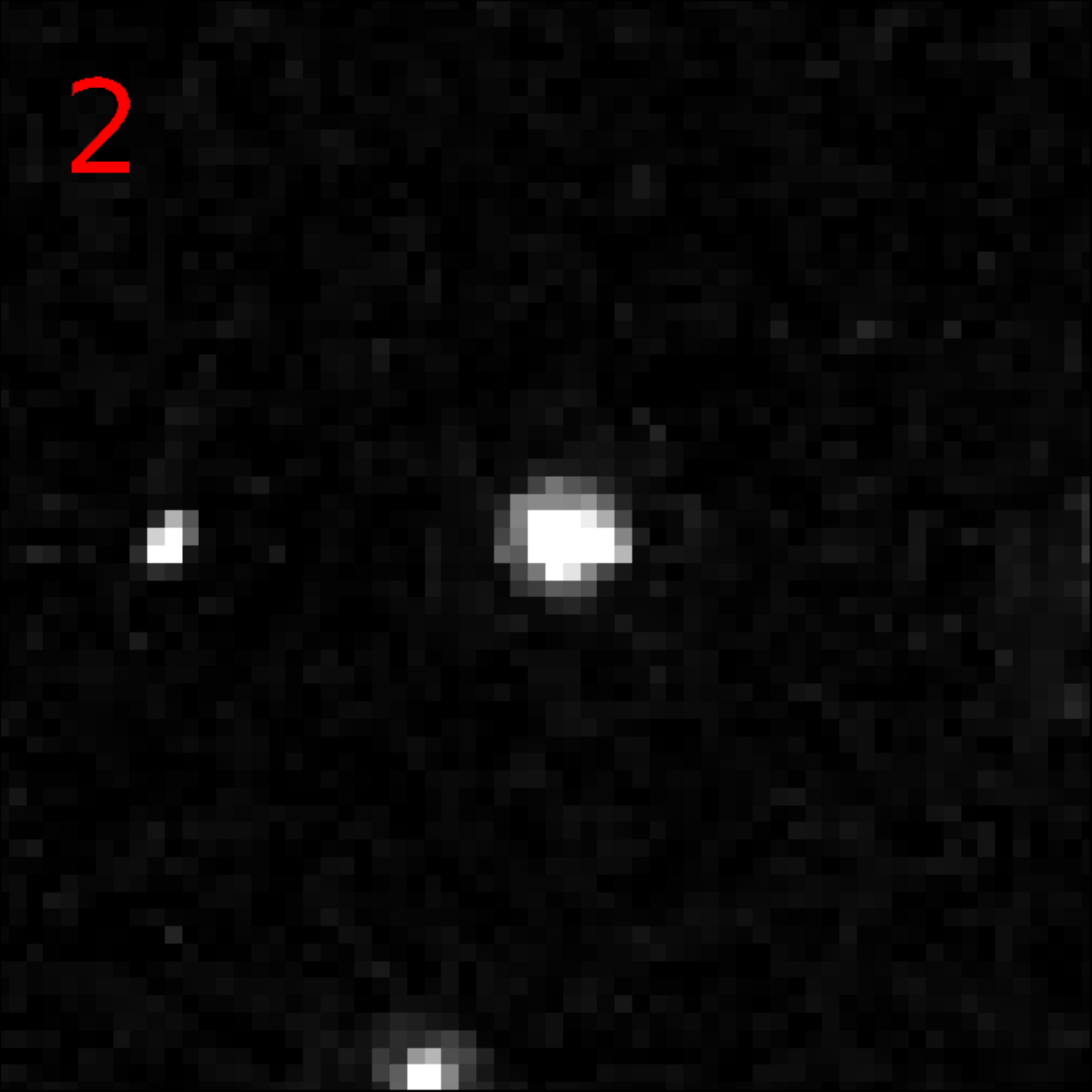}
\includegraphics[width=0.33\textwidth]{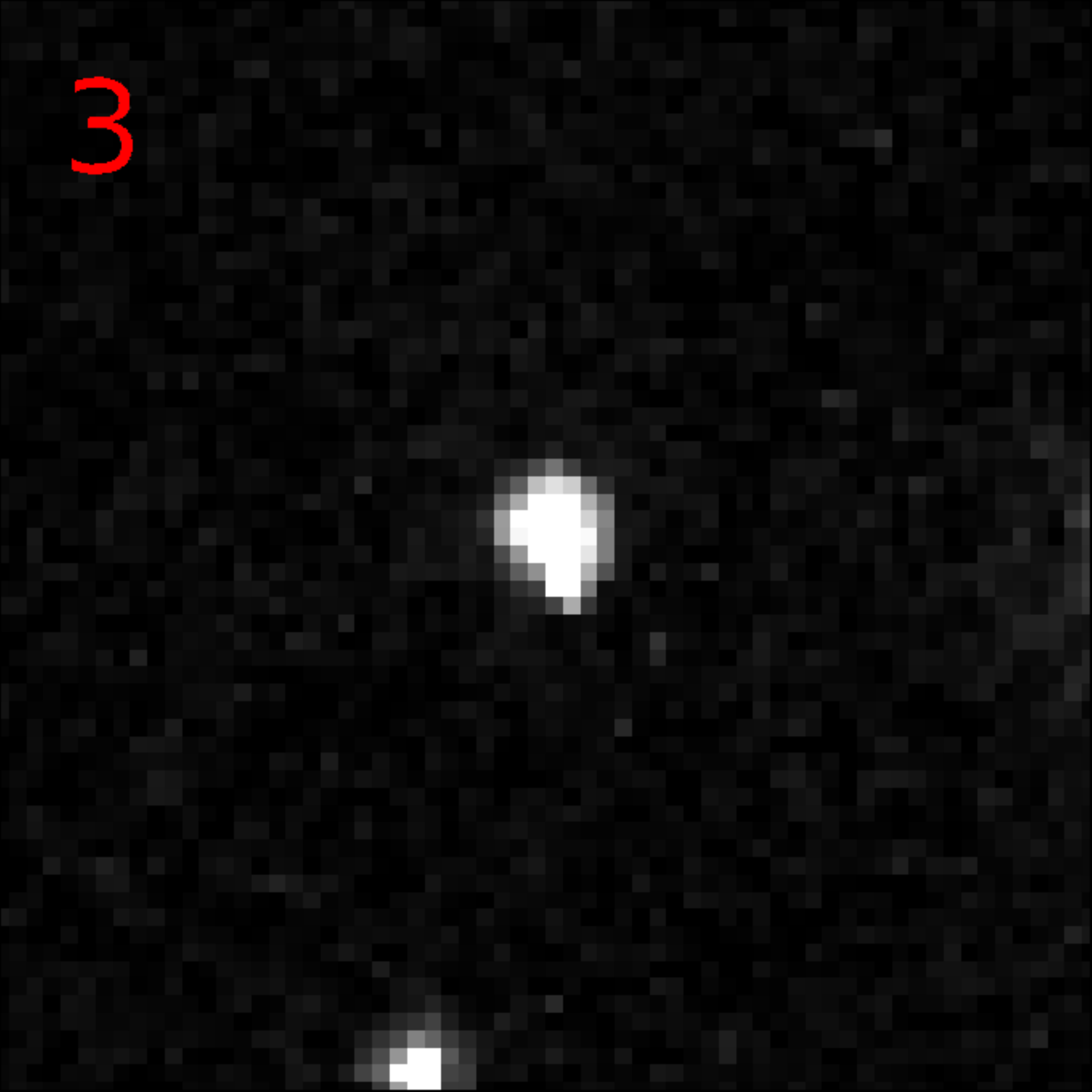}
\includegraphics[width=0.33\textwidth]{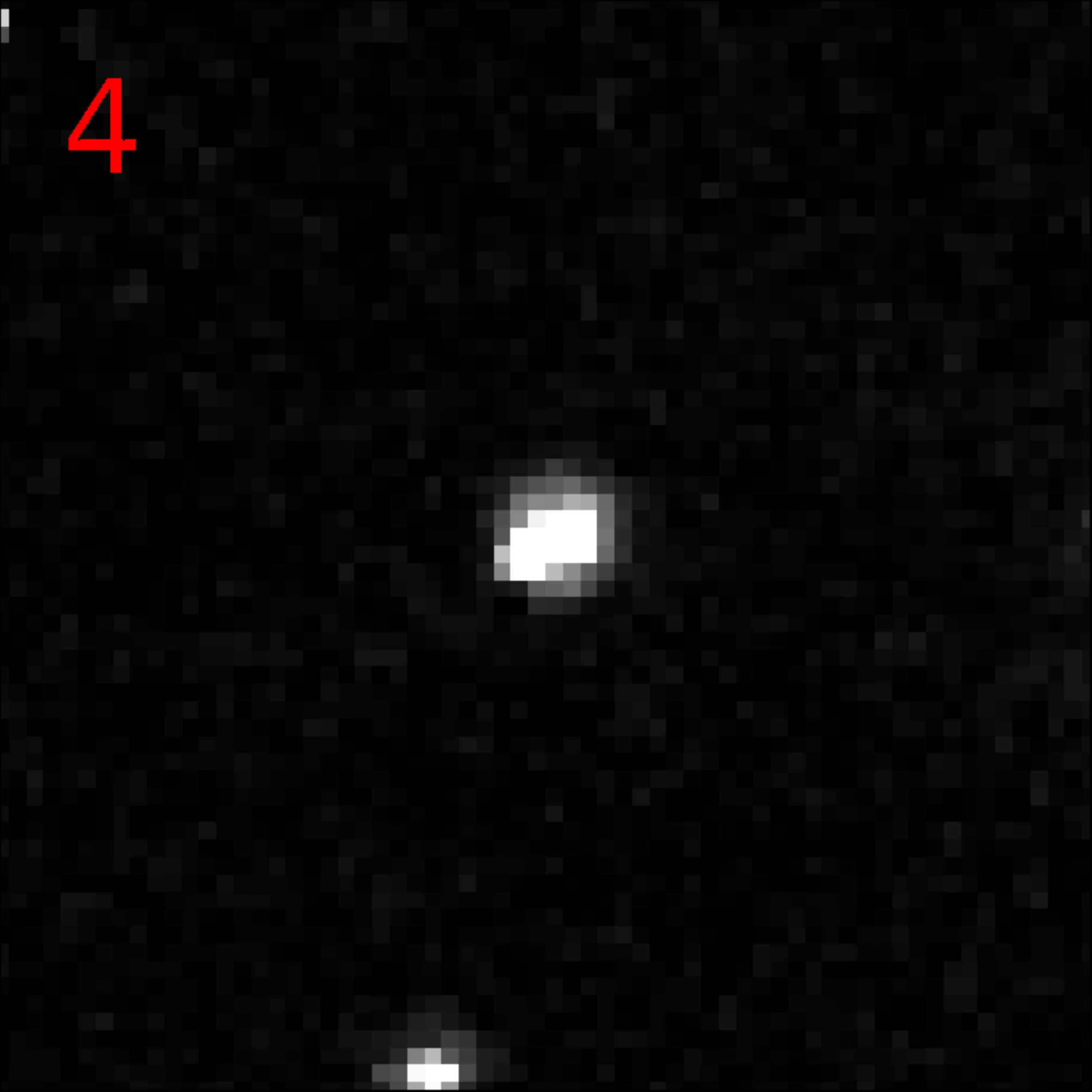}
\includegraphics[width=0.33\textwidth]{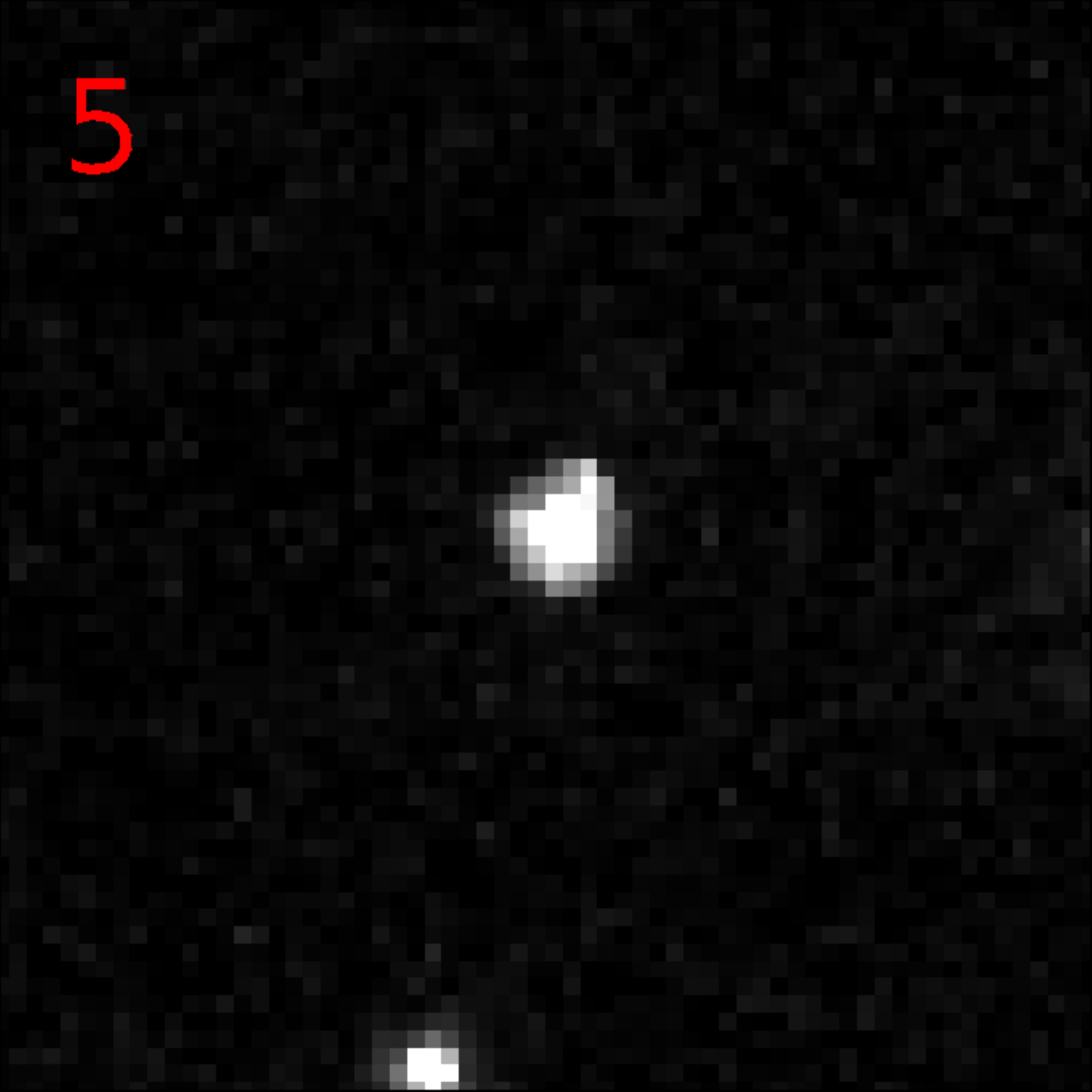}
\caption{WFC3 images of HSC~v3 source with \texttt{MatchID}=45740877 (centered) corresponding to the measurements highlighted in red in Fig.~\ref{fig:out45740877_WFC3_F775W}. Images 1--4 show obvious distortions of the object's shape due to uncleaned cosmic ray hits, resulting in outlier photometric measurements (Fig.~\ref{fig:out45740877_WFC3_F775W}). Image~0 is an example of a measurement unaffected by a cosmic ray hit. We note the additional uncleaned cosmic rays in images 0--2.
\label{fig:out45740877_WFC3_F775W_img}}
\end{figure*}

We follow the procedure for the simulations described in Sec.~3.8 of \cite{2017MNRAS.464..274S} 
to identify the index that is best suited for finding variable objects in the HSC data. 
The index should be sensitive to various types of variability and 
tolerate the wide range of observing cadence patterns found in the HSC, as 
well as occasional outlier measurements caused by cosmic ray hits and calibration errors.
The cosmic rays are a major problem for space-based photometry as
illustrated by the light curve in Fig.~\ref{fig:out45740877_WFC3_F775W}.
These observations consist of a single 20\,s exposure in the F775W filter per
HST visit, that is the exposures were not split into cosmic ray pairs
(Sec.~\ref{sec:hscintro}) making this data set heavily contaminated by
cosmic rays, despite the short exposure time. The outlier points in Fig.~\ref{fig:out45740877_WFC3_F775W}
are all caused by cosmic ray hits, as can be seen from inspecting the
corresponding images (Fig.~\ref{fig:out45740877_WFC3_F775W_img}).
Observations split into multiple exposures are less affected by cosmic
rays (Sec.~\ref{sec:hscintro}), however the probability of two cosmic rays hitting the same pixel in two images is non-negligible, and the edges of 
the stacked frame may be covered by only one image if dithering was applied.

To preserve the sampling and noise properties of the HSC data in our
simulations we use the HSC light curves of sources from selected control
sample fields \citep[][describe the use of non-variable object light curves
as a realistic photometric noise model]{2012A&A...548A..48E}.
We exclude the sources known to be variable and inject 1\% of
the non-variable sources with artificial variability. The variability is
modeled as a simple sine variation with an amplitude randomly selected for
each model variable source to be between 0 and 1\,mag (c.f. the experimental
amplitude distribution presented in Fig.~\ref{fig:NM_example}), 
the frequency of variations between 0.05 and 20.0\,cycles/day (periods in the range 0.05 to 20.0\,days;
the model distribution is uniform in frequency, so it has more short-period
variables compared to a uniform distribution in period). The initial phase of the sine variation is also chosen randomly for each object. Having the list
of objects in which we injected artificial variability, we perform
magnitude-dependent thresholding for each of the tested variability
indices. For each index we test a range of thresholds between 0 and
$50\sigma$ and adopt the threshold that provides the highest efficiency in
recovering the artificial variables. Changing the threshold is needed to
account for the fact that some variability indices have a non-gaussian
distribution of their values for non-variable sources. Different thresholds
may be optimal for different variability indices. 

The efficiency of variable source selection is quantified with the 
$F_1$ score\footnote{\url{https://en.wikipedia.org/wiki/F1_score}} \citep{evaluation} which is 
the harmonic mean of the precision and recall. It reaches the maximum value of 
1.0 when all the variable sources are recovered above the threshold while all 
the non-variable sources are below the threshold. 
We refer to the maximum $F_1$ score obtained over the trial thresholds as $F_{\rm 1\,max}$. 
The following indices were tested: 
reduced $\chi^2$ ($\chi_{\rm red}^2$) 
weighted standard deviation ($\sigma_w$), 
median absolute deviation (${\rm MAD}$), 
interquartile range (${\rm IQR}$), 
robust median statistic (${\rm RoMS}$), 
normalized excess variance ($\sigma_{\rm NXS}^2$), 
peak to peak amplitude $v_{\rm peak-to-peak}$, 
lag-1 autocorrelation $l_1$, 
Welch-Stetson variability index ($I$), 
Stetson's $J$ index and its variations 
($J({\rm time})$, 
$J({\rm clip})$, 
$L$), 
consecutive same-sign deviations from the mean magnitude (${\rm CSSD}$), 
excursions ($E_x$), 
inverse von~Neumann ratio ($1/\eta$), 
excess Abbe value ($\mathcal{E}_\mathcal{A}$), and 
$S_B$ statistic. The references to the previous uses of 
these indices to characterize photometric variability 
may be found in Table~\ref{tab:idxcmpsim}. The definitions of these indices 
may be found in \cite{2017MNRAS.464..274S} while \cite{2018MNRAS.475.2326P} 
discuss correlations between the indices (the degree of correlation depends  
on the data). The \texttt{VaST} code \citep{2018A&C....22...28S} was used to
perform the simulations.

The simulation results for six data sets from the Control Sample are presented in Table~\ref{tab:idxcmpsim}. The reported $F_{\rm 1\,max}$ are the median
values over 100 implementations of the procedure of injecting random 
amplitude/period/phase/ sine variability into a random set of light curves within the data set. The data sets were obtained with different cameras, cover a different time range (Table~\ref{tab:CS}), and differ in the number of observations (the median number of light curve points, $n_{\rm LC}$,
is reported in Table~\ref{tab:idxcmpsim}). One important difference between
the observations of M4 in F467M and F775W filters is that the latter were
taken with only one 20\,sec exposure per visit and are therefore have more cosmic
ray contamination (resulting in more photometric outliers) than the former
(five 392\,sec exposures per visit).
This results in a dramatic difference in $F_{\rm 1\,max}$ values for the
variability indices that are not robust to outliers (such as $\sigma_w$),
while the robust indices (like $MAD$ and $IQR$) remain efficient (retain
their high $F_{\rm 1\,max}$ scores). This is also evident from comparing 
Fig.~\ref{fig:out45740877_WFC3_F775W} and Fig.~\ref{fig:sim_npoints_WFC3_F467M}
discussed below.

The fact that the indices that characterize light curve smoothness
(Table~\ref{tab:idxcmpsim}) are not performing as well as the robust
measures of scatter is a feature of the simulation: 
we injected variability with timescales that, in most cases, are
shorter than the observing cadence, and therefore, the resulting
light curves are not going to be smooth. This relation between the
variability timescale and the observing cadence is expected to be common in the HCV catalog as many of the time-series {\em HST} observations are optimized for long-period (high-luminosity) Cepheids, while RR~Lyrae stars and eclipsing binaries that may be present in the same data tend to vary on shorter timescales.

\begin{figure}
\includegraphics[width=0.5\textwidth]{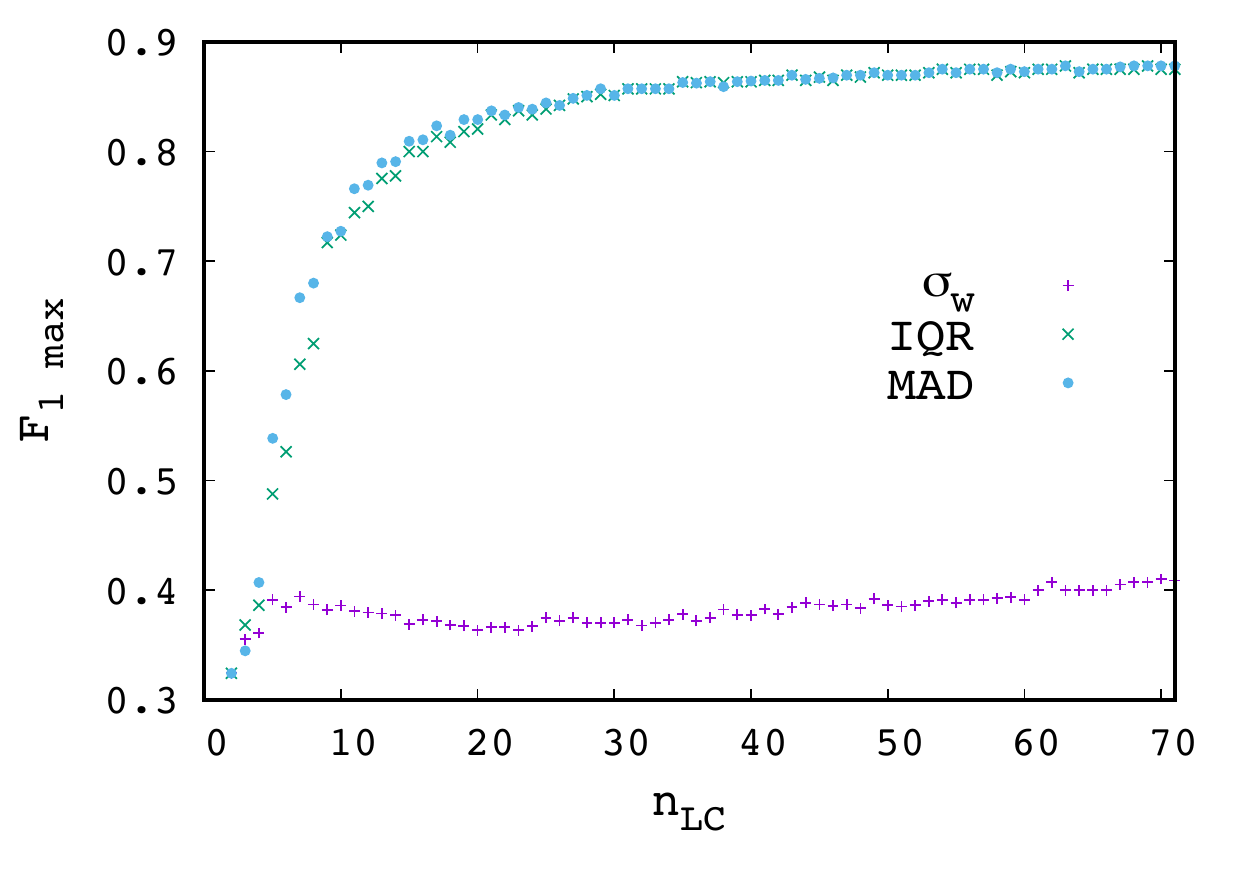}
\caption{Variability detection efficiency ($F_{\rm 1\,max}$ score) as 
a function of the number of points in the light curve. The simulated
variability was injected into the WFC3\_F775W filter observations of M4.
The data set is heavily affected by cosmic rays, resulting in the poor
performance of $\sigma_w$ compared to the robust indices MAD and IQR.
\label{fig:sim_npoints_WFC3_F775W}}
\end{figure}

\begin{figure}
\includegraphics[width=0.5\textwidth]{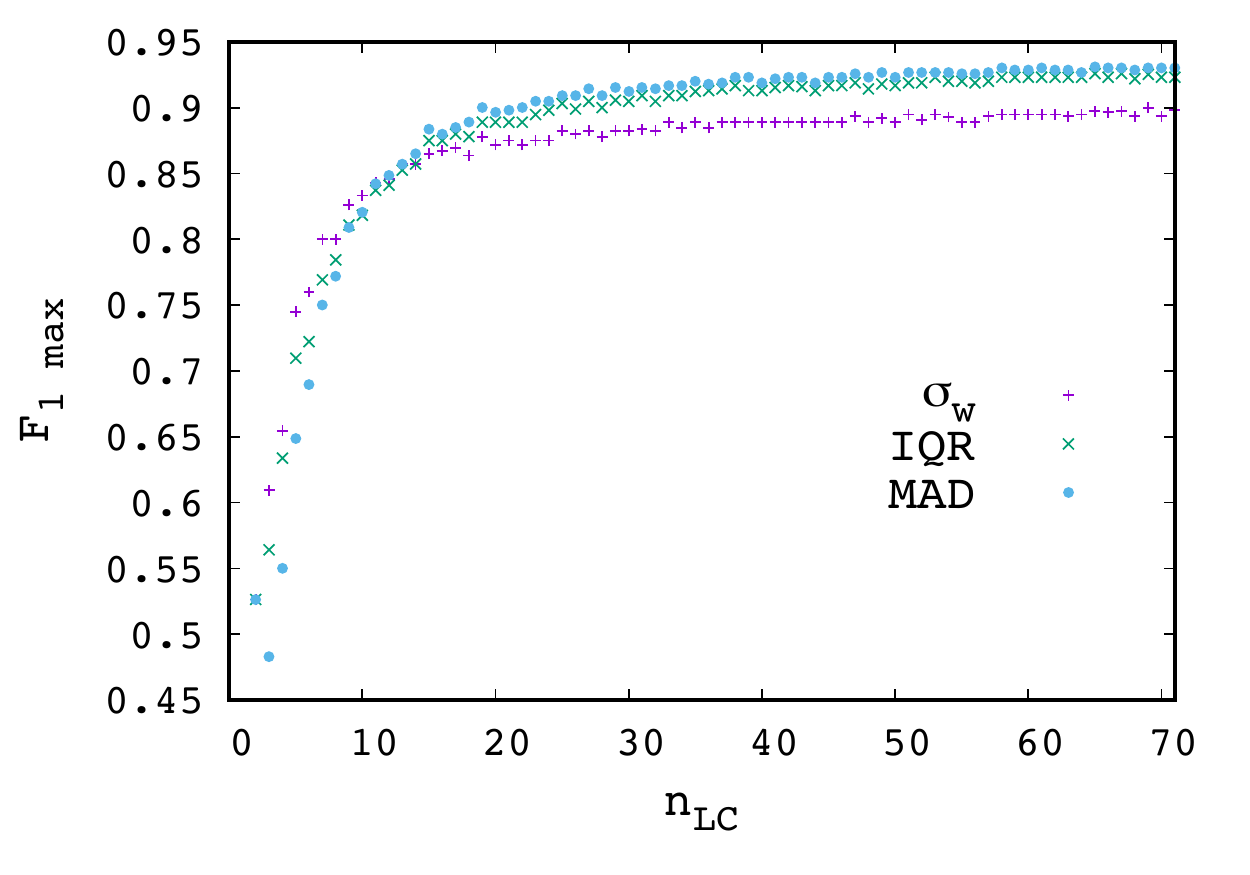}
\caption{Variability detection efficiency ($F_{\rm 1\,max}$ score) as 
a function of the number of points in the light curve. The simulated
variability was injected into the WFC3\_F467M filter observations of M4.
Unlike the F775W data (Fig.~\ref{fig:sim_npoints_WFC3_F775W}),
the F467M data are not heavily affected by cosmic rays. Under such conditions,
$\sigma_w$ performs better than the robust indices MAD and IQR for
light curves having a small number of points.
\label{fig:sim_npoints_WFC3_F467M}}
\end{figure}

In order to simulate how the variability detection efficiency changes with
the number of points in the light curve we repeat the above simulations for
the WFC3 F775W and F467M filter observations of M4, randomly selecting only $n_{\rm LC}$
observations from this data set and removing all others. The procedure is 
repeated 1000 times for each value of $n_{\rm LC}$ and the resulting median
$F_{\rm 1\,max}$ scores (for three variability indices quantifying scatter)  
are reported in Fig.~\ref{fig:sim_npoints_WFC3_F775W} and Fig.~\ref{fig:sim_npoints_WFC3_F467M}.
The F775W data set is heavily contaminated with cosmic rays, which results in a poor performance of $\sigma_w$ compared to the robust indices MAD and IQR.
In the presence of outliers, for the small number of points $4 < n_{\rm LC} < 10$ MAD provides noticeably higher $F_{\rm 1\,max}$ values compared to IQR, 
however, the opposite is true for the F467M data set where outliers are a
negligible problem (see also NGC\,4535 simulations in Table~\ref{tab:idxcmpsim}).
The simulation results support the expected trend: 
as long as there are no outlier measurements, non-robust measures of scatter
($\sigma_w$) should be more accurate than the robust ones for a given number
of points, while in the presence of outliers we expect the opposite.
The fact that for the F467M simulations
(Fig.~\ref{fig:sim_npoints_WFC3_F467M}) MAD and IQR outperform $\sigma_w$ for
$n_{\rm LC} \gtrsim 20$ suggests that this data set is not completely free
of outliers, they are just considerably less common than in the F775W data
set. In the regime $4 < n_{\rm LC} < 10$, where the
difference between the two robust indices, IQR and MAD is evident 
(Fig.~\ref{fig:sim_npoints_WFC3_F775W} and Fig.~\ref{fig:sim_npoints_WFC3_F467M}), the $F_{\rm 1\,max}$ score of the IQR (which can tolerate a smaller percentage of outliers
compared to MAD) lies between MAD and $\sigma_w$, providing a compromise
between robustness and sensitivity.

\begin{table*}
\caption{Comparison of $F_{\rm 1\,max}$ values for the tested variability indices.}             
\label{tab:idxcmpsim}      
{\centering                          
\begin{tabular}{c c c c c c c}        
\hline\hline                 
Field                    & M4    & M4    & M31-Halo11   & M31-Halo11    & NGC\,4535 & NGC\,4535 \\
Instrument               & WFC3  & WFC3  & ACS          & ACS           & WFPC2     & WFPC2     \\
Filter                   & F467M & F775W & F814W        & F606W         & F555W     & F814W     \\
$n_{\rm LC}$             & 82    & 55    & 32           & 28            &  9        & 7         \\ 
\hline
\multicolumn{7}{c}{Indices characterizing light curve scatter} \\
$\chi_{\rm red}^2$       & 0.87  & 0.31  & 0.86         & 0.83          & 0.33      & 0.27      \\
$\sigma_w$               & 0.89  & 0.47  & 0.87         & 0.84          & 0.55      & 0.39      \\
${\rm MAD}$              & 0.93  & 0.90  & 0.89         & 0.86          & 0.57      & 0.40      \\
${\rm IQR}$              & 0.92  & 0.90  & 0.89         & 0.86          & 0.67      & 0.50      \\
${\rm RoMS}$             & 0.91  & 0.86  & 0.89         & 0.85          & 0.43      & 0.35      \\
$\sigma_{\rm NXS}^2$     & 0.07  & 0.07  & 0.02         & 0.02          & 0.09      & 0.06      \\
$v_{\rm peak-to-peak}$   & 0.81  & 0.09  & 0.72         & 0.75          & 0.38      & 0.31      \\
\multicolumn{7}{c}{Indices characterizing light curve smoothness} \\
$l_1$                    & 0.20  & 0.16  & 0.03         & 0.03          & 0.04      & 0.03      \\
$I$                      & 0.56  & 0.48  & 0.40         & 0.36          & 0.02      & 0.02      \\
$J$                      & 0.59  & 0.59  & 0.81         & 0.76          & 0.44      & 0.33      \\
$J({\rm time})$          & 0.55  & 0.50  & 0.40         & 0.36          & 0.21      & 0.15      \\
$J({\rm clip})$          & 0.89  & 0.81  & 0.82         & 0.79          & 0.44      & 0.33      \\
$L$                      & 0.59  & 0.63  & 0.82         & 0.77          & 0.46      & 0.40      \\
${\rm CSSD}$             & 0.02  & 0.02  & 0.02         & 0.02          & 0.02      & 0.02      \\
$E_x$                    & 0.55  & 0.17  & 0.69         & 0.68          & 0.32      & 0.33      \\
$1/\eta$                 & 0.17  & 0.14  & 0.03         & 0.02          & 0.04      & 0.03      \\
$\mathcal{E}_\mathcal{A}$& 0.02  & 0.02  & 0.02         & 0.02          & 0.03      & 0.02      \\
$S_B$                    & 0.73  & 0.48  & 0.74         & 0.68          & 0.24      & 0.22      \\
\hline                                   
\end{tabular}
}
\small

References: 
$\chi_{\rm red}^2$~\citep{2010AJ....139.1269D},
$\sigma_w$~\citep{2008AcA....58..279K},
${\rm MAD}$~\citep{2016PASP..128c5001Z},
${\rm IQR}$~\citep{2017MNRAS.464..274S},
${\rm RoMS}$~\citep{2007AJ....134.2067R},
$\sigma_{\rm NXS}^2$~\citep{1997ApJ...476...70N},
$v_{\rm peak-to-peak}$~\citep{1989ApJ...340..150B},
$l_1$~\citep{2011ASPC..442..447K},
$I$~\citep{1993AJ....105.1813W},
$J$~\citep{1996PASP..108..851S},
$J({\rm time})$~\citep{2012AJ....143..140F},
$J({\rm clip})$~\citep{2017MNRAS.464..274S},
$L$~\citep{1996PASP..108..851S},
${\rm CSSD}$~\citep{2009MNRAS.400.1897S},
$E_x$~\citep{2014ApJS..211....3P},
$1/\eta$~\citep{2009MNRAS.400.1897S},
$\mathcal{E}_\mathcal{A}$~\citep{2014A&A...568A..78M},
$S_B$~\citep{2013A&A...556A..20F}.
\end{table*}

\clearpage
\section{Additional tables}

\onecolumn
\input{Table8.tex}
\onecolumn 
\input{Table9.tex}
\input{Table10.tex}

\end{appendix}

\end{document}

%% file: Table8.tex
\tiny
\begin{longtable}{rrrlrrrrr}
\caption{ \label{hcv_tableMFVCs} HCV multi-filter and single-filter variable candidates per \texttt{GroupID}.}\\
\hline\hline
R.A. (J2000)   &Dec. (J2000)   &\texttt{GroupID}	&Field Name    &\# Initial      &\# Final    &\# Filters        &\# MFVC        &\# SFVC   \\
(degrees)   &(degrees)   & 	&      &sources &sources  &        &       &  \\
\hline
\endfirsthead
\caption{continued.}\\
\hline\hline
R.A. (J2000)   &Dec. (J2000)   &\texttt{GroupID}	&Field Name    &\# Initial      &\# Final    &\# Filters        &\# MFVC        &\# SFVC   \\
(degrees)      &(degrees)      & 	                 &             &  sources        &sources     &               &               &         \\
\hline
\endhead
\hline
\endfoot
0.495079472	& -15.511714610	& 1062949	& WLM-STARCLUS	& 35990	& 31117	& 4	& 121	& 236	\\
3.475747041	& -30.390185660	& 1042688	& ABELL-2744-HFFPAR	& 1732	& 1495	& 7	& 13	& 11	\\
3.576352357	& -30.398039800	& 1043761	& AC118	& 2466	& 1688	& 7	& 23	& 39	\\
3.894313004	& -32.187015950	& 66669	& ESO410-005	& 17988	& 9633	& 1	& 0	& 19	\\
5.065006551	& 59.308725250	& 92609	& IC10-POS2	& 102640	& 44822	& 1	& 0	& 967	\\
5.660203860	& -72.068720590	& 439774	& NGC104-WFC-UPDATE	& 254025	& 77634	& 18	& 93	& 2026	\\
5.965410502	& -24.709494760	& 66514	& SCL-DE1	& 3732	& 1813	& 1	& 0	& 7	\\
6.544826341	& -11.079992920	& 1037352	& CETUS-DWARF	& 5342	& 4026	& 2	& 52	& 111	\\
6.636002643	& -41.857497960	& 66572	& ESO294-010	& 14545	& 7093	& 1	& 0	& 19	\\
8.199983458	& 48.378008730	& 1033762	& NGC147	& 36417	& 27339	& 2	& 32	& 186	\\
8.284913184	& 48.290157600	& 1010724	& ANY	& 3298	& 2424	& 2	& 0	& 13	\\
8.873996052	& 36.503287960	& 1045103	& ANDROMEDA-III	& 6614	& 5381	& 2	& 21	& 105	\\
8.954952633	& -43.206258020	& 1074873	& SPARCS-J003550-431210-COPY	& 711	& 519	& 2	& 7	& 1	\\
9.799728109	& 48.435035430	& 1033573	& NGC185	& 28758	& 21130	& 2	& 138	& 341	\\
9.932522332	& 48.415645960	& 1012451	& ANY	& 4056	& 3034	& 2	& 7	& 88	\\
10.134718080	& 40.746713140	& 1078908	& M31-B005	& 10432	& 1242	& 1	& 0	& 45	\\
10.942091470	& 41.001298780	& 1045904	& M31	& 11525935	& 551781	& 5	& 107	& 6386	\\
10.958906840	& 39.770107650	& 1033697	& ANY	& 1739	& 1376	& 2	& 0	& 17	\\
11.017427590	& -20.564792110	& 1064963	& XMM44	& 625	& 443	& 2	& 3	& 8	\\
11.082702210	& 39.784454780	& 1033421	& M31-TIDALSTREAM1	& 8957	& 6792	& 2	& 12	& 20	\\
11.296398110	& -73.211160770	& 56019	& HIGH-GALACTIC-LATITUDE	& 68231	& 49329	& 3	& 491	& 1641	\\
11.315025630	& 38.004159700	& 1071126	& ANY	& 1368	& 1122	& 2	& 2	& 23	\\
11.428953570	& 38.042769870	& 1064398	& ANDROMEDA-I	& 14801	& 12576	& 2	& 92	& 151	\\
11.526879280	& 40.698480400	& 1043756	& M31-B379	& 14359	& 11509	& 2	& 34	& 57	\\
11.575856070	& -73.342660500	& 1060093	& SMC-F18-WFC	& 87220	& 313	& 2	& 0	& 2	\\
11.639456390	& 40.671999130	& 1043755	& ANY	& 2073	& 1751	& 2	& 9	& 28	\\
11.658976770	& -73.176846010	& 289829	& LIST-1	& 6708	& 3740	& 3	& 1	& 80	\\
12.270891940	& 40.301782090	& 439858	& NGC224-22KPC	& 2026	& 1402	& 2	& 0	& 4	\\
12.288275700	& 42.759679010	& 1033359	& M31-OUTERDISK1	& 13974	& 10644	& 2	& 12	& 59	\\
12.404305260	& 42.707136260	& 1033692	& ANY	& 2291	& 1884	& 2	& 7	& 31	\\
13.370965580	& 39.832611450	& 445976	& NGC224-35KPCA	& 1456	& 1179	& 2	& 1	& 11	\\
13.537404890	& 39.797394510	& 439608	& NGC224-35KPCB	& 1333	& 1082	& 2	& 0	& 14	\\
13.641544760	& -72.648447860	& 1044880	& HIGH-GALACTIC-LATITUDE	& 65347	& 3872	& 1	& 0	& 73	\\
13.805802780	& -72.507935970	& 289947	& LIST-1	& 5902	& 828	& 3	& 3	& 32	\\
13.904171110	& -72.407267470	& 408245	& LIST-1	& 3231	& 549	& 1	& 0	& 13	\\
14.614070700	& -72.249865720	& 353029	& LIST-1	& 5771	& 2526	& 3	& 19	& 116	\\
14.780237720	& -72.188553230	& 1053537	& NGC346	& 46234	& 7918	& 2	& 10	& 134	\\
14.879852610	& 32.393442050	& 1045492	& ANDROMEDA-XVI	& 1743	& 940	& 2	& 4	& 11	\\
15.958660810	& 21.882082220	& 1063351	& PISCES-I	& 5233	& 3826	& 2	& 17	& 58	\\
16.046030500	& 2.220925169	& 1045442	& IC-1613-FIELD2	& 475	& 370	& 1	& 0	& 11	\\
16.120357510	& 2.156895687	& 69810	& IC1613	& 27045	& 23106	& 2	& 97	& 167	\\
17.376581170	& 35.725221740	& 36624	& NGC404	& 1884	& 862	& 1	& 0	& 9	\\
17.447468470	& -72.875866490	& 87786	& HIGH-GALACTIC-LATITUDE	& 20894	& 11501	& 1	& 0	& 187	\\
17.516163190	& -2.414328481	& 1011552	& HI-LAT	& 991	& 408	& 3	& 0	& 3	\\
17.768124330	& -72.891846220	& 55509	& HIGH-GALACTIC-LATITUDE	& 20675	& 15780	& 1	& 0	& 82	\\
17.939753130	& -35.062271290	& 1049880	& CLH12-WFC3POINTING	& 1310	& 898	& 1	& 0	& 5	\\
18.575030230	& 38.119420470	& 1045180	& ANDROMEDA-XV	& 2460	& 1599	& 2	& 9	& 69	\\
19.021788990	& 33.365106150	& 1058861	& ANY	& 1088	& 838	& 2	& 9	& 20	\\
19.098526490	& 33.434377400	& 1042143	& ANDROMEDA-II	& 12018	& 9470	& 2	& 72	& 166	\\
23.490565100	& 30.407033710	& 1062832	& M33	& 1060282	& 51548	& 4	& 2	& 562	\\
24.205936780	& 41.524189800	& 87237	& UNKNOWN-TARGET-1	& 590	& 352	& 1	& 0	& 2	\\
28.178193340	& -13.945233640	& 23565	& GAL-CLUS-015245-135737-POS08	& 1179	& 390	& 2	& 0	& 4	\\
31.442579700	& -58.483479470	& 1039295	& SPT0205-E08-418-MIDPOINT	& 756	& 617	& 2	& 13	& 8	\\
32.432429460	& -4.614702546	& 46172	& SGR-STREAM-1	& 1440	& 531	& 1	& 0	& 8	\\
34.357051550	& -5.204224549	& 1036556	& Z7-GIANTLAE	& 11352	& 3966	& 4	& 2	& 33	\\
35.280507730	& 35.941622150	& 64702	& QSO-022105+355613	& 1074	& 653	& 1	& 0	& 1	\\
35.716451440	& 42.488242950	& 1037724	& SN1986J	& 3113	& 1729	& 1	& 0	& 13	\\
36.111153410	& -3.391831535	& 1061394	& SPARCSJ0224	& 598	& 435	& 2	& 3	& 3	\\
36.780796660	& 33.590610420	& 538286	& NGC925	& 2132	& 1261	& 1	& 0	& 14	\\
39.539658600	& -1.329463556	& 1028931	& N1015	& 5380	& 3611	& 2	& 0	& 53	\\
39.971081180	& -1.586147087	& 1033838	& SNABELL370	& 3265	& 2053	& 7	& 45	& 73	\\
40.045881850	& -34.543154630	& 1059852	& FORNAX-CLUSTER4	& 8757	& 3371	& 3	& 23	& 58	\\
40.058950060	& -1.616596037	& 1083558	& ABELL-370-HFFPAR	& 1700	& 1455	& 7	& 11	& 14	\\
42.112940520	& -3.540953533	& 1021022	& CALIGULA3	& 732	& 319	& 1	& 0	& 1	\\
48.413834080	& -67.193566610	& 1079516	& ANY	& 555	& 339	& 3	& 2	& 7	\\
50.521822860	& -15.395683500	& 24452	& SN-2012Z	& 11519	& 8637	& 3	& 19	& 181	\\
51.276873770	& -36.358279700	& 483765	& NGC1326A	& 730	& 498	& 2	& 1	& 6	\\
52.727948060	& -28.722348560	& 1054528	& SPARCSJ0330	& 606	& 445	& 2	& 3	& 3	\\
53.177857170	& -27.922753570	& 1084533	& UDF$\_$(Merged)	& 25224	& 14278	& 9	& 90	& 164	\\
53.403340700	& -36.128855350	& 331331	& SN2001DU	& 9046	& 571	& 1	& 0	& 7	\\
55.553912440	& -29.899653340	& 483481	& NGC1425	& 661	& 388	& 1	& 0	& 8	\\
56.092058410	& -43.534428460	& 1075853	& ERIDANUS2	& 4005	& 2106	& 1	& 0	& 10	\\
56.102155050	& -44.599903700	& 1045280	& SN2003HN	& 13312	& 9228	& 2	& 1	& 145	\\
64.037020340	& -24.072588440	& 1042781	& SN-M0416-IR	& 2552	& 1830	& 7	& 33	& 49	\\
64.105624000	& -24.137125440	& 1040153	& MACSJ0416.1-2403-HFFPAR	& 1865	& 1404	& 7	& 13	& 10	\\
64.392534960	& -62.782648480	& 1084480	& SN-2009IB	& 32051	& 18461	& 1	& 0	& 269	\\
70.360306470	& -2.871318803	& 360365	& SN1999EM	& 1556	& 1054	& 2	& 1	& 21	\\
77.375965220	& -69.129122470	& 1048097	& NGC-1856	& 36678	& 11914	& 1	& 0	& 173	\\
78.517988600	& -40.054597500	& 33004	& NGC1851	& 28212	& 17926	& 2	& 13	& 546	\\
80.484860210	& -69.502344530	& 1056946	& OGLE052218.07-692827.4	& 172529	& 89657	& 1	& 0	& 585	\\
84.114945230	& -70.116525640	& 381717	& LIST-1	& 4372	& 1114	& 3	& 10	& 43	\\
84.245627080	& -66.363422320	& 1042179	& STAR-CLUS-053707-662203	& 2939	& 1588	& 1	& 0	& 16	\\
84.479026430	& -69.245704330	& 1039945	& STAR-0537-6910	& 580832	& 195362	& 27	& 468	& 4062	\\
86.752544600	& -34.281952750	& 177255	& NGC2090	& 1281	& 881	& 2	& 0	& 18	\\
86.978237130	& -70.230020860	& 56487	& HIGH-GALACTIC-LATITUDE	& 38162	& 25934	& 1	& 0	& 512	\\
106.021557800	& -3.845185765	& 1034275	& V838-MON-ECHO-COPY	& 1435	& 811	& 2	& 1	& 37	\\
109.395422800	& 37.760470440	& 62353	& MACSJ0717.5+3745-POS5	& 6078	& 3622	& 7	& 109	& 117	\\
110.830281800	& -73.463489250	& 1073046	& SMACSJ0723.3-7327	& 683	& 369	& 2	& 5	& 24	\\
114.119408100	& -69.554788310	& 73455	& SN1999GA	& 25654	& 17433	& 2	& 4	& 246	\\
114.609685100	& 65.464465330	& 25090	& NGC2403-HALO-2	& 2124	& 1192	& 2	& 1	& 21	\\
116.219973700	& 39.467756870	& 72495	& MACSJ0744.9+3927-Y1	& 1027	& 395	& 2	& 7	& 6	\\
122.229472100	& 6.728621118	& 81871	& GRB021211	& 1088	& 357	& 1	& 0	& 1	\\
123.659412000	& 49.056726260	& 483544	& NGC2541	& 1018	& 584	& 1	& 0	& 12	\\
138.027065000	& -64.864488840	& 33470	& NGC2808Y98B	& 75279	& 30350	& 3	& 85	& 1336	\\
138.272575500	& -64.844977390	& 25262	& NGC-2808	& 4424	& 2970	& 2	& 5	& 96	\\
140.524832700	& 50.953679330	& 18944	& SN1999BY-3	& 1366	& 755	& 1	& 0	& 15	\\
143.481802700	& 55.240553530	& 36988	& IZW18-MAINBODY	& 1542	& 1106	& 3	& 1	& 18	\\
143.656096600	& 17.125358980	& 1028455	& LEO-T	& 526	& 381	& 1	& 0	& 2	\\
143.728037200	& 17.047714070	& 1042327	& LEO-T	& 5444	& 3693	& 2	& 13	& 12	\\
147.753382800	& 33.559296090	& 24555	& NGC3021-ACS	& 3287	& 2345	& 2	& 2	& 38	\\
148.803039200	& 69.690576730	& 74633	& M82	& 21706	& 15801	& 2	& 4	& 71	\\
149.856321900	& 30.750169960	& 1037445	& LEOA-CENTER	& 24301	& 19744	& 2	& 79	& 308	\\
150.040369000	& 2.473513981	& 1081922	& cosmos	& 20657	& 6331	& 4	& 2	& 52	\\
152.089378700	& 12.300086950	& 25044	& LEO-I-DSPH	& 71454	& 37739	& 1	& 0	& 135	\\
153.522443000	& 0.642359289	& 1049163	& MOO1014-E08-470	& 806	& 643	& 2	& 7	& 4	\\
154.970955500	& 45.622203040	& 1038808	& NGC3198	& 1236	& 699	& 1	& 0	& 14	\\
155.450557800	& 18.092720520	& 1046053	& LEOP	& 2110	& 1438	& 2	& 7	& 17	\\
156.021671300	& -57.762214380	& 1054696	& WESTERLUND2	& 10687	& 5011	& 1	& 0	& 362	\\
159.292264000	& -27.662112840	& 38905	& NGC3314-UPLEFT	& 4106	& 3260	& 2	& 14	& 108	\\
159.799993500	& 41.681218920	& 483536	& NGC3319	& 988	& 589	& 1	& 0	& 12	\\
160.978310200	& 11.694576450	& 1030272	& SN2012AW	& 2206	& 885	& 1	& 0	& 23	\\
161.324267200	& -60.020004420	& 22917	& Eta-Car	& 50270	& 14325	& 3	& 0	& 550	\\
161.610152900	& -59.609894150	& 55731	& ANY	& 5007	& 3322	& 1	& 0	& 66	\\
161.777281200	& 11.827124280	& 1080182	& SN1998BU-3	& 666	& 314	& 1	& 0	& 11	\\
161.781170900	& 17.287822220	& 92880	& NGC3370-ACS	& 10033	& 7518	& 2	& 10	& 59	\\
161.933161600	& 13.988115580	& 27681	& NGC3377-NUC1	& 10270	& 1735	& 1	& 0	& 13	\\
162.053871700	& 12.602878380	& 1041148	& NGC3384-POS-W	& 27966	& 8732	& 2	& 7	& 182	\\
162.296210500	& -53.329749320	& 1083958	& LUHMAN16ABMINAPNE1	& 1038	& 784	& 2	& 0	& 14	\\
162.342403100	& 56.678447920	& 1047807	& SPARCSJ1049-REPOINT	& 482	& 329	& 2	& 3	& 2	\\
163.351519900	& 16.775237110	& 1028736	& N3447	& 9128	& 6964	& 2	& 0	& 58	\\
168.421973200	& 22.201180510	& 1014062	& LEOII-Q111340+221242	& 2593	& 1129	& 2	& 7	& 35	\\
169.587190100	& -32.780711850	& 1031228	& NGC3621-OFF	& 1836	& 1019	& 1	& 0	& 22	\\
170.083899500	& 12.983679220	& 445382	& SN2009HD	& 3553	& 637	& 1	& 0	& 8	\\
173.222913500	& -76.017929400	& 91445	& ANY	& 907	& 411	& 1	& 0	& 3	\\
175.692407600	& 15.456023330	& 1080745	& MOO1142	& 632	& 444	& 2	& 3	& 5	\\
177.387087900	& 22.398492700	& 66095	& REFSDALREAPPEARANCE	& 6130	& 3877	& 9	& 42	& 63	\\
178.930006300	& 55.322777590	& 1083372	& NGC3972	& 8085	& 5492	& 2	& 3	& 143	\\
179.126633000	& 55.125475070	& 73298	& NGC4051	& 736	& 350	& 1	& 0	& 4	\\
180.485472500	& -18.883810650	& 1043713	& NGC4038-S-TAIL	& 1595	& 422	& 1	& 0	& 12	\\
182.648904000	& 39.403874870	& 1017300	& NGC4151	& 13752	& 5789	& 1	& 0	& 59	\\
184.838133500	& 47.391646000	& 1043384	& UGC-07356	& 114212	& 61491	& 8	& 230	& 830	\\
185.791870500	& 15.804409580	& 47524	& SN2006X	& 2721	& 868	& 1	& 0	& 23	\\
186.599851500	& 31.221500200	& 499217	& SN2013DF	& 804	& 461	& 1	& 0	& 16	\\
186.797143900	& 9.462037454	& 1045340	& SN2012CG	& 11400	& 6814	& 2	& 1	& 46	\\
186.804972300	& 23.858403990	& 1046067	& COMA-BERENICES-V22	& 5680	& 1062	& 2	& 0	& 5	\\
187.048255300	& 12.560581790	& 1934	& VIRGO-ICFIELD3	& 1456	& 1201	& 2	& 0	& 9	\\
187.168581400	& 62.637148600	& 1084534	& HDF$\_$(Merged)	& 23045	& 13837	& 10	& 75	& 143	\\
187.468856200	& 7.981483036	& 1040498	& VCC1226	& 1837	& 820	& 4	& 0	& 5	\\
187.713602300	& 12.390529780	& 1072340	& M87	& 26817	& 15731	& 4	& 31	& 157	\\
187.914199500	& 3.928083565	& 1051133	& SDSS-588010877688807475	& 1943	& 1233	& 1	& 0	& 27	\\
188.484162300	& 2.608183274	& 508907	& NGC4527	& 930	& 371	& 1	& 0	& 2	\\
188.569682600	& 8.165917216	& 209749	& NGC4535	& 1537	& 1032	& 2	& 5	& 21	\\
188.594096100	& 2.143856732	& 1079995	& NGC4536-WFPC2	& 1279	& 767	& 1	& 0	& 6	\\
188.867305200	& 14.469535620	& 531865	& NGC4548	& 1343	& 968	& 2	& 0	& 15	\\
189.018156800	& 26.033051320	& 46574	& NGC4565-HALO3	& 31651	& 1848	& 2	& 5	& 26	\\
189.233096200	& 14.214497220	& 529814	& NGC4571	& 1119	& 691	& 1	& 0	& 10	\\
189.955780500	& -11.441077530	& 1074339	& ANY	& 4640	& 407	& 1	& 0	& 4	\\
190.220145100	& -40.965968070	& 531716	& SN2008CN	& 1162	& 639	& 1	& 0	& 21	\\
190.724637900	& 13.255459210	& 1009478	& NGC4639	& 1071	& 624	& 1	& 0	& 8	\\
190.917643000	& 11.594597720	& 1038219	& VCC1978-OFFSET	& 1773	& 684	& 2	& 0	& 11	\\
192.659387700	& 25.546854610	& 555853	& NGC4725	& 704	& 393	& 1	& 0	& 7	\\
194.286273300	& 34.321497660	& 1076197	& CVNII-Q125704+341920	& 2433	& 1278	& 1	& 0	& 7	\\
195.354923200	& 27.884096870	& 74366	& NGC4921-PA299	& 4387	& 2812	& 2	& 0	& 21	\\
195.480154400	& 27.881910250	& 1039468	& ANY	& 1671	& 325	& 3	& 0	& 14	\\
197.871282200	& -1.346418497	& 1033313	& ABELL1689-POS3	& 2005	& 1442	& 2	& 0	& 10	\\
198.205971300	& -47.475380210	& 45491	& ER8-FIELD	& 1486	& 434	& 1	& 0	& 12	\\
201.375683100	& -42.981138010	& 1055708	& SN2016ADJ	& 2159	& 375	& 1	& 0	& 7	\\
201.379131000	& -47.577621440	& 1080307	& WCEN	& 57830	& 26673	& 9	& 40	& 744	\\
201.781488400	& -47.499462370	& 12611	& OMEGACEN-3	& 246460	& 166087	& 13	& 879	& 6685	\\
201.969708500	& -31.506322180	& 21359	& ESO444-G046	& 6035	& 2124	& 1	& 0	& 8	\\
203.817517400	& 40.995825340	& 1077193	& ACO-1763	& 728	& 417	& 2	& 3	& 13	\\
206.886618400	& -11.754619380	& 23539	& RXJ1347.5-1145	& 628	& 383	& 3	& 15	& 12	\\
209.276881100	& 4.515200147	& 35730	& REF13	& 903	& 321	& 3	& 1	& 0	\\
210.817624000	& 54.159625820	& 1053852	& M101-A	& 280111	& 136077	& 9	& 669	& 1803	\\
211.208077800	& 54.472818010	& 1043478	& M101-NE-PLUME3	& 4276	& 2643	& 2	& 1	& 30	\\
211.991244600	& -11.411374500	& 37745	& FIELD-140801-1126H	& 6822	& 5242	& 1	& 0	& 11	\\
215.110454900	& 52.971386610	& 1045196	& SN1997A2	& 19733	& 5749	& 5	& 10	& 50	\\
215.588734400	& -0.399217810	& 1013227	& NGC5584	& 16211	& 13660	& 3	& 32	& 105	\\
215.950589100	& 24.090980410	& 62371	& ROSE-GRISM	& 757	& 432	& 3	& 4	& 4	\\
216.447362500	& 35.543002490	& 82373	& FIELD-142549+353248	& 611	& 311	& 1	& 0	& 1	\\
216.633965900	& 35.151219170	& 1030110	& ISCS25.687	& 802	& 488	& 2	& 12	& 2	\\
218.111565900	& 32.827098630	& 1034184	& ISCSJ1432.4+3250	& 623	& 471	& 2	& 7	& 3	\\
218.649094600	& 59.333769890	& 1028696	& U9391	& 3003	& 2341	& 2	& 0	& 21	\\
227.458545100	& 67.295365940	& 1052878	& URSA-MINOR-GALAXY-4-COPY-1	& 9771	& 671	& 2	& 4	& 24	\\
229.640406500	& 2.080274917	& 33584	& M5	& 61211	& 18872	& 3	& 63	& 940	\\
230.375962800	& -7.386011459	& 1027929	& SN2005CF	& 3464	& 2075	& 2	& 1	& 72	\\
244.159754000	& -22.920672690	& 1039049	& M-80	& 952	& 476	& 2	& 0	& 3	\\
244.258760500	& -22.978397670	& 33079	& M80-WFPC2	& 20832	& 12857	& 4	& 4	& 299	\\
245.896921400	& -26.526456380	& 33675	& M4-CORE	& 11299	& 8460	& 2	& 10	& 120	\\
245.984119900	& -26.533600770	& 304489	& M4-4-CORE-RADII	& 711	& 493	& 2	& 7	& 19	\\
246.445580100	& -72.202364880	& 65764	& NGC6101	& 5451	& 3249	& 3	& 1	& 68	\\
247.825014800	& -40.323817340	& 516320	& LOW	& 1378	& 843	& 1	& 0	& 40	\\
247.891084800	& 37.613912910	& 1009308	& PAR	& 694	& 376	& 4	& 1	& 6	\\
248.131039900	& -13.042828370	& 33308	& NGC6171-M107	& 3760	& 1278	& 1	& 0	& 48	\\
250.793840200	& 39.339911080	& 1030681	& ANY	& 452	& 322	& 1	& 0	& 1	\\
259.276716700	& 43.138331520	& 33610	& NGC6341-POS1	& 48146	& 11185	& 1	& 0	& 336	\\
264.070239500	& -44.725694360	& 33109	& NGC6388	& 78750	& 38240	& 3	& 53	& 316	\\
265.157698900	& -53.681256880	& 33688	& NGC6397-HH	& 19462	& 10711	& 5	& 9	& 306	\\
265.266884700	& -53.741198200	& 1033498	& NGC6397-WFPC2	& 48150	& 32139	& 2	& 117	& 1309	\\
265.627413000	& -40.260208250	& 1030256	& LOW	& 4723	& 778	& 1	& 0	& 65	\\
266.356899900	& -29.030646960	& 88697	& SGRA-NC	& 89152	& 6695	& 1	& 0	& 260	\\
267.147272100	& -20.312820830	& 1046062	& ANY	& 45958	& 26078	& 2	& 4	& 288	\\
267.227849200	& -20.375595770	& 1024360	& PSR6440B	& 86322	& 67371	& 3	& 404	& 2328	\\
267.602621300	& -37.131468260	& 1033918	& NGC-6441-OFF	& 39823	& 16139	& 1	& 0	& 168	\\
268.128047600	& -17.688308660	& 521507	& SAKURAI-NOVALIKE-VAR	& 3541	& 2922	& 2	& 28	& 117	\\
269.020511000	& -21.955747840	& 63563	& LOW	& 5820	& 2577	& 1	& 0	& 103	\\
269.676660500	& -29.142625410	& 1063416	& WFPC2-2	& 394202	& 277258	& 2	& 1098	& 5961	\\
269.817975500	& -29.194997840	& 1040910	& SWEEPS-R1	& 689565	& 548735	& 2	& 2374	& 13683	\\
269.833593400	& -29.271668850	& 1061725	& ANY	& 171177	& 132584	& 2	& 810	& 2981	\\
269.842928000	& -29.327223720	& 1047823	& ANY	& 170242	& 134197	& 2	& 835	& 2997	\\
271.836126900	& -24.999712230	& 17120	& NGC6544	& 26359	& 14421	& 2	& 18	& 292	\\
274.687931600	& -13.747139180	& 1045014	& M16-A	& 3345	& 468	& 1	& 0	& 19	\\
275.923454300	& -30.360921720	& 68312	& NGC6624-WFPC2-POS4	& 2932	& 1181	& 1	& 0	& 1	\\
280.733373100	& -32.220545050	& 1039204	& PAR	& 2748	& 409	& 1	& 0	& 7	\\
280.801687700	& -32.294429130	& 33390	& NGC6681-WFPC2-POS4	& 2583	& 1040	& 1	& 0	& 1	\\
280.960699300	& -32.334952990	& 423081	& ANY	& 2035	& 960	& 1	& 0	& 29	\\
283.767304900	& -30.271751110	& 510437	& SGR-DSPH2	& 3186	& 399	& 1	& 0	& 11	\\
283.801435200	& -31.346838320	& 507380	& ANY	& 1522	& 781	& 1	& 0	& 19	\\
284.066234000	& -37.853507310	& 519756	& LOW	& 588	& 423	& 1	& 0	& 25	\\
286.980282200	& -18.768425360	& 525031	& LIST-1	& 804	& 540	& 2	& 8	& 37	\\
287.722874300	& -59.974237780	& 33701	& PSR-J1911-5958A	& 53907	& 25691	& 5	& 44	& 870	\\
290.212620800	& 37.800541050	& 1008520	& NGC6791	& 2778	& 1381	& 2	& 0	& 14	\\
291.088415100	& 9.895533570	& 1063228	& VY2-2-COPY	& 6343	& 2187	& 4	& 1	& 23	\\
292.296639300	& -6.869730730	& 93958	& ANY	& 2616	& 1728	& 1	& 0	& 31	\\
292.501623500	& -17.669640940	& 22844	& SAGITTARIUS-DWARF-IRREGULAR	& 13602	& 7671	& 1	& 0	& 195	\\
292.853101400	& -26.536974030	& 1024716	& MACS1931-WFC3PAR1	& 835	& 378	& 2	& 2	& 30	\\
292.982474000	& 11.036191290	& 381893	& ANY	& 2324	& 476	& 1	& 0	& 30	\\
295.659026100	& -10.329599350	& 1014035	& NGC6814	& 7331	& 5757	& 2	& 3	& 32	\\
296.659407700	& -19.605604900	& 82136	& GRB020813	& 1441	& 351	& 1	& 0	& 1	\\
299.679758600	& 35.247485890	& 420853	& ANY	& 1437	& 572	& 1	& 0	& 27	\\
299.838458900	& 40.852085320	& 509388	& PARALLEL-FIELD	& 709	& 385	& 1	& 0	& 10	\\
303.791751800	& 6.210771995	& 1024567	& GRB120923A	& 596	& 389	& 1	& 0	& 3	\\
308.811011400	& 60.195994110	& 1037524	& SPIRITS-15AFP	& 20341	& 5989	& 1	& 0	& 80	\\
310.249431200	& -44.860242880	& 1044811	& SPT2040	& 868	& 708	& 2	& 13	& 14	\\
311.715749600	& -12.835157210	& 1041481	& DDO210	& 11216	& 9450	& 2	& 75	& 92	\\
316.524228400	& -58.749529310	& 1039240	& SPT-CLJ2106-5844	& 831	& 675	& 2	& 15	& 10	\\
322.337010400	& 0.172586888	& 1040914	& RXJ2129-WFC3PAR2	& 1029	& 667	& 2	& 0	& 17	\\
322.364992000	& -7.681219071	& 62377	& MACSJ2129-0741	& 792	& 399	& 2	& 8	& 12	\\
322.427334200	& 12.222673330	& 353549	& ANY	& 3800	& 2576	& 1	& 0	& 139	\\
322.579603100	& 12.115746100	& 25292	& M15-SECOND	& 6044	& 1958	& 1	& 0	& 106	\\
325.094528200	& -23.185945480	& 33571	& NGC7099-WFPC2-POS4	& 6480	& 779	& 1	& 0	& 3	\\
325.150647700	& -23.236204030	& 514176	& NGC7099-OUTER	& 1247	& 639	& 1	& 0	& 21	\\
334.366518600	& 0.134914196	& 19330	& TARGET6	& 3510	& 953	& 2	& 0	& 6	\\
334.423026200	& 0.834759540	& 19400	& SA22POINTL36	& 1241	& 763	& 5	& 5	& 13	\\
334.585476300	& 40.550329410	& 1027220	& SN2013DY	& 4741	& 2921	& 2	& 0	& 93	\\
338.175007100	& 31.211115750	& 1045157	& ANDROMEDA-XXVIII	& 4028	& 3078	& 2	& 17	& 56	\\
338.247644200	& -60.546434810	& 79757	& HDFS	& 789	& 420	& 2	& 0	& 26	\\
339.266996300	& 34.468875420	& 555871	& NGC7331-POS1	& 996	& 641	& 1	& 0	& 27	\\
340.450104200	& -64.420934510	& 1037453	& TUCANA-DWARF	& 6210	& 4838	& 2	& 69	& 234	\\
342.170209300	& -44.530294380	& 1042675	& RXJ2248-ROT	& 2751	& 1768	& 7	& 44	& 44	\\
342.318216500	& -44.548460310	& 1075304	& RXJ2248-WFC3PAR2	& 1723	& 1434	& 7	& 17	& 17	\\
350.777554300	& 58.787328770	& 53275	& CAS-A-WEST	& 5091	& 560	& 3	& 13	& 27	\\
352.062055100	& 14.780585490	& 1051244	& ANY	& 2880	& 2574	& 2	& 0	& 29	\\
352.156042200	& 14.724993450	& 1064674	& PEGASUS-B	& 55108	& 45352	& 2	& 228	& 747	\\
\end{longtable}                                               

%% file: Table9.tex
\begin{landscape}
\begin{table*}
\tiny
      \begin{minipage}[t]{\textwidth}
      \caption[]{The HCV catalog.}
      \label{hcv_tableHCV} 
      \begin{tabular}{llrlllllcllrllrr}
\hline\hline
R.A. (J2000)	  & Dec. (J2000)     &\texttt{MatchID}  &\texttt{GroupID}     &Subgroup$^a$   &Pipeline 	      &Expert	        &\# Filters    & Filter	& FilterDetection	   &VarQuality	 &n$_{LC}$ 	 & <m$_{\rm HSC}$> & <m$_{\rm HCV}$> & MAD & $\chi_{\rm red}^{2}$	\\
(degrees)	  &(degrees)	     &		        &      		      &          &Class.$^b$          &Class.$^c$	&	    &                   &  Flag$^d$        &Flag$^e$	 &		 &(mag)	      &(mag)	 &      &               \\
\hline
0.4281922281	&-15.4549541473		&51033165 	&1062949	 	&-5	&1	 	 	&0	 	&2	 	&WFC3\_F814W	&1	&- - - - -		&12	&24.165	 	&24.166	 	&0.043	 	&4.4       \\
0.4620466232	&-15.5108289719		&17833555 	&1062949	 	&-5	&1	 	 	&0	 	&2	 	&ACS\_F814W	&1	&CAAAB		&12	&24.897	 	&24.897	 	&0.023	 	&3.7       \\
0.4634322822	&-15.5074214935		&44149982 	&1062949	 	&-5	&2	 	 	&1	 	&2	 	&ACS\_F814W	&1	&AABAA		&11	&25.713	 	&25.713	 	&0.148	 	&29.4      \\
0.4645933211	&-15.5040855408		&44006736 	&1062949	 	&-5	&1	 	 	&0	 	&2	 	&ACS\_F814W	&1	&AAAAB		&12	&26.013	 	&26.013	 	&0.177	 	&7.1       \\
0.4657302499	&-15.5397748947		&104466632	&1062949	 	&-5	&1	 	 	&0	 	&1	 	&ACS\_F814W	&1	&AAAAC		&6	&24.642	 	&24.642	 	&0.063	 	&5.0       \\
0.4685320258	&-15.5385761261		&49380750 	&1062949	 	&-5	&1	 	 	&0	 	&2	 	&ACS\_F814W	&1	&AAAAC		&10	&23.449	 	&23.448	 	&0.035	 	&38.5      \\
0.4685845673	&-15.5160827637		&41579373 	&1062949	 	&-5	&1	 	 	&0	 	&2	 	&ACS\_F814W	&1	&BAAAB		&11	&24.528	 	&24.527	 	&0.021	 	&1.2       \\
0.4689802527	&-15.5137805939		&86716251 	&1062949	 	&-5	&1	 	 	&0	 	&2	 	&ACS\_F814W	&1	&AAABB		&12	&23.734	 	&23.733	 	&0.020	 	&15.7      \\
0.4693635106	&-15.5307703018		&105204801	&1062949	 	&-5	&2	 	 	&1	 	&2	 	&ACS\_F814W	&1	&BABAB		&10	&25.750	 	&25.750	 	&0.177	 	&14.7      \\
0.4695036709	&-15.5268716812		&12663862 	&1062949	 	&-5	&1	 	 	&0	 	&2	 	&ACS\_F814W	&1	&AAAAC		&12	&21.971	 	&21.971	 	&0.026	 	&104.1     \\
\hline
\end{tabular} 
\end{minipage}
\tablefoot{This table is available in its entirety in machine-readable and Virtual Observatory (VO) forms in the online journal. A portion is shown here for guidance regarding its form and content.
\tablefoottext{a}{A value of "-5" denotes that the \texttt{GroupID} was processed as one subgroup.}
\tablefoottext{b}{A pipeline classification ``1'' corresponds to a single-filter variable candidate (SFVC), while ``2'' corresponds to a multi-filter variable candidate (MFVC).}
\tablefoottext{c}{The expert-validation classification flag: ``0'' indicates a variable candidate ``not classified by the experts'', ``1'' indicates a ``high confidence variable'', ``2'' indicates a ``probable variable'', ``4'' indicates a ``possible artifact''.}
\tablefoottext{d}{The filter detection flag indicates whether the object is variable [``1''] or is not variable [``0''] in the preceding instrument and filter combination.}
\tablefoottext{e}{The variability quality flag in the preceding instrument and filter combination as defined in Section~\ref{sec:validationalgorithm}. The five letters correspond to \texttt{CI}, \texttt{D}, \texttt{MagerrAper2}, \texttt{MagAper2-MagAuto}, p2p; AAAAA corresponds to the highest quality flag.}
}
\end{table*}
\end{landscape}

%% file: Table10.tex
\begin{landscape}
\begin{table*}
\small
      \begin{minipage}[t]{\textwidth}
      \caption[]{Catalog of sources that were not classified as variable candidates by the HCV pipeline. }
      \label{hcv_tableSourcesBelowThreshold} 
      \begin{tabular}{llrlllcrlllc}
\hline\hline
R.A.		& Dec.	       &\texttt{MatchID} &\texttt{GroupID}  &Subgroup$^a$	&\# Filters & Filter   &n$_{LC}$         &<m$_\textrm{HSC}$>    &<m$_\textrm{HCV}$> &MAD &$\chi_{\rm red}^{2}$ \\
(J2000)	 	&(J2000)       &		 &		    &	        &	     &                & 	        &(mag)	       &(mag)	   &    &         \\
\hline				
0.3962652087	&-15.4694614410	&38868203	&1062949	&-5	&2	&WFC3\_F475W	&12	&26.224	&26.224	&0.050	&0.2   \\
0.3999678493	&-15.4831781387	&41866145	&1062949	&-5	&2	&WFC3\_F475W	&12	&25.431	&25.431	&0.027	&0.1   \\
0.4000068605	&-15.4822616577	&30810204	&1062949	&-5	&1	&WFC3\_F475W	&9	&26.158	&26.158	&0.051	&0.3   \\
0.4021478295	&-15.4797763824	&95901547	&1062949	&-5	&2	&WFC3\_F475W	&12	&25.889	&25.889	&0.024	&0.1   \\
0.4022069871	&-15.4806766510	&100130418	&1062949	&-5	&2	&WFC3\_F475W	&12	&26.216	&26.214	&0.046	&0.2   \\
0.4026675820	&-15.4820632935	&37320200	&1062949	&-5	&2	&WFC3\_F475W	&12	&26.001	&26.000	&0.033	&0.2   \\
0.4029722214	&-15.4806251526	&42893121	&1062949	&-5	&2	&WFC3\_F475W	&11	&25.806	&25.806	&0.083	&2.5   \\
0.4030649364	&-15.4813785553	&46076923	&1062949	&-5	&1	&WFC3\_F814W	&10	&25.499	&25.499	&0.015	&0.2   \\
0.4032958448	&-15.4865312576	&37088064	&1062949	&-5	&1	&WFC3\_F814W	&10	&25.144	&25.144	&0.023	&0.2   \\
0.4034545422	&-15.4482450485	&77175405	&1062949	&-5	&2	&WFC3\_F475W	&11	&26.233	&26.233	&0.021	&0.0   \\
\hline
\end{tabular} 
\end{minipage}
\tablefoot{This table is available in its entirety in machine-readable and Virtual Observatory (VO) forms in the online journal. A portion is shown here for guidance regarding its form and content.\\
\tablefoottext{a}{A value of "-5" denotes that the \texttt{GroupID} was processed as one subgroup.}\\
}
\end{table*}
\end{landscape}